\documentclass{article}
\usepackage{epsfig}
\def\ordermb2{\mbox{${\cal O}(1/m_b^2)$}}
\def\eps{\epsilon}
\def\Diamond{\diamond}
\def\dilog{{\mbox {Li}}_2\,}
\def\xb{{\bar{x}}}
\def\xbar{{\bar{x}}}

\def\lnlamg{{\log\lambda_G}}
\def\diff{\mbox{d}}
\def\Lbar{\bar{\Lambda}}
\def\lam1{\lambda_1}
\def\logpm{\ln\frac{1+t}{1-t}}
\def\LBar{\mbox{$\bar{\Lambda}$}}
\def\Ljeden{\mbox{$\lambda_1$}}
\def\GeV{\mbox{ GeV}}
\def\xnu{{x_{\nu}}}
\def\xnuz{{x_{\nu0}}}
\def\kmod{|\vec{k}|}
\def\BaBar{Harrison:1998yr}
\def\JM{Jezabek:1996ia,Jezabek:1996db}

\def\FLNN{Falk:1994gw}
\def\BKPS{Balk:1994sz}

\begin{document}
\pagestyle{empty}
\begin{center}
{\Huge\bf A study of semileptonic $B$ decays}
\end{center}
\vspace{2cm}
\begin{center}
\Large {Doctoral Thesis of Piotr Urban}
\end{center}
\begin{center}
\large{June 2000}\\
\vspace{1cm}
\Large {Institute of Physics, University of Silesia, Katowice, Poland,\\
	Institute of Nuclear Physics, Cracow, Poland.}
\end{center}
\pagebreak
\ \ \ \\
\begin{flushright}
{\Large I would like to thank my advisor,\\
 Professor Marek Je\.zabek,\\
for essential help with this work.}
\end{flushright}
\ \ \ \\
\pagebreak
\begin{titlepage}
\title{\bf A study of semileptonic $B$ decays.}
\author{Piotr Urban}
\date{}
\maketitle
\begin{center}
Institute of Physics, University of Silesia,\\
 ul. Uniwersytecka 4, 
PL-40007 Katowice, Poland.
\end{center}
\vspace{2cm}
\abstract{Perturbative and nonperturbative QCD corrections to
semileptonic $B$ decay rates are  calculated and given in analytical
and numerical forms. We present the analytic expression for the double
distribution in hadronic mass and electron energy in $B\rightarrow X_u
e \bar{\nu_e}$ decays through one loop order of QCD. We then study the
leading twist corrections to that quantity within the formalism of
HQET and employ it in a proposed method of measuring $V_{ub}$ with an
accuracy of $10\%$. Our other point of interest is the polarisation of
the $\tau$ lepton in $B\rightarrow X_c \tau \bar{\nu_{\tau}}$
processes. Analytic formulae are given for the one-loop QCD correction
to the longitudinal polarisation. The resulting moments of
distributions are discussed, incorporating the HQET corrections found
by Falk et al. The polarisation receives only a very small
perturbative correction. It can be used in extracting the quark
masses. Experimental considerations led us to the calculation of 
corrections to the polarisation of $\tau$ with respect to the momentum
of the virtual $W$ boson. Analytical results for the ${\cal
O}(1/m_b^2)$ terms have been combined with numerically evaluated
perturbative contributions to give the corrected polarisation.}
\end{titlepage}

\tableofcontents
\pagestyle{plain}
\pagebreak
\pagestyle{plain}
\section{Introduction}
%
%
Semileptonic $B$ decays form one of the most vigorously expanding
branch of the studies of the Standard Model (SM), the current theory of elementary
interactions \cite{Glashow:1961tr,Weinberg:1967tq,Salam} that has
stood the test of some 20 years of experimental verification. This
model is based on the concept of quantum field theory, developped at
the time of creation  
of quantum electrodynamics 
\cite{Feynman:1949zx,Schwinger:1948yk,Schwinger:1948yj}. It has thus
built on  and extended the perturbative
\cite{Behrends:1956mb,Berman:1958ti,Kinoshita:1959ha,Kinoshita:1962ur,Lee:1964is}
as well as nonperturbative  \cite{Shifman:1987sm,Shifman:1988rj,Politzer:1988wp,Politzer:1988bs,Isgur:1989vq,Isgur:1990ed,Eichten:1990zv,Eichten:1990vp,Grinstein:1990mj,Georgi:1990um,Neubert:1994ch}
methods of field theory. For all its success, however, SM is far from
completely describing the fundamental laws of nature. Even if we
refrain from listing the nondiscovery of the Higgs boson as a
liability, we are nonetheless left with a plethora of questions of
theoretical nature. One of them is the origin of the values of the
masses entering the model as well as the no less enigmatic entries of
the Cabibbo-Kobayashi-Maskawa mixing matrix. The latter are connected
with the topical problem of $CP$ violation. Whereas there exist theories that
attempt to provide a deeper insight on these, like supersymmetry and
superstrings, the main obstacle often
does not seem to lie so much with any lack of ideas as with exact
measurement and precise theoretical prediction. The latter issue is
due to the high level of complication characterising SM (not to
mention any extension thereof) as well as the 
relatively large value of the strong coupling constant. The strong
interactions cause trouble even in the description of the so-called
weak processes due to the quantum
structure of the theory, making their appearance as higher order
corrections.	 It is there that the semileptonic $B$ decays 
belong. The QCD corrections  bring about a change of the decay rate of 
up to $20\%$, so it is important to take them into account in any
description aspiring to realism.
%
%
\par
This paper is concerned with making some progress towards the
determination of some of the SM parameters by calculating QCD
corrections to semileptonic $B$ decays, which serve as the field of
many tests of the model and are currently under heavy experimental
investigation. Before reviewing the methods and previous results, let
us  delineate  the objectives of these calculations. 
\par
One of them is the determination of the $V_{ub}$ matrix element. The
entries of the $CKM$ matrix are of great import for the understanding
of the physics behind the weak processes and theories have been
proposed to disentagle the observed structure \cite{Froggatt:1979nt,Fritzsch:1979zq,Dimopoulos:1992yz,Fritzsch:1997fw}. If the matrix one day proves 
not to be unitary, which assertion would require precision measurements, it will 
be an undeniable sign of the existence of hitherto unknown species of particles. A
more tangible problem related to these elements is the $CP$
violation. It is due to none other than precise measurements
\cite{Fanti:1999nm,Alavi-Harati:1999xp}  that we now know that $CP$ is 
directly violated. The value of $V_{ub}$ has always been deducted from 
semileptonic $B$ decays and we propose a way to establish it up to an
uncertainty of $10\%$. 
\par
Our effort went also into finding the corrections to the polarisation
of the $\tau$ lepton produced in $B$ decays. Polarisation is a
quantity blessed, a least potentially, with a smaller
dependence on parameters compared to total rates. This is gorgeously
shown in the case of the longitudinal polarisation of the charged
lepton: even though the rate gets corrected by a fifth of its tree
level value, the
polarisation is never affected by more than one percent. Such a
quantity is an excellent candidate for determining the quark
masses. These  masses have been notorious for their haziness but the
difference between the beautiful and charmed quarks is rather well
defined. On the other hand, making predictions on absolute rates
involves the masses themselves and is rather problematic. We will have 
more to say about that in the discussion of nonperturbative
corrections.
%
%
\par
Since we are unable to perform an exact calculation of the
semileptonic decay rate (or any, for that matter), we must resort to
some approximation or other. The zeroth order approximation results in 
the Born term. This is supplemented by corrections which can be
divided into perturbative and nonperturbative. The perturbative ones
are valid as long as the decaying particle can be viewed as almost
free. Such an assumption is viable in $B$ decay processes because of
the high momenta transferred and the asymptotic freedom of partons in
QCD. As mentioned above, the ensuing corrections are substantial and
their inclusion is a must for an accurate prediction. The technique
employed in these calculation is that of Feynman diagrams. The
accuracy achieved in such a way can be estimated by looking at the
relative size of some higher order quantities or at the various scale
prescriptions giving what they claim to be the optimal value of the
gauge coupling. Unfortunately, the evidence is not uplifting for $B$
decays. The total rate gets a two loop correction of a size comparable
to that of the one loop one, and the scale predicted by BLM scheme
\cite{Brodsky:1983gc}  
is rather low, too.
\par
At the same time, use has been made of the natural small parameter
available in the description of the $B$ decays, which is the ratio of
the QCD scale to the released energy, the latter of the order of the
$b$ quark mass. The methods exploiting this fact are known as
nonperturbative. The early phenomenological  approach to explicitly
deal with $B$ decays was proposed by Altarelli et
al. \cite{Altarelli:1982kh}, who considered the probability of the $b$ 
quark inside the meson to have a given kinetic energy as a function of 
this energy. The decay was to be calculated incoherently, which is
plausible due to the large energy released during decay. This attempt, 
referred to as ACCMM,
was later criticised by the proponents of the heavy quark effective
theory, HQET, as failing to capture QCD-specific features of the decay 
process. HQET was developped \cite{Chay:1990da} as a systematic
expansion in terms of phenomenological parameters of the order of
inverse quark mass.  Later on, another type of expansion
appeared \cite{Neubert:1994ch}, with the first nontrivial  term corresponding to
the leading twist approximation as used in deep inelastic
scattering. The latter method emerged after the realisation of the
difficulties that the mass expansion had had. In the end, the discrepancy
between the phenomenological method and HQET has been seen
diminishing. In our calculations we employ the HQET methods, but we take
an interesting look at the kinematics of the process, using a
generalised ACCMM model. Neither method is too illuminating about the
mass of the quark in question. In HQET, the mass expansion can be
challenged by questioning the notion of quark mass  as a physical quantity. In
ACCMM, this problem is even less well defined because of the arbitrary 
character of the incoherent integration. Of course we have to make
specific predictions so we will state our assumptions clearly in
Sec.\ref{Extraction}. 
%
%
\par
Our work is a continuation of the research done in this field. The
semileptonic decays of $B$ mesons have been investigated for some
time using both perturbative and nonperturbative methods. As regards
the total decay rate, the analogy to muon decay made it possible to
use the old results \cite{Berman:1958ti}, which were based on the
calculation of radiative corrections performed in
\cite{Behrends:1956mb}.
The ${\cal
O}(\alpha_s)$ corrections to the lepton spectra were first given in a
correct, analytical  form in \cite{Jezabek:1989ja} (while numerical
estimates were given in \cite{Ali:1979is}) and then
distributions taking into account the $b$ quark polarisation were
found \cite{Czarnecki:1991pe}. The lepton mass was taken into account
in \cite{Czarnecki:1995bn,Jezabek:1996ia,Jezabek:1996db}, allowing to consider semitauonic
decays. In  \cite{Czarnecki:1995bn}, the one-loop order correction was
also found to the longitudinal polarisation of the  $\tau$ lepton in
the rest frame of the virtual $W$ boson. The polarisation turned out
to be only slightly affected by the radiative corrections. We have
found the analytic formulae for the longitudinal polarisation in the
rest frame of the decaying quark in an analytic
form, which is presented in this paper.
\par
 At the same time, the
non-perturbative corrections were calculated. As the heavy quark
effective theory arose together with the application of the operator
product expansion, inclusive lepton spectra were considered, first in
\cite{Chay:1990da}. The calculation of the average hadronic mass
squared discussed in that paper already suggested that the expansion
in inverse powers of the decaying quark mass may start with the
quadratic, rather than linear, terms. This statement, which in fact
needed to be 
qualified and is now known as Luke's theorem \cite{Luke:1992cs}, is true for the
inclusive leptonic spectra, as is seen in particular from the results
presented here. The actual non-trivial HQET corrections to the
leptonic spectra were first found in \cite{Bigi:1993fe} and the work
was then extended to double differential spectra in electron energy
and leptonic invariant mass squared
\cite{Blok:1994va,Manohar:1994qn}. All these results neglected the
mass of the final charged lepton. The latter was considered in
\cite{\BKPS,Koyrakh:1994pq,Koyrakh:1996fr}, where the inclusive lepton spectra were
calculated. In \cite{Falk:1994gw}, the longitudinal polarisation of
the $\tau$ lepton was found through order $1/m_b^2$. 
We apply the inverse mass expansion to evaluating the nonperturbative
corrections to the polarisation of the tau lepton with respect to the
momentum of the virtual $W$ boson. The latter is advantageous from
the experimental point of view
\cite{Rozanska:1998me,Kiers:1997zt}. This HQET calculation shows some
apparent divergencies, which however cancel. A similar, although
different, problem was addressed in \cite{Gremm:1995rv}.
\par
The determination of the $V_{ub}$ matrix element is also about finding
the perturbative and nonperturbative corrections. We have already
mentioned the basic papers
\cite{Berman:1958ti,Behrends:1956mb,Jezabek:1989ja} concerning the
perturbative part. On the other hand, the HQET provided the inverse
mass expansions
\cite{Chay:1990da,Bigi:1993fe,Blok:1994va,Manohar:1994qn}. However, it
was clear from the beginning \cite{Bigi:1993fe} that this kind of
expansion suffers from singularities close to the high endpoint of the
lepton energy spectrum. Usually this problem may be partially solved
by resorting to moments. However, in the case of the charmless decays, it is
exactly this part of the spectrum that is cleanly distinguishable from
the much more copious charmed decays and therefore we would rather it were
treatable on a more detailed basis. One method designed to sidestep these
difficulties was the so-called leading twist expansion. We take this
approach and propose a quantity that will enable measuring $V_{ub}$
with an accuracy of $10\%$.
%
%
\par
This dissertation comprises several results regarding the polarisation in
semileptonic $B$ decays and the determination of the $V_{ub}$ matrix
element. Apart from the leading twist convolution of the $b\rightarrow 
ue\bar{\nu_{e}}$ rate, these results have been published before. The
one-loop QCD correction to the longitudinal $\tau$ polarisation has
been published in \cite{Jezabek:1997rk}. The paper \cite{Urban:1998zs}
contains the Born term of the polarisation of the $\tau$ lepton with
respect to the momentum of the virtual $W$ boson, while the ${\cal
O}(m_b^2)$ corrections to it have been calculated in
\cite{Jezabek:1999vg}. Lastly, the double differential distribution in
terms of the hadronic mass and electron energy has been found through
one loop order in \cite{Jezabek:1999mn}.
\par
The paper is organised as following.  We start with the kinematics of
the considered processes, in   
Sec.\ref{Kinematics}. Then we discuss the types of corrections involved in the
calculation, Sec. \ref{EvalofCorr}. At the end of that section, we
devote some space to the discussion of the Fermi motion inside a $B$
meson. In Sec. \ref{Double} we present
the one loop correction to the double distribution in hadronic mass
and electron energy. Sec. \ref{Extraction} is devoted to the
discussion of this quantity in the context of measuring $V_{ub}$. It
presents results taking into account the leading twist
corrections. The polarisation of the $\tau$ lepton in $B\rightarrow
X_c \tau \bar{\nu_{\tau}}$ decays is considered in
Secs.\ref{Longitudinal} and \ref{WPOL}. First, we give the one loop
corrections to the longitudinal polarisation and then, in
Sec. \ref{WPOL}, the nonperturbative corrections to the polarisation
with respect to the momentum of the virtual $W$ boson are given in an
analytic form, while the inclusion of the corresponding perturbative
contribution is performed numerically.
\pagebreak
\section{Kinematics}\label{Kinematics}
In this section, we wish to exhaust all the kinematic information about the processes
that are going to be discussed further on. Except for the calculation of
the \ordermb2 corrections to the polarisation of the $\tau$ (see
Sec.\ref{WPOL}), the results presented below are found in the rest frame
of the decaying $b$ quark. While this is rather obvious for the
perturbative corrections, the leading twist corrections which are used
also assume a static quark. The Fermi motion is reflected solely in
the variable effective mass of the quark. One notes, moreover, that
even the inverse mass expansion leads ultimately to kinematical ranges 
identical to the partonic ones.

\subsection{Variables}
The four-momenta of  the particles are denoted as following:\\
\begin{tabular}{ll}
$Q$ & for the decaying quark,\\
$q$ & for the final quark (charm or up),\\
$l$ & for the charged lepton, or\\
$\tau$ & specifically for the $\tau$ lepton,\\
$\nu$ & for the antineutrino, and\\
$G$ & for the final real gluon.
\end{tabular} 
\par
The system of the final quark and gluon has its momentum denoted as
$P=q+G$, while the leptonic momentum is written as $W=l+\nu$.
Then the scaled masses of the final particles are written in terms of
the parameters defined as
\begin{equation}
\eta=\frac{\tau^2}{m_b^2},\qquad \rho\equiv\eps^2=\frac{q^2}{m_b^2}.
\end{equation}
The  particles are assumed to be on shell.
 The following scaled variables have been employed in the
partonic part of the calculations:
\begin{equation}
x=\frac{2 v\cdot l}{m_b},\qquad t=\frac{(l+\nu)^2}{m_b^2},\qquad
z=\frac{(q+G)^2}{m_b^2},\qquad x_p= \frac{2v\cdot P}{m_b},
\end{equation}
where $v$ stands for the $B$ meson four-velocity in its rest
frame. For the perturbative calculation, the rest frame of the meson
and the $b$ quark coincide and then one can write,   
\begin{equation}
x=\frac{2(Q\cdot l)}{m_b^2},\qquad t=\frac{(l+\nu)^2}{m_b^2},\qquad
z=\frac{(q+G)^2}{m_b^2},\qquad x_p=\frac{2Q\cdot P}{m_b^2}
\end{equation}
where $x$ stands for the charged lepton energy, $t$ for the
invariant leptonic mass,  $z$ for the hadronic mass, and $x_p$ for the
hadronic energy,  all scaled to the units of $b$ quark mass. We also
define the scaled neutrino energy, $x_{\nu}$, but we reserve this
symbol for the energy in the $B$ meson rest frame. Denoting the meson
four-velocity as $v$, we have
\begin{equation}\label{xnudef}
x_{\nu}=\frac{2 v\cdot \nu}{m_b^2}.
\end{equation}
\par
 The charged lepton
 is described by the light-cone variables:
\begin{equation}
\tau_{\pm}={1 \over 2}(x \pm \sqrt{x^{2}-4\eta}).
\end{equation}
 Thus, the system of the $c$ quark and the real gluon is described by the following 
quantities:
\begin{eqnarray}
P_{0}&=&{1 \over 2}(1-t+z)\\
P_{3}&=&\sqrt{P_{0}^{2}-z}={1 \over 2}[1+t^{2}+z^{2}-2(t+z+tz)]^{1 \over 2},\\
P_{\pm}(z)&=&P_{0}(z)\pm P_3(z),\\
{\cal Y}_p&=&{1 \over 2}\ln{{P_+(z)} \over {P_-(z)}}=\ln{{P_+(z)} \over {\sqrt{z}}}
\end{eqnarray}
where $P_0(z)$ and $P_3(z)$ are the energy and the length of the momentum vector
of the system in the $b$ quark rest frame, ${\cal Y}_p(z)$ is the corresponding 
rapidity. Similarly for the virtual boson $W$:
\begin{eqnarray}
W_0(z)={1 \over 2}(1+t-z),\\
W_3(z)=\sqrt{W_0^2-t}={1 \over 2}[1+t^2+z^2-2(t+z+tz)]^{1/2},\\
W_{\pm}(z)=W_0(z)\pm W_3(z),\\
{\cal Y}_w(z)={1 \over 2}\ln{{W_+(z)} \over {W_-(z)}}=\ln{{W_+(z)} \over \sqrt{t}}.
\end{eqnarray}
 Kinematically, the three body decay is a special case of the four body one, 
with the four-momentum of the gluon set to zero, thus resulting in simply
replacing $z=\rho$. The following variables are then useful:
\begin {eqnarray}
p_0=P_0(\rho)={1 \over 2}(1-t+\rho)&,& p_3=P_3(\rho) = 
\sqrt{p_0^2-\rho},\\
p_{\pm}=P_{\pm}(\rho) = p_0\pm p_3&,&  w_{\pm}=W_{\pm}(\rho) = 1-p_{\mp},\\
Y_p={\cal Y}_p(\rho) = {1 \over 2}\ln{{p_+} \over {p_-}}&,&  Y_w={\cal 
Y}_w(\rho) = {1 \over 2}\ln{{w_+} \over {w_-}}. \end{eqnarray}
 We also express the scalar products in terms of the variables used above, so
in the units of the $b$ quark mass one gets:
\begin{eqnarray}
Q\cdot P={1 \over 2}(1+z-t)&,& \tau\cdot\nu ={1 \over 2}(t-\eta),
\nonumber \\
Q\cdot \nu={1 \over 2}(1-z-x+t)&,& \tau \cdot P={1\over 2}(x-t-\eta),
\nonumber \\
Q\cdot \tau={1 \over 2}x&,&\nu\cdot q={1 \over 2}(1-x-z+\eta).
\end{eqnarray}
\subsection{Kinematic boundaries}
The boundaries of the kinematic variables depend on the process which we
are dealing with. For the partonic part, one can distinguish between
three and four body final states, the latter including a real gluon. A
still different set of limits is obtained if one allows for Fermi
motion, so that the hadronic entities enter the game. This motion is
taken into account in the computation of the \ordermb2 and leading
twist corrections within
the heavy quark effective theory. It is worth noting that these
expansions have the net effect of formally leaving the $b$ quark static.
\subsubsection{Partonic limits}
The scaled electron energy $x$ can vary within the limits
\begin{equation}
2\sqrt{\eta}\le x \le 1.
\end{equation}
With a fixed value of $x$, the limits of $t$ are
\begin{equation}
\eta \le t\le {\tau}_+(1-{\rho \over 
{1-{\tau}_+}})\equiv t_2.
\end{equation}
However, for the case of a final charm quark this range is
conveniently split into
two regions. In one of them, which we call $A$, both three-body and
four-body processes are allowed, while in the other, $B$, only four-body
final states are possible. The region $A$ is delimited by
\begin{equation}
t_1\equiv {\tau}_-(1-{\rho \over {1-{\tau}_-}})\le t\le t_2,
\end{equation}
so that the remaining room of the phase space, $B$, corresponds to
\begin{equation}
\eta \le t\le t_1.
\end{equation}
Conversely, if x should vary at a fixed value of t, the boundaries read:
\begin{equation}
\eta\le t\le (1-\sqrt{\rho})^2,\qquad w_-+{\eta \over {w_-}}\le x\le 
w_++{\eta \over {w_+}}, \end{equation}
for region A, and
\begin{equation}
\eta\le t\le \sqrt{\eta}(1-{\rho\over{1-\sqrt{\eta}}}),\qquad 
2\sqrt{\eta}\le x\le w_-+{\eta\over{w_-}}. \end{equation}
for region B.
The upper limit of the mass squared of the final hadronic system is in 
both regions given by
\begin{equation}
z_{max}=(1-{\tau}_+)(1-t/{\tau}_+),
\end{equation}
whereas the lower limit depends on the region:
\begin{equation}
z_{min}=\left\{
\begin{array}{ll}
\rho                        & \rm{ in\: Region\: A}\\
(1-{\tau}_-)(1-t/{\tau}_-)  & \rm{ in\: Region\: B}
\end{array}
\right.
\end{equation}
\subsubsection{Partonic limits - massless case}
For the $b\rightarrow ue\bar{\nu_e}$ decay, the masses of the final
particles may be treated as vanishing. Then the above kinematic limits
simplify considerably. The electron energy $x$ varies within the range
of
\begin{equation}
0\le x \le 1,
\end{equation}
and for a fixed $x$, the invariant mass of hadrons can be
\begin{equation}
0\le z \le 1-x.
\end{equation}
If we fix $z$ first, it can take on values from
\begin{equation}
0 \le z \le 1,
\end{equation}
and then the limits on $x$ are
\begin{equation}
0 \le x \le 1-z.
\end{equation}
We will also use a triple distribution in terms of electron energy $x$,
hadronic energy $x_p$ and hadronic invariant mass $z$. With a fixed
value of $x$,
\begin{equation}
1-x \le x_p \le 2-x,\qquad {\mbox max}\{0,x_p-1\} \le z \le
(1-x)(x_p+x-1).
\end{equation}
As mentioned above, for massless final particles,  the distinction
between the region where only four 
body final state is allowed and the rest disappears when speaking in terms of
variables $x,t,z$ fixed in this order. However, with the hadronic energy
$x_p$ used as a variable, an analogous division holds. It is now
reflected in the fact that the values of $x_p>1$ correspond to
four-body final state. 
\subsubsection{Meson kinematic boundaries}
In the evaluation of the \ordermb2 corrections, it is necessary to
work out the kinematics in the rest frame of the decaying meson. Let
us denote the four-velocity of the decaying quark by $v$. The neutrino
energy is no longer fixed by the other variables, as it is in the
partonic calculation. Rather, it becomes another kinematic
variable, see \cite{Balk:1994sz} for the boundaries. The partonic value of
the neutrino energy is given by the relation
\begin{equation}
x_{\nu 0}=1+t-x-\rho.
\end{equation}
\pagebreak
\section{Evaluation of corrections}\label{EvalofCorr}
\subsection{Perturbative corrections}\label{PertCorr}
Let us denote the final $c$ or $u$ quark generically as $q$ and the
charged
lepton as $l$.
The QCD-corrected differential rate for the $b\rightarrow q+l_-+\bar{\nu_l} $ reads:
\begin{equation}
d\Gamma ^{\pm}=d{\Gamma}_0^{\pm}+d{\Gamma}_{1,3}^{\pm}+d{\Gamma}_{1,4}^{\pm},
\end{equation}
where
\begin{equation}
d{\Gamma}_{0}^{\pm}=G_F^2m_b^5|V_{CKM}|^2{\cal M}_{0,3}^{\pm} d{\cal 
R}_3(Q;q,\tau,\nu)/{\pi}^5 \end{equation}
is the Born approximation, while
\begin{equation}
d{\Gamma}_{1,3}^{\pm}={2 \over 3}{\alpha}_sG_F^2m_b^5|V_{CKM}|^2{\cal 
M}_{1,3}^{\pm}d{\cal R}_3(Q;q,\tau,\nu)/{\pi}^6 \end{equation}
comes from the interference between the virtual gluon and Born amplitudes. 
Then, \begin{equation}
d{\Gamma}_{1,4}^{\pm}={2\over 3}{\alpha}_sG_F^2m_b^5|V_{CKM}|^2{\cal 
M}_{1,4}^{\pm}d{\cal R}_4(Q;q,G,\tau,\nu)/{\pi}^7 \end{equation}
is due to the real gluon emission, $G$ denoting the gluon 
four-momentum. $V_{CKM}$ is the Cabibbo-Kobayashi-Maskawa matrix
element corresponding to the $b$ to $c$ or $u$ quark weak transition. The Lorentz 
invariant $n$-body phase space is defined as 
\begin{equation}
d{\cal R}_n(P;p_1,...,p_n)={\delta}^{(4)}(P-\sum{p_i}){\prod}_i{{d^3{\bf p}_i} \over {2E_i}}.
\end{equation}
The superscript $\pm$ refers to the polarisation of the charged
lepton when it is taken into account.
\par

The three-body phase space is parametrized by Dalitz variables:
\begin{equation}
d{\cal R}_3(Q;q,\tau,\nu)={{{\pi}^2} \over 4}dx dt. 
\end{equation}
To be specific, consider the unpolarised rates.
The evaluation of the virtual gluon exchange matrix element yields:
\begin{eqnarray}
{\cal M}_{1,3}^{un}(\tau)=-\left[ H_0 q\cdot\tau Q\cdot\nu +H_+\rho Q\cdot\nu Q\cdot\tau +H_- q\cdot\nu q\cdot\tau \right. \nonumber  \\
+{1 \over 2}{\rho}(H_++H_-)\nu\cdot\tau +{1\over 
2}\rho(H_+-H_-+H_L)[\tau\cdot{(Q-q-\nu)}] (Q\cdot\nu)\nonumber\\ 
\left.-{1\over 2}H_L[\tau\cdot{(Q-q-\nu)}] (q\cdot\nu) \right],
\end{eqnarray}
where
\begin{eqnarray}
\nonumber H_0=4(1-Y_pp_0/p_3)\ln{\lambda}_G+(2p_0/p_3)[Li_2(1-{{p_-w_-}\over {p_+w_+}})\\
\nonumber -Li_2(1-{{w_-}\over {w_+}})
-Y_p(Y_p+1)+2(\ln\sqrt{\rho}+Y_p)(Y_w+Y_p)]\\
+[2p_3Y_p+(1-\rho-2t)\ln\sqrt{\rho}]/t+4,\\
H_{\pm}={1\over 2}[1\pm (1-\rho)/t]Y_p/p_3\pm {1\over t}\ln\sqrt{\rho},\\
H_L={1\over t}(1-\ln\sqrt{\rho})+{{1-\rho}\over {t^2}}\ln\sqrt{\rho}+{2\over {t^2}}Y_pp_3+{{\rho} \over t}{{Y_p} \over {p_3}},
\end{eqnarray}
and the superscript ''un'' stands for the unpolarised rate.
After renormalization, the virtual correction ${\cal M}_{1,3}^{un}$ is ultraviolet convergent. However, the infrared 
divergences are left. They are regularized by a small mass of gluon denoted
as ${\lambda}_G$. In accordance with the Kinoshita-Lee-Nauenberg
theorem \cite{Kinoshita:1962ur,Lee:1964is},this
divergence cancels out when the real emission is taken into account.
 The rate from real gluon emission is evaluated by integrating the expression
\begin{equation}
{\cal M}_{1,4}^{un}(\tau)={{{\cal B}_1(\tau)}\over {(Q\cdot G)^2}}-{ {{\cal B}_2(\tau)}\over {Q\cdot G P\cdot G}}+{ {{\cal B}_3(\tau)} \over {(P\cdot G)^2}},
\end{equation}
where
\begin{eqnarray}
{\cal B}_1(\tau)=q\cdot \tau[Q\cdot\nu(Q\cdot G-1)+G\cdot\nu-Q\cdot\nu Q\cdot G],\\
{\cal B}_2(\tau)=q\cdot \tau[G\cdot\nu -q\cdot\nu Q\cdot G+Q\cdot\nu(q\cdot G-Q\cdot G-2q\cdot Q)]\nonumber\\
+Q\cdot\nu(Q\cdot\tau q\cdot G-G\cdot\tau q\cdot Q),\\
{\cal B}_3(\tau)=Q\cdot\nu(G\cdot\tau q\cdot G-\rho\tau\cdot P).
\end{eqnarray}
\par
The polarized case requires an appropriate substition of the lepton
four-momentum $\tau$ present in the formulae above by a combination of
four-momenta specific for a sought polarisation. The expressions
defining these relations are given in the following sections. It will
also be seen that this is true both for the Born approximation and the
corrections affecting the hadronic tensor.
\par
The four-body phase space is decomposed as follows:
\begin{equation}
d{\cal R}_4(Q;q,G,\tau,\nu)=dz d{\cal R}_3(Q;P,\tau,\nu)d{\cal R}_2(P;q,G).
\end{equation}
After em\-ploy\-ing the Da\-litz pa\-ra\-metr\-iz\-at\-ion of the three body pha\-se spa\-ce 
${\cal R}_3$
and integration we arrive at an infrared-divergent expression. 
 
The method used in these calculations is the same as the one employed in the previous ones \cite{Jezabek:1996ia,Jezabek:1996db,Jezabek:1989ja,Czarnecki:1994pu}. The infrared-divergent
part is regularized by a small gluon mass ${\lambda}_G$ which enters into
the expressions as $\ln({\lambda}_G)$. When the three- and four-body contributions
are added the divergent terms cancel out and then the limit ${\lambda}_G \rightarrow 0$
is performed.This procedure yields well-defined double-differential distributions 
of lepton spectra as described in the following sections.
\subsection{Nonperturbative corrections}
The $b$ quark is confined within a $B$ meson so its decay cannot be completely
described if one sticks with the perturbative, partonic picture where the 
quark is treated as an almost free particle. While the large momentum transfer
accompanying the decay renders it suitable for such an analysis in a theory
that is asymptotically free, some effects remain and have to be addressed via 
nonperturbative techniques. The HQET methods have established their status as
{\em the} tool to deal with semileptonic $B$ decays and we are using them here. They
provide a systematic expansion around the limit of infinite quark masses, where
new symmetries arise, those of flavour and of spin. The expansion comes in two
kinds. It can be done in terms of consequent orders of the inverse mass of
the decaying quark, or according to the twist of the operators included. 

\subsubsection{Corrections of order $1/m_b^2$}
Using the operator expansion technique, one can obtain corrections to 
the decay widths of heavy hadrons which effectively lead to new terms in
the
hadronic tensor appearing in the triple differential decay width,
\begin{equation}\label{d3}
{{d\Gamma}\over{dx_{\nu} dt dx}}={{|V_{cb}|^2 G_F^2}\over{2\pi ^3}}
{\cal L}_{\mu\nu}{\cal W}_{\mu\nu} \quad .
\end{equation}
The hadronic tensor ${\cal W}$, related to an inclusive decay of a beautiful
hadron $H_b$,
\begin{equation}
{\cal W}_{\mu\nu}=(2\pi)^3\sum_{X}\delta ^4(p_{H_b}-q-p_X)
<H_b(v,s)|J_{\mu}^{c\dagger}|X><X|J_{\nu}^c|H_b(v,s)>
\end{equation}
can be expanded in the form
\begin{equation}\label{Wn}
{\cal W}_{\mu\nu}=-g_{\mu\nu}W_1+v_{\mu}v_{\nu}W_2-i\eps_{\mu\nu\alpha\beta}
v^{\alpha}q^{\beta}W_3+q_{\mu}q_{\nu}W_4
+(q_{\mu}v_{\nu}+v_{\mu}q_{\nu})W_5\ .
\end{equation}
The form factors $W_n$ can be determined by using the relation between the 
tensor ${\cal W}$ and the matrix element of the transition operator 
\begin{equation}
T_{\mu\nu}=-i\int d^4xe^{-iqx}T[J_{\mu}^{c\dagger}(x)J_{\nu}^c(0)]\ ,
\end{equation}
which is
\begin{equation}
{\cal W}_{\mu\nu}=-{1\over\pi} Im <H_b|T_{\mu\nu}|H_b>\ .
\end{equation}
The coefficients $W_n$ of (\ref{Wn}) have all been found elsewhere, see
eg. \cite{Balk:1994sz} for a complete list.
Then the distribution (\ref{d3}) can be schematically cast in the following
form:
\begin{equation}\label{d3a}
{{d\Gamma}\over{d\xnu dt dx}}=
f_1\delta(x_{\nu}-x_{\nu 0})+f_2\delta '(\xnu-\xnuz)+f_3\delta ''(\xnu-\xnuz)\ ,
\end{equation}
where
\begin{equation}
\xnuz=1+t-\rho-x
\end{equation}
is the value of the neutrino energy in the parton model kinematics.
The triple differential distribution must be integrated 
over the neutrino energy to give
meaningful results. The final lepton energy distribution obtained on two 
subsequent integrations may be trusted except for the endpoint region where 
the operator product expansion fails.
In the present paper we give the double differential distribution so that
the lepton energy distribution has to be obtained numerically. The calculation
does not show any features unfamiliar from the cases of the other known 
polarisations, although apparent divergences pop up.

\subsubsection{Leading twist approximation}
In certain kinematical regions, the inverse mass expansion does not work
due to the fact that the formally higer terms do not get smaller. Then it
may still be possible to perform an expansion in terms of increasing twist
of the retained operators. This technique is taken over from the deep inelastic
scattering methods. It is applicable when the assumption is valid that the
four-momentum of the final state hadrons is almost light-like. This is the
large momentum that has to occur in a twist expansion.
\par
Under that assumption, the generic quantity $\Gamma$ characterizing the $B$
 decay
can be written as
\begin{equation}\label{convolution}
\Gamma(\Xi_i) = \int dk_+ f(k_+) \Gamma^{partonic}(\xi_i)\delta[\Xi(\xi)].
\end{equation}
In the above formula, $\Gamma^{partonic}$ is the sought quantity as evaluated
within the parton model with the $b$ quark mass set to $m_b+k_+\equiv m_b^*$. The variables
characterizing the partonic system are generically denoted as $\xi$ while those
corresponding to the hadronic system by $\Xi$. The relation between
the hadron and parton system variables involves kinematics in which
the $b$ quark is viewed as static but has a variable effective mass
$m_b^*$. Thus the expression has the
form of a convolution of the partonic rate with the so-called {\em shape
 function } $f(k_+)$. This function is a nonperturbative entity, defined as
\begin{equation}
f(k_+)=<B|b \delta(D_+ - k_+) \bar{b} |B>,
\end{equation}
with $D_+$ defining the light-cone component of the covariant derivative.
\par
In the expression (\ref{convolution}) the partonic rate is evaluated
for a quark mass equal $m_b^*$, but this replacement is not made
everywhere. The prefactor $m_b^5$ is left constant. We will not
discuss this prescription, although it seems admittedly {\em ad
hoc}. Let us just note that in the ratio of decay rates, which is what
we study, this point may be believed not to cause too much trouble.
\subsubsection{Discussion of Fermi motion}
In this paper, what we use are the perturbative corrections and the
HQET methods of including Fermi motion effects. The latter are worth
discussing for a while. When it comes to differential spectra, the
leading twist approximation is the realistic way of using heavy quark
theory. In this approximation, the dependence of the kinematics on the
Fermi motion is rather formal, with the quark effectively staying at
rest and having a variable mass. We therefore think it an interesting
point to make to consider the kinematics of the Fermi motion from a
general point of view. Even though we abandon the leading twist
approximation and the kinematic conditions it depends upon, we are
still led to non-trivial conclusions.
\par
The products of the decay, the hadronic system and the two leptons, can
be characterised by three variables, which one may choose to be the
hadronic invariant mass squared $M_X^2$ and the energies of each lepton,
$E_l$ for the electron and $E_\nu$ for the neutrino. The neutrino
energy can be measured indirectly using energy-momentum
conservation. The task of including the Fermi motion consists in
performing a boost from the rest frame of the $b$ quark to that of the
decaying meson. Then one can sum incoherently over the various
configurations of the quark in the meson. This probabilistic approach is
 based on the fact that the momentum transfer in the decay is large
 enough to make an impulse approximation valid. 
We will thus relate the above introduced variables to the
ones defined in the $b$ quark rest frame. Quantities in this frame will be denoted with an
asterisk: $P^*$ for the partonic four-momentum, $E_l^*$ and $E_\nu^*$
for the lepton energies. Because of the form of the distribution
formulae, it is convenient to also employ the scaled variables,
\begin{equation}
x_l=\frac{2E_l^*}{m_b},\qquad x_\nu=\frac{2E_\nu^*}{m_b},\qquad
z=\frac{P^2}{m_b^2}.
\end{equation}
In the above definition, the mass $m_b$ of the $b$ quark will in general
vary as it is the effective quark mass which defines the kinematics of
the process. However, we will not vary the factor of $m_b^5$
multiplying the decay width since  this factor,
reflecting the phase space suppression, is probably compensated for by
the Coulomb enhancement. As a matter of fact, this mass
dependence will drop out of the ratio of the charmless to charmed
decays. 
\par
In order to specify the relation between the two frames of interest, one
needs to parametrise the Fermi motion. We do this in a general way,
making no assumptions on the specific form of the distribution of the
quark or the light degrees of freedom.
\par
Working in the rest frame of the $B$ meson, assume that the heavy quark
has three-momentum $\vec{k}$ and energy $E_b$. Forming a bound state,
the quark is generally off mass shell and its effective mass is then 
$\mu = \sqrt{E_b^2-|\vec{k}|^2}$. Neither are the light degrees of
freedom assumed to have the mass of the light quark. In fact, using
energy and momentum conservation, we write the four-momentum $K$ of the spectator as   
\begin{equation}
K=(M_B-E_b,-\vec{k})\equiv(E_s,-\vec{k}).
\end{equation}
Then we have set up the framework for performing the boost. Starting
from the system where the $b$ quark stays at rest, the boost needs to
endow it with momentum $\vec{k}$, which defines its parameters to be
\begin{equation}
\gamma = \frac{E_b}{\mu} \quad {\rm and\ \ }\vec{\beta} = \frac{\vec{k}}{E_b}.
\end{equation}
It is instructive to look closely at the transformation formulae for the
different quantities. The electron energy in the $B$ meson frame can be
written as 
\begin{equation}\label{ElEn}
E_l={1\over2}x_l(M_B-E_s+\kmod\cos \theta_l)
\end{equation}
where $\theta_l$ is the angle between the directions of the electron and
the momentum $\vec{k}$. Similarly, the neutrino energy is transformed
according to the formula
\begin{equation}\label{NeEn}
E_\nu={1\over2}x_l(M_B-E_s+\kmod\cos \theta_\nu)
\end{equation}
with $\theta_\nu$ denoting the corresponding angle for the
neutrino. Note that in each of Eqs.(\ref{ElEn},\ref{NeEn}) the reference to
the Fermi motion is encoded in one rather than two variables. These
are, respectively for the electron and neutrino, 
\begin{equation}
k_{l,+}\equiv E_s-\kmod\cos \theta_l, \qquad {\rm and\ \ }
k_{\nu,+}\equiv E_s-\kmod\cos \theta_\nu.
\end{equation}
It is easy to verify that this structure is due to the fact that the
corresponding particles are massless. The other case is relevant for the
lepton pair, which usually has a non-vanishing invariant mass, yielding
the following relation between the $B$ meson energy, $E_W$, and the
scaled energy in the quark rest frame,
\begin{equation}
x_W={1\over2}(x_l+x_\nu)=\frac{E_W}{\mu},
\end{equation}
\begin{equation}
E_W=M_B\, x_W - E_s\,x_W-\kmod\cos\theta_{W}\sqrt{(1-x_W^2)-z}.
\end{equation}
Clearly, different values of $z$ make different combinations of $E_s$
and $\kmod\cos\theta_{W}$ contribute to this expression. In the case of
the massless leptons, therefore, we can reduce the Fermi motion
dependence to a function of the variable $k_{X,+}$, $X$ denoting a
generic particle. This justifies the use of the convolution
with the light-cone shape function. On the contrary, the invariant mass
of the lepton system cannot be calculated in this way. It also obvious
that the double differential spectrum of lepton energies also has a more
general dependence since it involves two different combinations of
$\kmod$ and $E_s$. 
\par
Let us now have a look at the energy spectrum of the hadron system. It
might at first seem that as long as it can be viewed as massless the
light-cone dominance is in effect. However, one realizes easily that
this distribution is directly related to that of the $W$ energy by the
relation $M_B=E_W+E_H$, $E_H$ standing for the hadronic energy in $B$
meson rest frame. The arguments applied to massless leptons  do not work
here since the energy of the hadrons is the sum of the parton and
spectator energy. The latter inevitably is disconnected from the boost
and contributes pure energy rather than the required combination:
\begin{equation}
E_H=M_B\,x_H+E_s-x_H\,k_{+,H}.
\end{equation}
For the hadronic invariant mass, the reduction fails to hold if this
mass does not vanish:
\begin{equation}
M_X^2=E_b^2\,z+x_H\,E_b\,E_s+|\vec{p}_H|\,E_b\,\kmod\cos\theta_H+{\cal O}(\kmod^2).
\end{equation} 
\pagebreak
\section{Spectrum of hadronic  mass and electron energy}\label{Double}
\subsection{Introduction}
The spectrum of hadronic mass and electron energy is a quantity that
can be helpful in extracting the $V_{ub}$ matrix element. This section
shows the perturbative corrections to this distribution which are essential
due to their numerical value of approximately $20\%$. We will later on take
a more comprehensive approach at the spectrum of hadron mass and electron 
energy, but this partonic distribution can be useful to let us see clearly
what effects come from nonperturbative regime and is also evidently a check
on the more sophisticated calculations. Moreover, while the more
realistic calculations rely on modelling in one way or another,
moments of distributions are believed to be correctly given by
partonic results due to parton-hadron duality.
\par
We postpone the discussion of the current status of the $|V_{ub}|$
related calculations until the next section, where the leading twist
approximation is discussed. Here, we present the way we evaluate the
perturbative corrections in Subsection \ref{Double:Evaluation}, then
show the analytical results, Subsec. \ref{Double:Results} and use them 
to evaluate a few moments of the hadronic mass distribution,
Subsec. \ref{Double:Moments}. The contents of this section has been
published in \cite{Jezabek:1999mn}.
\subsection{Evaluation}\label{Double:Evaluation}
The tree level result for this spectrum can be read off from the
previously given formulae, while the one-loop corrections are evaluated
following an analogous procedure. This calculation, however, required
the evaluation of the integral over the three-body phase space of the
vertex correction as well as the integration of the four-body decay
rate. 
\par
As regards the virtual gluon contribution, integration over the invariant mass of the intermediate $W$ boson gives the 
desired contribution to the double differential distribution,
\begin{eqnarray}
\frac{d\Gamma_{1,3}}{12\Gamma_0\,dxdz}\!\!\!\!&=&\!\!\!\!\delta(z)\big\{  {1\over3}\lnlamg \big[ 10x 
- 25x^2 + {34\over3}x^3 
          +( 10 - 24x + 18x^2 - 4x^3)\log\xb \nonumber\\
&&        - 6x^2\log\eps + 4x^3\log\eps \big]
        +{1\over3}(2x^3-3x^2)\dilog(x)
        +{1\over18}\log\xb\big[121\nonumber\\
&&-276x+195x^2-40x^3           -\log\xb(30-72x+54x^2-12x^3 )\big] \nonumber\\
&&  + {121\over18}x- {443\over36}x^2+ {128\over27}x^3 +{1\over6}\log\eps(-3x^2+2x^3+6x^2\log\eps\nonumber\\
&&-4x^3\log\eps)\big\},
\end{eqnarray}
where we have denoted
\begin{equation}
\xb=1-x.
\end{equation}
While we present the decay rate assuming massless final particles, it is not
possible to eliminate the final quark as well as gluon mass dependence out of
 the matrix element alone. This is removed once the integrated real
 radiation term is added. 
 This rate
is conveniently split into terms according as they are infrared convergent or
 divergent. Hence we can write,
\begin{equation}
\frac{d\Gamma_{1,4}}{12\Gamma _0\,dx\,dt\,dz}={\cal F}_{conv}+{\cal F}_{div}.
\end{equation}
Integration over $t$ gives
\begin{equation}
\int_{0}^{t_m}{{\cal F}_{div}\,dt}={\cal F}_{div,s}\delta(z)
+{\cal F}_{div,c}\theta(z-\lambda _G^2),
\end{equation}
and
we obtain the following expressions for the above integral:
\begin{eqnarray}
{\cal F}_{div,c}&=&\log z\big[ \xb^{-1}(8z+6z^2)+\xb^{-2}(-z-3z^2)+2\xb^{-3}z^2/3\nonumber\\
&&      +\log\xb(16 - 4xz - 14x + 2x^2 + 18z + 2z^2)
        + 12xz + 16x - x^2/z \nonumber\\
&&- 7x^2 + {2\over3}x^3/z - 7z - {{11}\over3}
         z^2 \big]\nonumber\\
&&+\log\xb \big[\xb^{-1}(-16z-12z^2)+\xb^{-2}(2z+6z^2)-4\xb^{-3}z^2/3\nonumber\\
&&      +22 - 8x/z - 12xz - 36x + 6x^2/z + 10x^2 - 
         {4\over3}x^3/z + {{10}\over3}z^{-1} - 4z\big]\nonumber\\
&& + \log^2\xb  (  - 16 + 4xz + 14x - 2x^2 - 18z - 2z^2 )
       + \xb^{-1}  (  - 20z - 14z^2 )\nonumber\\
&& 
       + \xb^{-2}  ( 2z + 5z^2 )      -\frac{10}{9}\xb^{-3}z^2
        + {10\over3}xz^{-1} - 2xz + 22x - {22\over3}x^2z^{-1}\nonumber\\
&& - 9x^2  
        + {28\over9}x^3z^{-1} + 18z + {91\over9}z^2,
\end{eqnarray}
\begin{eqnarray} 
{\cal F}_{div,s}&=&-{1\over9}\big\{ 9x^2\eps^2\lnlamg 
       + \lnlamg[log(1-x)]  (  - 30 + 72x - 54x^2 + 12x^3 )\nonumber\\
&&       + \lnlamg  (  - 30x + 57x^2 - 22x^3 )
       + \lnlamg\log\frac{\lambda_G}{\eps}  ( 18x^2 - 12x^3 )\nonumber\\
&&       + \log\eps(1+\log\eps)  ( 9x^2 - 6x^3 )
       +   (  - 64 + 156x - 120x^2 + 28x^3 )\log\xb\nonumber\\
&&       +  ( 15 - 36x + 27x^2 - 6x^3 )\log^2\xb
       - 64x + 3\pi^2x^2 + 100x^2 - 2\pi^2x^3 \nonumber\\
&&- 100x^3/3\big\}.
\end{eqnarray}
\par
The infrared finite part has been integrated with the help of FORM. It gives,
in the same notation,
\begin{equation}\label{Fconvi}
\int_{0}^{t_m}{{\cal F}_{conv}\,dt}={\cal F}_{conv,s}\delta(z)
+{\cal F}_{conv,c}\theta(z-\lambda _G^2), 
\end{equation}
with
\begin{eqnarray}
{\cal F}_{conv,s} &=& \lnlamg  (  - 3x^2/2 + x^3 )
 + {1\over48}(173-360x+240x^2-44x^3)\log\xb\nonumber\\ 
&&       + {1\over8}(3x^2-2x^3)\log\eps
       + {1\over96}(346x - 657x^2 + 294x^3),
\end{eqnarray}
\begin{eqnarray}
{\cal F}_{conv,c}&=& \log z( \xb^{-1}(2z-3z^2)+\xb^{-2}z^2+9xz+16x-9x^2-2z+2z^2)\nonumber\\
&&      + \log\xb \big[\xb^{-1} ( -4z + 6z^2 )-2z^2\xb^{-2}+57/2
         - 5xz - 45x + 25x^2/2 \nonumber\\
&& - 27z - 3z^2/2 \big] + \log\xb \log\frac{\xb}{z}  (  - 16 + 2xz + 16x - 2x^2
        - 11z + z^2 )\nonumber\\
&& 
        +1/4\big[ \xb^{-1}  ( -58z + 6z^2 )      + \xb^{-2}  (  + 3z + 5z^2)- 2z^2\xb^{-3} 
      - 72xz\nonumber\\
&&  + 114x - 3x^2z^-1 - 61x^2 + 2x^3z^{-1} + 55z
       - 9z^2\big].
\end{eqnarray}
The formulae above suffer from infrared and collinear
divergences. Thence the gluon and final quark masses  
subsist as regulators in spite of the limit we have taken. Of course,
both remnant dependences vanish after the integration over the hadronic
system mass is performed, which involves summing the virtual and real
contributions. As a matter of fact, the part we have termed as
convergent, given by Eq. (\ref{Fconvi}), may be integrated on its own
and then only the logarithm of the final quark mass still remains,
 reflecting the
collinear divergence to be cancelled against a similar term in 
the virtual correction.  
\par
\subsection{Analytical results}\label{Double:Results}
The first order QCD corrected double differential decay rate can be written in
the form,
\begin{equation}\label{LoopResult}
\frac{d\Gamma}{12\Gamma _0\,dx\,dz}=f_0(x)\delta (z)+\frac{2\alpha _s}{3\pi}\big[f_1^s \delta(z)+f_1^c(x,z) \theta(z-\lambda _G^2)\big],
\end{equation}
with
\begin{equation}
\Gamma _0=\frac{G_F^2 m_b^5}{192 \pi^3},
\end{equation}
where the first term on the right hand side is the Born approximation, given by
\begin{equation}\label{BornTerm}
f_0(x)={1\over6}x^2(3-2x),
\end{equation}
while
\begin{eqnarray}\label{f1s}
f_1^s(x)&=&-4f_0(x)\log^2\lambda_G+\frac{1}{18}(120x-291x^2+130x^3)\lnlamg\nonumber\\
&& +{1\over3}(-10+24x-18x^2+4x^3)\log\xb \log\left(\frac{\xbar}{\lambda_G^2}\right)\nonumber\\
&&      +{1\over6}(83-196x+145x^2-32x^3)\log\xb\nonumber\\
&&      +{1\over{18}}(249x-426x^2+155x^3)
        -{2\over3}f_0(x)\left[\pi^2+3\dilog(x)\right],
\end{eqnarray}
and
\begin{eqnarray}\label{f1c}
f_1^c(x,z)&=&{1\over z}\big[-2f_0(x)\log z+{1\over36}\left(120x-291x^2+130
x^3\right)+{1\over3}(10-24x\nonumber\\
&&+18x^2-4x^3)\log\xb\big]+\big[(10z+3z^2)/\xb-(z+2z^2)/\xb^2\nonumber\\
&&+{2\over3}z^2/\xb^3\big]\log\frac{z}{\xb^2}
       + {1\over2}\xb^{-1}  (  - 69z - 25z^2 )
       + {1\over4}\xb^{-2}  (   11z + 25z^2 )\nonumber\\
&& 
       + \xb^{-3}  ( - 29z^2/18 )+  ( 101/2  - 17xz - 81x  + 45x^2/2\nonumber\\
 && - 31z - 3z^2/2 ) \log\xb
       +  (   32 - 6xz - 30x + 4x^2 + 29z\nonumber\\ 
&&      + z^2 )\log\xb \log\frac{z}{\xb}+\log z (21xz+32x-16x^2-9z-5z^2/3)\nonumber\\
&&       
         - 20xz + 101x/2  - 97x^2/4  + 127z/4 + 283z^2/36.
\end{eqnarray}
The above formula is easily integrated over either of the variables
 to give the single differential distributions in hadronic system 
mass or charged lepton energy.
Then expressions confirming previous calculations
\cite{Jezabek:1989ja,Falk:1996kn,Neubert:1997tv}  are found. 
The evident singularity of this distribution disappears after
 integration over the hadronic system mass. That this indeed is so,
 can be seen by expressing it in terms of the following distributions,
\begin{eqnarray}
\left( \frac{1}{z}\right)_+&=&\lim_{\lambda\rightarrow 0} \left
(  \frac{1}{z}\theta(z-\lambda) + \log \lambda \,\delta(z)\right),\\
\left( \frac{\log z}{z} \right)_+ &=& \lim_{\lambda\rightarrow 0}
\left(\frac{\log z}{z} \theta(z-\lambda) +\frac{1}{2}\log^2\!\lambda\, \delta(z)\right).
\end{eqnarray}
The substitution of these to Eqs.(\ref{f1s},\ref{f1c}) results in the formal
identification,
\begin{eqnarray}\label{Drules}
\theta(z-\lambda_G^2)\frac{1}{z}&=&\theta(z-\lambda_G^2)\left
( \frac{1}{z}\right)_+ - \delta(z)\log\lambda_G^2,\\
\label{DruleB}\theta(z-\lambda_G^2)\frac{\log z}{z}&=&\theta(z-\lambda_G^2)\left
( \frac{\log z}{z} \right)_+ - {1\over2}\log^2\lambda_G^2\delta(z).
\end{eqnarray}
Upon application of Eqs. (\ref{Drules},\ref{DruleB}) to the correction terms, the
latter take on the following form:
\begin{eqnarray}\label{f1sD}
f_1^s(x)&=& {1\over3}(-10+24x-18x^2+4x^3)\log^2\xb \nonumber\\
&&      +{1\over6}(83-196x+145x^2-32x^3)\log\xb\nonumber\\
&&      +{1\over{18}}(249x-426x^2+155x^3)
        -{2\over3}f_0(x)\left[\pi^2+3\dilog(x)\right],
\end{eqnarray}
\begin{eqnarray}\label{f1cD}
f_1^c(x,z)&=&-2f_0(x)\left( \frac{\log z}{z} \right)_+ +\left
( \frac{1}{z}\right)_+\big[{1\over36}\left(120x-291x^2+130 
x^3\right)\nonumber\\
&&+(\frac{10}{3}-8x+6x^2-\frac{4}{3}x^3)\log\xb\big]+\big[(10z+3z^2)/\xb-(z+2z^2)/\xb^2\nonumber\\
&&+{2\over3}z^2/\xb^3\big]\log\frac{z}{\xb^2}
       + {1\over2}\xb^{-1}  (  - 69z - 25z^2 )
       + {1\over4}\xb^{-2}  (   11z + 25z^2 )\nonumber\\
&& 
       + \xb^{-3}  ( - 29z^2/18 )+  ( 101/2  - 17xz - 81x  + 45x^2/2\nonumber\\
 && - 31z - 3z^2/2 ) \log\xb
       +  (   32 - 6xz - 30x + 4x^2 + 29z\nonumber\\ 
&&      + z^2 )\log\xb \log\frac{z}{\xb}+\log z (21xz+32x-16x^2-9z-5z^2/3)\nonumber\\
&&       
         - 20xz + 101x/2  - 97x^2/4  + 127z/4 + 283z^2/36.
\end{eqnarray}

Clearly, the gluon mass does not enter the integrated distribution, defined as
\begin{equation}
F(x,z)={1\over{12\Gamma
_0}}\int_0^{z}dz'{\frac{d\Gamma}{dx\,dz'}}=f_0(x)
+\frac{2\alpha_s}{3\pi}F_1(x,z)\,,
\end{equation}
for which we obtain,
\begin{equation}
F_1(x,z)=(c_1+ c_2\,z +c_3\,z^2+c_4\,z^3)\log z 
         +c_5\log ^2 z + c_6 z +c_7 z^2 + c_8 z^3 +c_9.
\end{equation}
The coefficients $c_1$ to $c_9$ are as follows,
\begin{eqnarray}
c_1 &=& {1\over{36}}\big[ ( 120 - 288x+ 216x^2 - 48x^3 )\log\xb 
               + 120x  - 291x^2 + 130x^3 \big],\nonumber\\
c_2 &=& (-30x+ 4x^2+ 32) \log\xb+ 32x- 16x^2,\nonumber\\
c_3 &=& 5{\xb}^{-1}-{1\over2}{\xb}^{-2}+{1\over2}(-6x+29)\log\xb+(21x-9)/2,\nonumber\\
c_4 &=& \xb^{-1}-{2\over3}{\xb}^{-2}+{2\over9}{\xb}^{-3}+{1\over3}\log\xb - 5/9,\nonumber\\
c_5 &=& - x^2/2 + x^3/3,\nonumber\\
c_6 &=& {1\over2}(-102x + 37x^2+ 37)\log\xb +  ( 30x -
        4x^2- 32) \log^2\xb+\nonumber\\
&& (74x- 33x^2)/4,\nonumber\\
c_7 &=& {1\over4}(-28x-91)\log\xb + {1\over2}(6x-29)\log^2\xb
        -10{\xb}^{-1}\log\xb
        -{79\over4}{\xb}^{-1}+\nonumber\\
&& {\xb}^{-2}(\log\xb+{13\over8})- {61\over4}x+
       {145\over8},\nonumber\\
c_8 &=& {1\over{36}}\big\{-22\log\xb-12\log^2\xb-72{\xb}^{-1}\log\xb
        -162{\xb}^{-1}\nonumber\\
&&+48{\xb}^{-2}\log\xb+83{\xb}^{-2}
         -16{\xb}^{-3}\log\xb-22{\xb}^{-3}+101\big\},\nonumber\\
c_9 &=& {1\over{36}}\big\{6  ( 83  - 196x + 145x^2 - 32x^3    )\log\xb
       + 12  (  - 10  + 24x \nonumber\\
&&- 18x^2 + 4x^3   )\log^2\xb       -  12\pi^2 x^2 + 8\pi^2x^3   + 498x
\nonumber\\
&&
        - 852x^2 + 310x^3   - 36x^2\dilog(x) + 24x^3\dilog(x)\big\}.
\end{eqnarray}
\subsection{Moments}\label{Double:Moments}
\begin{figure}[!]
\epsfig{file=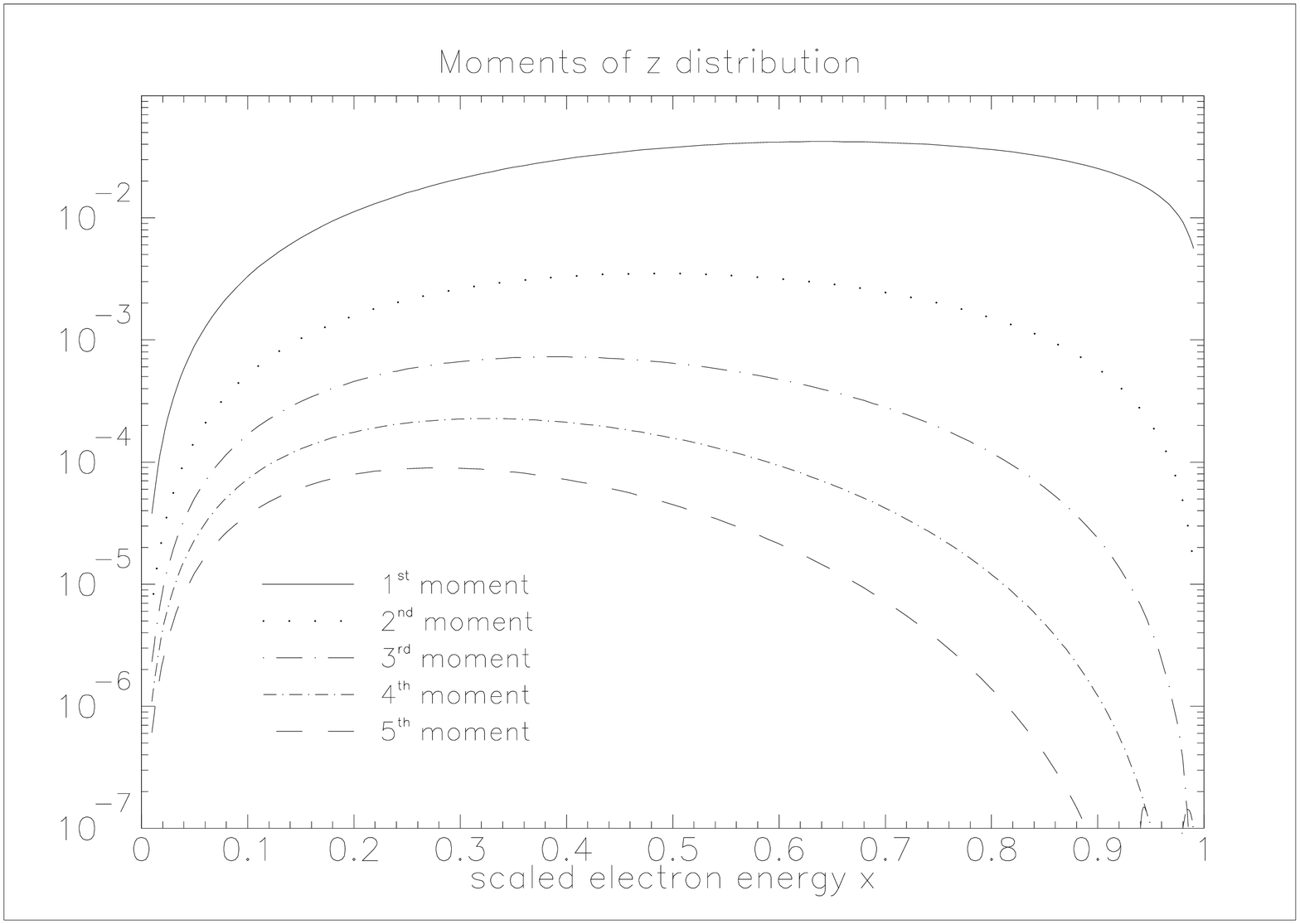,height=250pt,width=350pt}
\caption{\label{FigMoments}Moments of hadron mass distribution $m_n$ as a function of the scaled
electron energy $x$. The first five moments are shown as indicated by the legend.}
\end{figure}

One way of making comparison between the parton model predictions and the 
resonance ridden experimental data is to consider moments of distribution.
We define those as
\begin{equation}
M_n(x)={1\over{12\Gamma_0}}\int_0^{1-x}z^n\frac{d\Gamma}{dx\,dz}dz.
\end{equation}
While the zeroth moment corresponds to the electron energy distribution itself,
the singular part of the double distribution leaves no trace on the higher 
moments. In fact, they are then vanishing in the Born approximation. The first
five moments are expressed in terms of the following functions,
\begin{equation}
M_n(x)=\frac{2\alpha_s}{3\pi}m_n(x),\qquad n\ge 1,
\end{equation}
which are given by the formulae beneath (see also Fig. \ref{FigMoments}):
\begin{eqnarray}
m_1 &=&
         (  - 35/144 + 5x^2/8 - 4x^3/9 + x^4/16 ) \log\xbar \nonumber \\
&&       - 35x/144 + 19x^2/72 - x^3/48, \\
m_2 &=&   (  - 449/3600 + 7x/24 - 61x^2/360 - 2x^3/45 + 13x^4/240\nonumber \\
&&
         - 13x^5/1800 )\log\xbar    - 449x/3600 + 533x^2/1800 - 9x^3/40\nonumber \\
&& + 109x^4/1800 - x^5/144,\\
m_3 &=&   (  - 103/1800 + 119x/600 - 29x^2/120 + 19x^3/180 + x^4/
         120\nonumber\\
&& - 3x^5/200 + x^6/600 )\log\xbar      - 103x/1800 + 697x^2/3600\nonumber\\
&& - 653x^3/2700 + 49x^4/360 - 7x^5/200 + 
         47x^6/10800,\\
m_4&=&  \log\xbar  (  - 1313/44100 + 2x/15 - 81x^2/350 + 59x^3/315 - 5x^4/
         84\nonumber\\
&& - x^5/175 + 2x^6/315 - 2x^7/3675 )      - 1313x/44100 + 5729x^2/44100\nonumber\\
&&  - 11x^3/49 + 857x^4/4410  - 781x^5/8820
         + 311x^6/14700\nonumber\\
&& - 19x^7/7350,\\
m_5&=&  \log\xbar  (  - 485/28224 + 23x/245 - 211x^2/1008 + 151x^3/630\nonumber\\
&& 
          - 95x^4/672 + 2x^5/63 + 29x^6/5040 - x^7/294 + 31x^8/141120 )\nonumber\\
&&  - 485x/28224 + 1321x^2/14400 - 15879x^3/78400\nonumber\\
&& + 33599x^4/141120  - 
         4523x^5/28224 + 979x^6/15680\nonumber\\
&& - 9809x^7/705600 + 163x^8/100800.
\end{eqnarray}
\subsection{Summary}
In this section we have presented the results from
\cite{Jezabek:1999mn}, where they were found for the first time. The
analytical formulae have been shown for the  
double differential distribution in hadronic mass and electron energy
in semileptonic $B$ decays. These corrections can be meaningfully used 
on their own by calculating moments of the hadronic mass
distributions, which we have presented, too. We will extend this study 
to a more specific method of determining $|V_{ub}|$ in the next section.
\pagebreak
\section{Extraction of $|V_{ub}|$ from B decays}\label{Extraction}
\subsection{Introduction}
%
%
The elements of the Cabibbo-Kobayashi-Maskawa mixing matrix are
central to the physics of electroweak interactions. A precise
determination of these is essential in establishing the unitarity of
this matrix and so the completeness of the Standard Model, as well as
in unveiling the nature of the $CP$ violating interactions that remain 
puzzling. The beauty to up transition element, $V_{ub}$, is  one of the 
smallest and least known of the matrix. The current PDG estimate
ranges from $|V_{ub}|=0.002$ to $|V_{ub}|=0.005$. It is therefore
currently a significant problem of particle physics to obtain a
better estimate of it. As $B$ factories are now under operation, the
experimental data can be expected to offer greater accuracy so that
theoretical prediction should also aim at a higher level of sophistication.
\par
%
%
The $V_{ub}$ matrix element has long been extracted from semileptonic
$B$ decays owing to their relative cleanness. They are basically due to the weak interactions, but are
affected by chromodynamical quantum corrections. The inclusion of the
latter is vital due to their significant size of up to $20\%$. The
process at hand has been studied using perturbative and
nonperturbative techniques to take QCD effects into account.
In our present case, QED calculations are of use
\cite{Behrends:1956mb} due to the topology of the diagrams
involved. While the total decay rate of $b$ to $u$ can be inferred
from an analogous analysis of the muon decay, due to Berman
\cite{Berman:1958ti}, numerical calculations of spectra appeared in
\cite{Ali:1979is}, and the first correct analytical one-loop results
were published in \cite{Jezabek:1989ja}. 
\par
All of the calculations
mentioned so far dealt with the perturbative corrections, which are
valid due to the large momentum transfer occuring in the $B$
decays. On the other hand, nonperturbative corrections are not
negligible, but due to their relative smallness can be evaluated using 
an expansion in the scale provided by the ratio of the QCD scale to
the transferred energy. A first, phenomenological  approach was
proposed in \cite{Altarelli:1982kh}, where the $b$ quark inside a
meson was assigned a probability distribution describing its Fermi
motion. The decay was then calculated on an incoherent basis. This
method, referred to as ACCMM after the names of the article's authors,
has later been dismissed by the proponents of a more systematic
formulation, the heavy quark effective theory (HQET), as falling short 
of describing some aspects of the decays, allegedly accessible with
HQET. With a systematic expansion proposed in 
\cite{Chay:1990da}, the new method was used to calculate the first
nontrivial corrections \cite{Bigi:1993fe} to lepton spectra. These
corrections were expressed as an expansion in inverse powers of the
heavy quark mass and involved terms viewed as breaking the heavy quark 
and spin symmetry, two symmetries of the effective lagrangian. Double
differential spectra were considered in
\cite{Manohar:1994qn,Blok:1994va} while the mass of the final lepton
began to be included with the papers of Balk {\em et al.}
\cite{Balk:1994sz} and Koyrakh \cite{Koyrakh:1994pq}. 
\par
Aside from these 
inverse mass power corrections, another technique was developped
\cite{Neubert:1994ch} consisting in a leading twist approximation as
familiar from deep inelastic scattering. This new approach
stemmed from the observation that the operator product expansion (OPE),
forming the basis of the power corrections, fails at the high endpoint 
of the electron energy spectrum. This regime turned out to require a
separate treatment with new nonperturbative input which was
introduced in the form of the so-called shape function. This function
describes the light-cone distribution of the $b$ quark within $B$
meson. The light cone approximation holds as long as the final hadrons 
can be treated as nearly massless. It is worth noting that at this
point the effective theory made contact with the 
phenomenological approach. Indeed, analyses
have been done \cite{Bigi:1994it} to establish the connexion between the two.
At any rate, these essentially model-like descriptions appear to be
necessary whenever one focusses on spectra rather than restricting
oneself to moments, which are believed to be reproduced correctly from 
the mass expansion. The shape function does not depend on the final
states but rather is inherent to the $B$ meson itself. This in
particular means that information gathered from other processes can be 
used. Along these lines, the case of semileptonic $B$ decays is
related \cite{Mannel:1999gs} to radiative $b\rightarrow s\gamma$
decays.  Formally, the leading twist expansion corresponds
to resumming the most singular terms in the inverse mass series so
that it is perfectly possible to calculate both kinds of corrections
and disentangle the different terms in order to avoid double-counting
any part. One may think of those two methods as summing the
terms appearing in the full result along different paths, and we know
which they are. 
\par
In the following section we explain our choice of the method of
extracting $|V_{ub}|$. Then Sec. \ref{Combine} is devoted to the
articulation of both corrections in the desired approximation.
 Subsequently, we discuss
the shape function characterising the nonperturbative input used in
our calculations in Sec.\ref{Extraction:ShapeFunction} and then go over to the
results in Sec. \ref{Extraction:Results}.
\subsection{Method of measuring $V_{ub}$.}
\par
In our paper, we propose to use the semileptonic $B$ decays in the
determination of the $V_{ub}$ element. One of the difficulties to be
negotiated is the prevalence of charmed final states, which exceed the
charmless ones by a factor of $100$. We try to address this problem by 
imposing an upper cut on the invariant mass of the final
hadrons. Since the lowest charmed state has a mass squared of
approximately $3.5 \GeV^2$, this value suggests the placement of our
cut. Whatever the exact value chosen, this kind of cut has consequences on the methods we have to
apply. As regards perturbative QCD corrections, we use the one-loop
level spectra, which is the best result available. The question is
what nonperturbative effects should be included. As mentioned above,
there are at least two approaches to the problem. It turns out that
the cut on the hadronic mass makes the resulting distributions
sensitive to the Fermi motion of the $b$ quark within the decaying
meson. Due to this fact, we include the leading twist corrected
terms. As explained in the section discussing the HQET expansions,
this effectively consists in performing a convolution with the shape
function. Thus both the tree level and one-loop terms of the partonic
distribution are weighed with this function with the mass of the $b$
quark effectively varied.
\par
The convolution needed to obtain the leading twist approximation
requires a choice of the shape function. Since it is of a
nonperturbative nature, we need to decide on a model with parameters to be
fitted from experimental data. As regards our present knowledge about
the shape function, it is usually expressed in terms of its moments
and the difference between the meson and quark mass $\LBar=m_B-m_b$. 
These are not entirely independent quantities, however. The best known 
moment is the upper limit of the support of the shape function,
related to $\LBar$.   
The first moment
vanishes by equation of motion, while the second corresponds to the
kinetic energy of the $b$ quark in the meson and is proportional to
the parameter $\Ljeden$, which is rather poorly known to date. The
third moment is considered in the literature, but we ignore it
altogether as little is known about it quantitatively. 
Summarizing the properties of the shape function $f(m_b^*)$, we can
write the following conditions, of which the first is normalisation.
\begin{eqnarray}
&&\int_{-\infty}^{M_B}f(m_b^*)dm_b^*=1,\\
&&\int_{-\infty}^{M_B}f(m_b^*)(m_b^*-m_b)dm_b^*=0,\\
&&\int_{-\infty}^{M_B}f(m_b^*)(m_b^*-m_b)^2 dm_b^*=-\frac{1}{3}\Ljeden,\\
&&\int_{-\infty}^{M_B}f(m_b^*)(m_b^*-m_b)^2 dm_b^*=-\frac{1}{3}\rho_1.
\end{eqnarray}
In these formulae, $m_b^*$ denotes the effective mass of the $b$
quark.
Certainly,
 as one gets to know the moments of the shape function better, it
will be of 
use to include them all. This is technically reduced to examining 
 the shape function to be used in the convolution for an extended range
of parameters.
\par
For our analysis, we have chosen to use the form of the shape function 
suggested in the BaBar physics book \cite{\BaBar}, which reads 
\begin{equation}
f(m^*)=\widetilde{f}(\frac{k_+}{\Lbar})\equiv \widetilde{f}(\frac{m^*-m_b}{\Lbar}),
\end{equation}
with
\begin{equation}
\widetilde{f}(x)=N\theta(1-x)e^{c(x-1)}(1-x)^{(c-1)}.
\end{equation}
We have scanned the results obtained for a couple of parameter sets
corresponding to realistic shape functions. The error estimate of
$10\%$ which we have arrived at is based upon the range within which the
result falls subject to the choice of the nonperturbative and
perturbative parameters.
\par 
In making our choice of the quantity to be studied, we were also
guided by consideration of the mass definiton. The quark masses affect 
the decay widths rather substantially, entering with a fifth power in
the overall prefactor. In the prescription of effective mass
convolution we have skipped the point of this prefactor. Although it
is by no means a proved statement, an analogy to muon capture suggests 
that these factors, determining the total rate, should be left
constant and that is what we do. Note that this is not the case in the 
original ACCMM model. Whatever the consequences of the treatment of
mass be, however, we hope that we can ease this dependence by
considering the ratio of charmless to charmed events. 
\par
As already mentioned, our analysis involves a few parameters and two cuts.
The most important cut that is freely adjustable is that on the
invariant hadronic mass. As will be seen from the results, the greater the cut
the weaker the sensitivity to the shape function parameters gets. On
the other hand, the very reason for placing this cut, the removal of
charmed states, requires it to fall in the vicinity of $3.5
\GeV$. These two arguments are clearly contradictory and we would not
feel comfortable picking one value for all the subsequent
analysis. Rather, we have decided to consider this cut as a variable
and study the abovementioned ratio for different values of the cut. It 
will be seen that too low a cut would mar the result with wild
dependence on fine details of the nonperturbative shape function,
which we do not know that well. Eventually, the range of $2\GeV^2$
through $4\GeV^2$ will be studied.
\subsection{Combining perturbative and nonperturbative
corrections}\label{Combine} 
In order to provide means of extracting the ratio of $V_{CKM}$ elements,
one has to define a quantity that is to be matched to experimental results.
We examine the ratio
\begin{equation}\label{Ratio}
R=\frac{\Gamma_u(M_{cut}^2,E_{cut})}{\Gamma_c(E_{cut})}\equiv r\frac{|V_{ub}|^2}{|V_{cb}|^2},
\end{equation}
where the rates are defined as follows:
\begin{equation}
\Gamma_u(M_{cut}^2,E_{cut})=\int_0^{M_{cut}^2}\diff M_X^2
\int_{E_{cut}}^{M_B/2} \diff E_l \quad \frac{\diff \Gamma(B\rightarrow
  X_u e\bar{\nu_e})}{\diff M_X^2 \diff E_l},
\end{equation}
\begin{equation}
\Gamma_c(E_{cut})=\int_{E_{cut}}^{E_{max}}\frac{\diff
  \Gamma(B\rightarrow X_c e \bar{\nu_e})}{\diff E_l}.
\end{equation}
In the above formulae, $E_{cut}$ denotes the minimal electron energy,
while $M_{cut}^2$ the maximal hadronic invariant mass squared. For the
charmed decay, the maximal electron energy is given by
\begin{equation}
E_{max}=\frac{m_b^2-m_c^2}{2m_b},
\end{equation}
as it is treated in the perturbative approximation to order ${\cal
  O}(\alpha_s)$. Thus the expression for the decay to the charm quark is
  in fact taken as
\begin{equation}
\Gamma_c(E_{cut})=\int_{2E_{cut}/m_b}^{1-\rho} \frac{\diff \Gamma}{\diff
  x} \diff x,
\end{equation}
which involves partonic distributions only. The mass of the $b$ quark
which is taken for this formula is related to the parameter $\Lbar$
describing the shape function using the approximation
\begin{equation}
m_b=M_B-\Lbar,
\end{equation}
valid up to ${\cal O}(1/m_b^2)$ corrections. Thus the dependence on the
shape function actually enters both decay rates defining the ratio
(\ref{Ratio}). The rate of the decay to $X_u$ states is computed taking
into account both perturbative corrections to order $\alpha_s$ and the
leading twist non-pertrubative effects due to the Fermi motion, which
are expressed in terms of the shape function $f(m^*)$. The resulting
distribution is 
\begin{eqnarray}
\frac{\diff \Gamma}{\diff M_X^2 \diff E_l}&=&\int_0^1 \diff x
\int_{1-x}^{2-x} \diff x_p \int_{z_{min}}^{z_{max}} \diff z
\int_{-\infty}^{M_B} \diff m^* 
f(m^*) \frac{\diff \Gamma^{parton}}{\diff x \diff x_p \diff z}\times\nonumber\\
&& \delta(M_X^2-m^{*2}z-x_p\Lbar^* m^* - \Lbar^{*2}) \delta(E_l -
 \frac{m^* x}{2}).
\end{eqnarray}
The partonic rate can be split into the tree level term and the ${\cal
  O}(\alpha_s)$ correction,
\begin{equation}
\diff\Gamma^{parton}=\diff\Gamma^{parton,0}+\frac{2\alpha_s}{3\pi}
\diff\Gamma^{parton,1},
\end{equation}
and then the convoluted rate can be divided up accordingly. Using the
delta functions, the integrations involved in these rates can be
simplified, yielding,
\begin{eqnarray}
&&\int_{E_{cut}}^{M_B/2} \diff E_l \int_0^{M_{cut}^2}\diff M_X^2
 \frac{\diff \Gamma^{(0)}(B\rightarrow
  X_u e\bar{\nu_e})}{\diff M_X^2 \diff E_l}=\nonumber\\
&&\int_{E_{cut}}^{M_B/2}\diff E_l \int_{2E_l/M_B}^1 \diff x
 \int_{y_{min}}^x\diff y\,
\frac{2}{x}f\left( \frac{2E_l}{x} \right)\frac{\diff\Gamma^{parton,0}}{
 \diff x \diff y},
\end{eqnarray}
\begin{equation}
y_{min}=\max \{ 0,
\frac{1-M_{cut}^2-(M_B-2E_l/x)^2}{(2E_l/x)(M_B-2E_l/x)}\}
\end{equation}
for the tree level contribution, while the $\alpha_s$ correction
contributes
\begin{eqnarray}\label{CorrConv}
&&\int_{E_{cut}}^{M_B/2} \diff E_l \int_0^{M_{cut}^2}\diff M_X^2
 \frac{\diff \Gamma^{(1)}(B\rightarrow
  X_u e\bar{\nu_e})}{\diff M_X^2 \diff E_l}=\nonumber\\
&&\int_{E_{cut}}^{M_B/2}\diff E_l\int_{2E_l/M_B}^1\diff x \int_{1-x}^{2-x} \diff x_p
 \int_{z_{min}}^{z_{max}} \diff z\, \frac{2}{x} f\left(\frac{2E_l}{x}\right)
 \frac{\diff\Gamma^{parton,1}}{\diff x \diff x_p \diff z},
\end{eqnarray}
where 
\begin{equation}
z_{min}=\left\{
\begin{array}{cc}
  0,&{\mbox for\ } x_p \le 1,\\
  x_p-1&{\mbox for\ } x_p > 1,
\end{array}\right.
\end{equation}
and
\begin{equation}
z_{max}=\min\{ \frac{M_{cut}^2-x_p (M_B-2E_l/x)-(M_B-2E_l/x)^2}{(2E_l/x)^2},
(1-x)(x_p+x-1)\}.
\end{equation}
The integral over $z$ in Eq. (\ref{CorrConv}) has been performed
analytically so that we use the prime function $F(x,x_p,z)$ defined as
\begin{equation}
F(x,x_p,z)=\int_{z_{min}}^z \diff z' \frac{\diff \Gamma
  ^{parton,1}}{\Gamma_0\, \diff x \diff x_p \diff z'}.
\end{equation}
 \begin{equation}
F=\left\{
\begin{array}{ll}
        F_1(z)-F_1(0)+F_2(z)+F_3,&x_p<1,\nonumber\\
        F_1(z)-F_1(z_{min})+F_2(z)-F_2(z_{min})&x_p>1,
\end{array}
\right.
\end{equation}
where
\begin{equation}
z_{min}=x_p-1,
\end{equation}
and
\begin{eqnarray}
F_1&=&\frac{1}{8}\left[{1\over t}\logpm [x_p^2-4(x-1+x_p/2)^2/t^2]
+8(x-1+x_p/2)^2/t^2\right]\nonumber \\
&&\times(36-39 x_p + 12 x_p^2 - 3 x_p^3/4)+{1\over t}\logpm(x-1+x_p/2)[48 - 78 x_p\nonumber \\
&& + 45 x_p^2 - 21 x_p^3/2 + 3x_p^4/4
 + (x-1+x_p/2)  (  - 42 - 15 x_p + 27x_p^2 \nonumber\\
&&- 21/8 x_p^3 )]
        +(A_1 t + A_2)\logpm
        +A_3\log(1-t)   
        +A_4.\\
F_2&=& 12[2\ln^2 z-(8\ln x_p-7)\ln z]v(\xb-x_p),\\
F_3 &=&(x_p-\xb)\left\{ 24\ln x_p \left(-1+5v-4v\ln x_p\right)\right.\nonumber \\
&&\left.
        +v[-16\pi^2-60-48 {\mbox Li_2}(1-x_p)]\right\}. 
\end{eqnarray}
In the above formulae,
\begin{eqnarray}
A_1&=&(x_p-\xb)  ( x_p z \xb/2 + 21 x_p \xb/2 - 6 x_p 
         v - 12 x_p^2 \xb + 3 x_p^2 v + 7x_p^3\xb/4 )\nonumber\\
&&+ ( 2 x_p z v - 9 x_p z/2 - 9/10x_pz^2 - 
         x_p^2 z v + 7x_p^2z + 12x_p^2 v+ 15x_p^2/2 \nonumber \\
&&- 21x_p^3z/20 + 17x_p^3 v/2 - 105/4x_p^3 - 5/4x_p^4 v + 53/4
         x_p^4\nonumber \\
&& - 219/160x_p^5 ),\\
A_2&=&(x_p-\xb)  (  -36 x_p (1-x_p x) + 6x_p^2v + 4x_p^3\xb + 
12\xb - 96v\ln2 + 48v )\nonumber \\
&&- 24x_p + 48x_p^2 + 16x_p^3v - 60
         x_p^3 - 2x_p^4v + 26x_p^4 - 12/5x_p^5,\\
A_3&=&2A_2+96(x_p-\xb)(1+\xb-x_p)\ln(1+t),\\
A_4&=&  (x_p-\xb)  ( 5x_p z\xb - 24z\xb + 12zv )
       + 32x_pzv- 81x_pz - 21/5x_pz^2 \nonumber \\
&& - 4x_p^2zv+ 40x_p^2z - 81/20x_p^3z + 36zv + 12z - 6z^2v + 24
         z^2,
\end{eqnarray}
and
\begin{equation}
\xb=1-x,\qquad v=2-x-x_p,\qquad t=\sqrt{1-4z/x_p^2}.
\end{equation}
\subsection{Shape function parameters}\label{Extraction:ShapeFunction}
The decay rate distributions depend on the strong coupling constant, $\alpha_s$, as well as the parameters characterizing the shape function. The latter are
identified with the meson-quark mass difference $\Lbar$ and the quark kinetic 
energy inside the meson, $-{1\over3} \lambda_1$. These physical quantities
make their appearance in the shape function according to the BaBar
ansatz \cite{\BaBar},
\begin{equation}
f(m^*)=\widetilde{f}(\frac{k_+}{\Lbar})\equiv \widetilde{f}(\frac{m^*-m_b}{\Lbar}),
\end{equation}
with
\begin{equation}
\widetilde{f}(x)=N\theta(1-x)e^{c(x-1)}(1-x)^{(c-1)}.
\end{equation}
The normalizing factor $N$ is given by
\begin{equation}
N=\frac{c^c}{\Gamma(c) \Lbar},
\end{equation}
and the parameter $c$ is related to the physical ones by
\begin{equation}
c=\frac{-3\Lbar^2}{\lambda_1}.
\end{equation}
We have considered a set of parameters $\Lbar,\lam1$. They are believed to
fall in the ranges
\begin{equation}
0.3 \GeV \le \Lbar \le 0.5 \GeV, \qquad -0.6 \GeV^2 \le \lam1 \le -0.1 \GeV^2.
\end{equation}
However, we have also imposed the condition that the mean kinetic energy of
the quark should not exceed considerably the meson-quark mass difference. 
For $\alpha_s$ we have used
\begin{equation}
\alpha_s \in \{ 0.2, 0.3\},
\end{equation}
For each value of $\alpha_s$, the following sets of non-perturbative 
parameters have been examined: 
\begin{eqnarray}
\Lbar &=& 0.3 \GeV,\qquad \lam1 = -0.1 \GeV^2,\label{Set1} \\
\Lbar &=& 0.3 \GeV, \qquad \lam1 = -0.2 \GeV^2, \\
\Lbar &=& 0.3 \GeV, \qquad \lam1 = -0.25 \GeV^2, \\
\Lbar &=& 0.5 \GeV,\qquad \lam1 = -0.1 \GeV^2, \\
\Lbar &=& 0.5 \GeV, \qquad \lam1 = -0.4 \GeV^2, \\
\Lbar &=& 0.5 \GeV, \qquad \lam1 = -0.6 \GeV^2.\label{Set6}
\end{eqnarray}
The parameters of the shape function as well as the value of the
coupling constant have a twofold effect on the examined ratio $r$
since both the numerator and the denominator of the expression depend
on those numbers. The decay rate to charmless states has the
dependence encoded in the convolution with the shape function and the
perturbative corrections. The charmed rate depends on the
$\bar{\Lambda}$ parameter through the ratio of $b$ to $c$ quark masses.
The Born rate normalised to $\Gamma_0$ of $b\rightarrow c$ decay is
\begin{eqnarray}
\Gamma = 0.44064 & \mbox{\ for\ } \Lbar = 0.3 \GeV,\label{bc1},\\
\Gamma = 0.49938 & \mbox{\ for \ } \Lbar= 0.5 \GeV,\label{bc2}
\end{eqnarray}
where the electron energy cut was set to $E_{cut}=0.5 \GeV$. 
\par
\begin{figure}
\epsfig{file=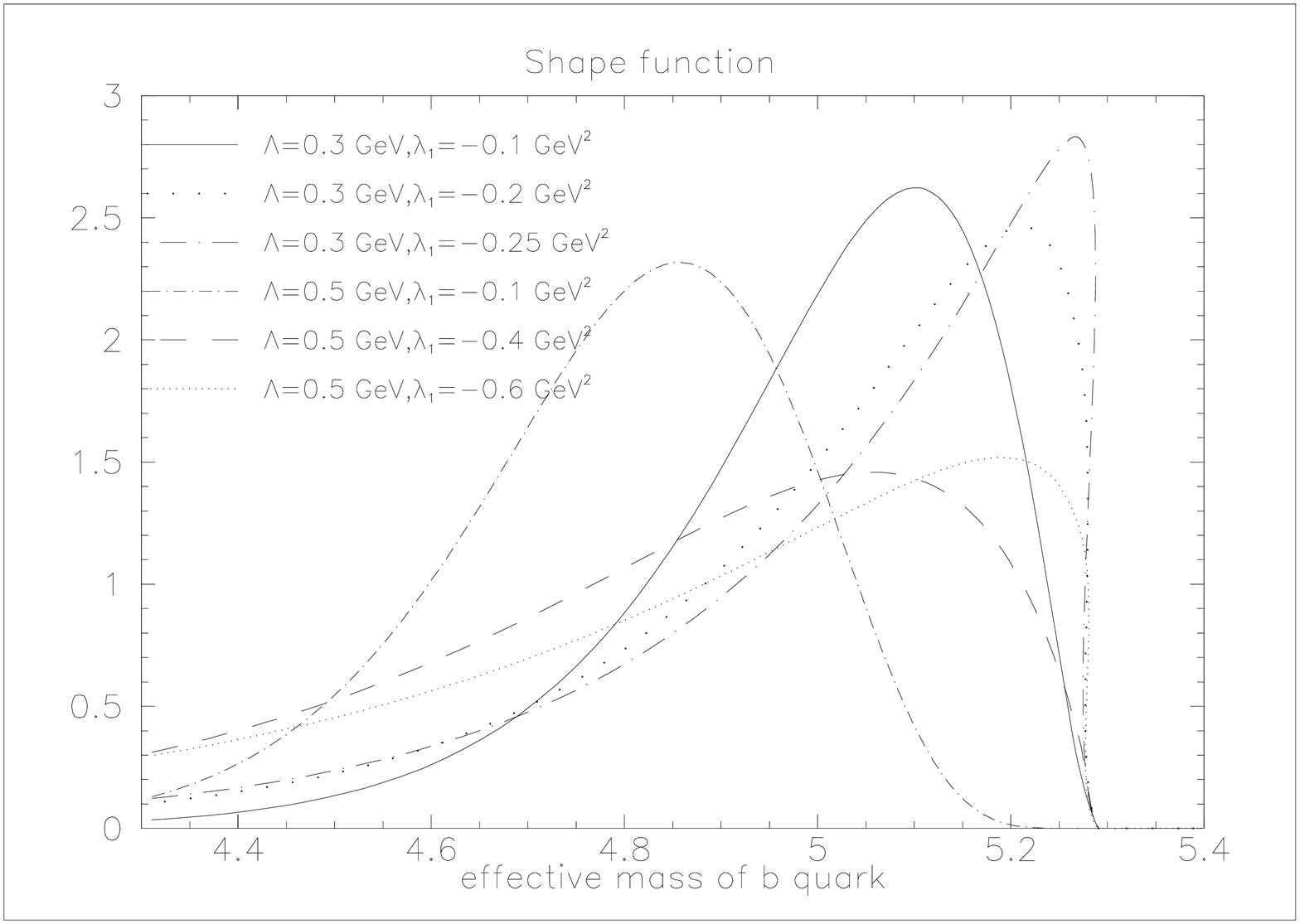,height=250pt,width=350pt}
\caption{Shape function $f(m^*)$ for different values of $\Lbar$ and $\lam1$.}\label{AllShapes}
\end{figure}
\begin{figure}
    \epsfig{file=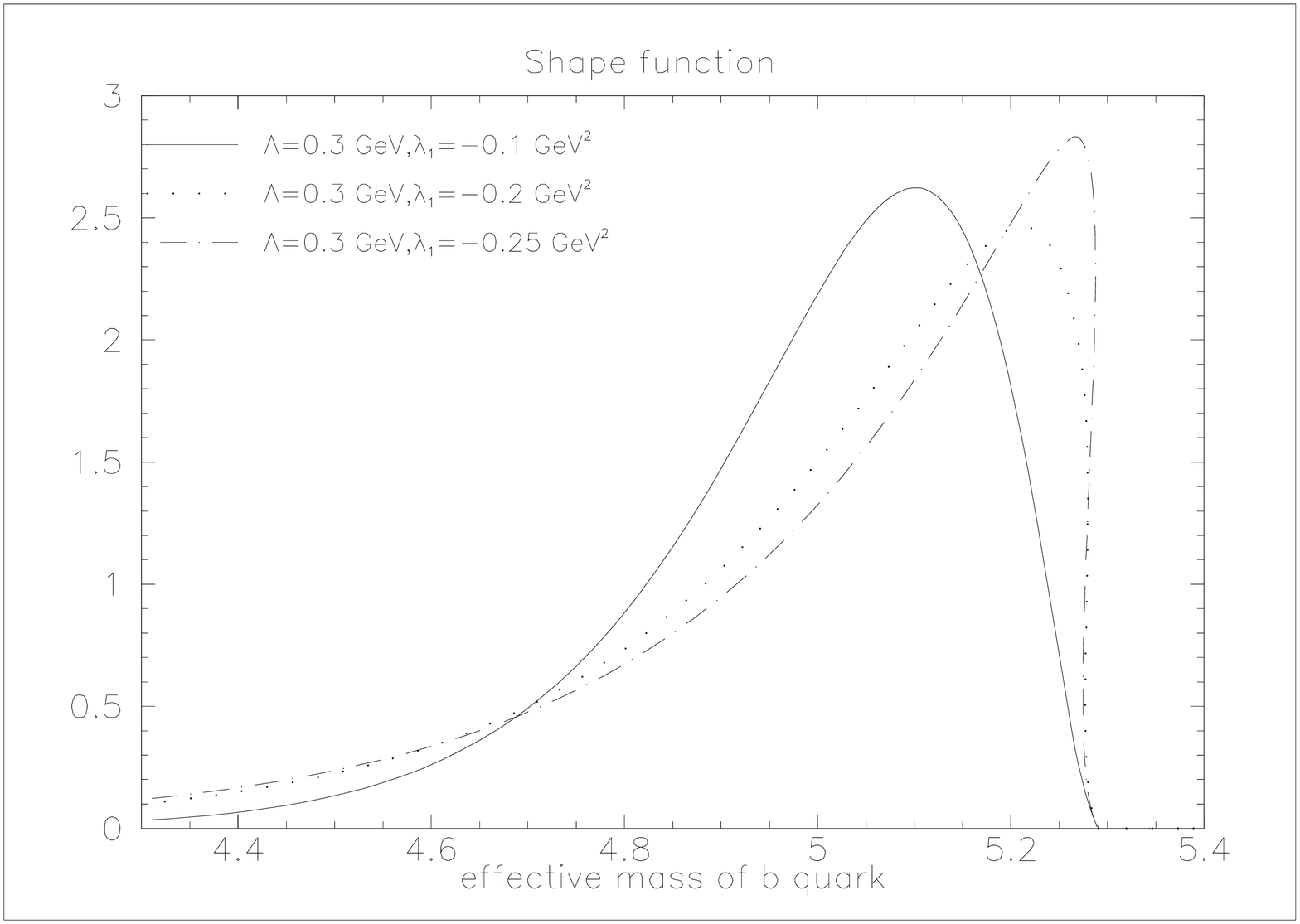,height=250pt,width=350pt}
  \caption{Shape function for $\Lbar=0.3\ \GeV$.}\label{ShapesLbar3}
\end{figure}
\begin{figure}
    \epsfig{file=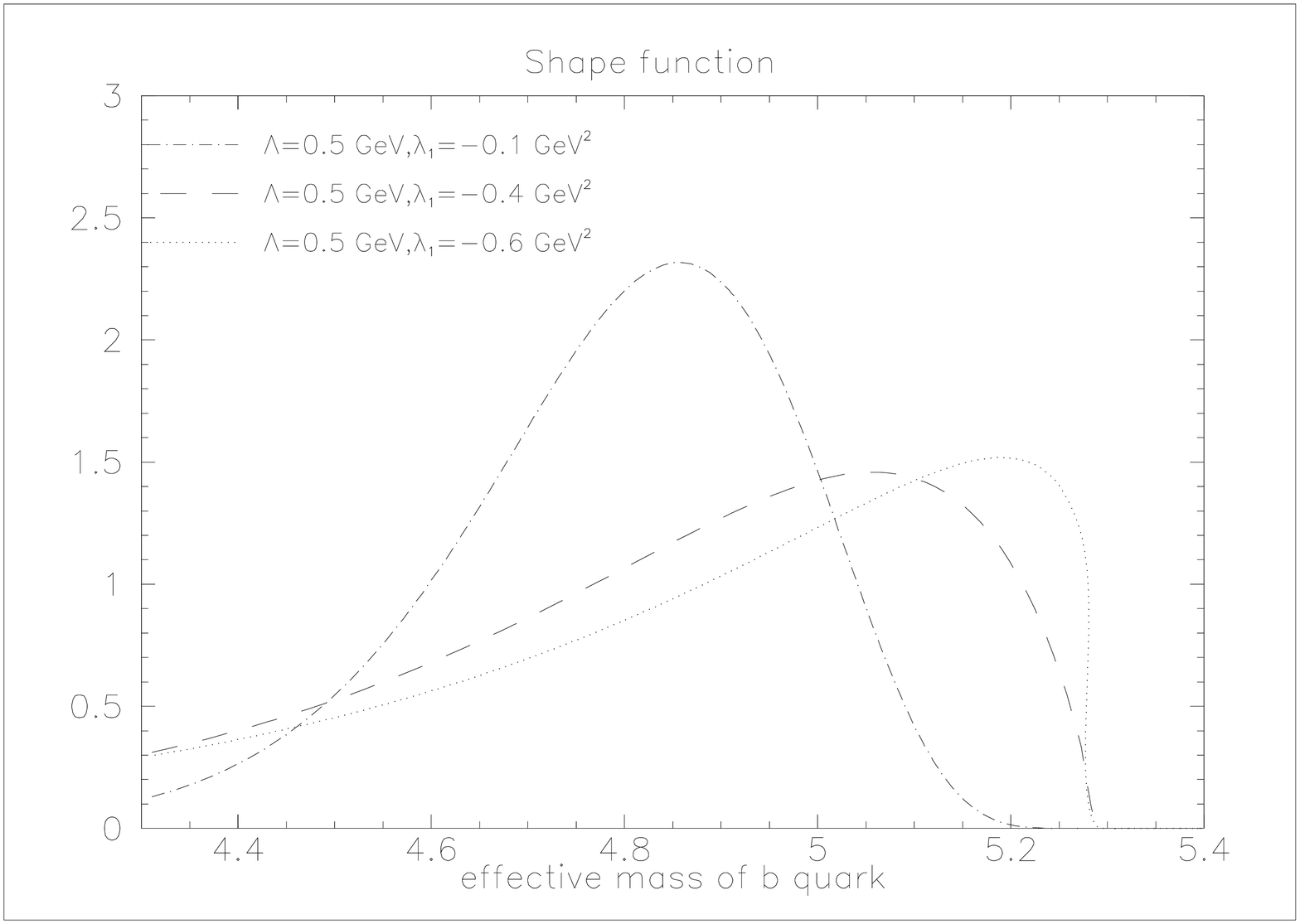,height=250pt,width=350pt}
  \caption{Shape function for $\Lbar=0.5 \GeV$.}\label{ShapesLbar5}
\end{figure}

The shape function corresponding to the $6$ sets of parameters is plotted
as a function of the effective b quark mass in Fig.\ref{AllShapes}. For clearer
distinction see Figs. \ref{ShapesLbar3} and \ref{ShapesLbar5}.
\begin{figure}
\epsfig{file=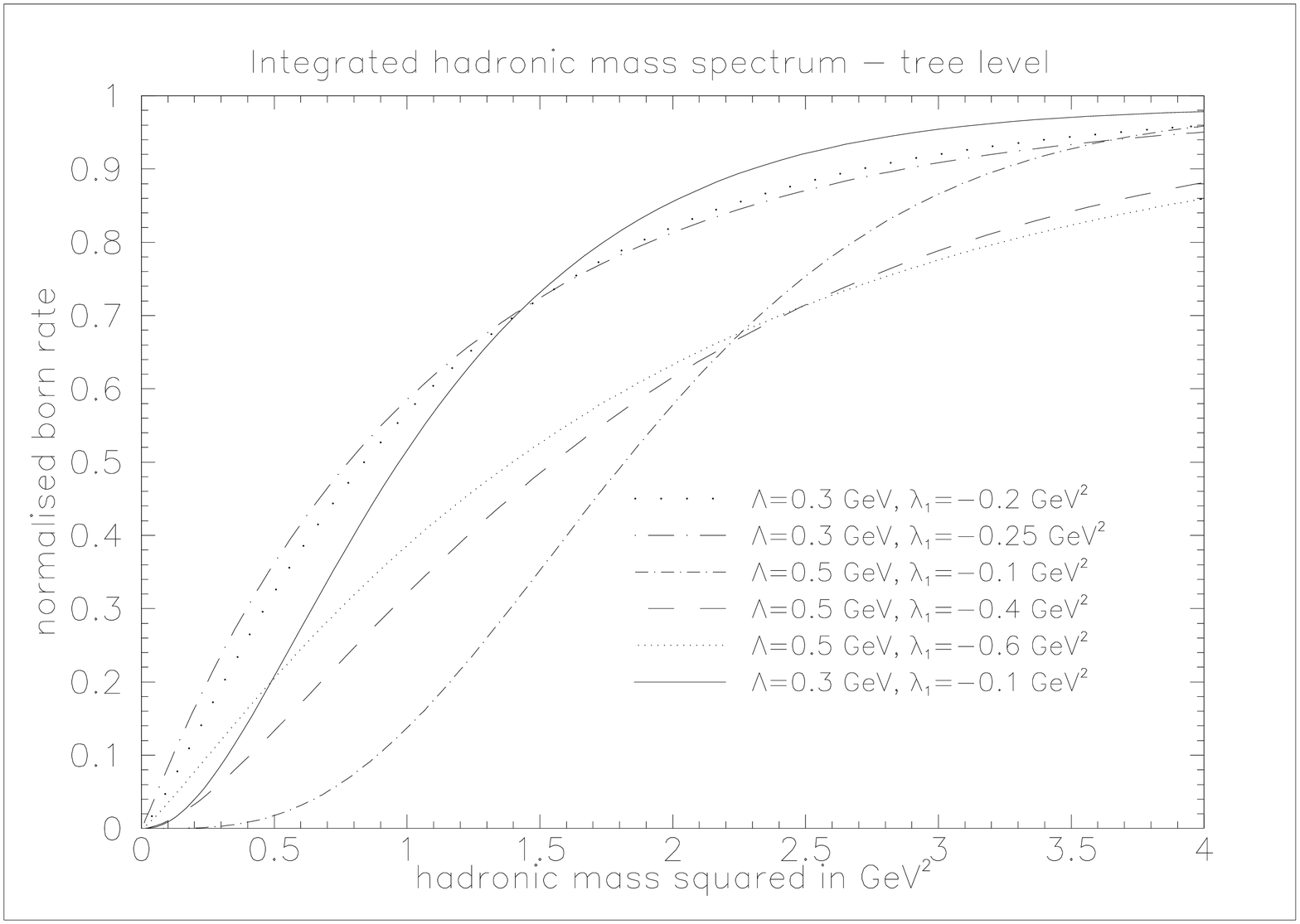,height=250pt,width=350pt}
\caption{$B\rightarrow X_u$ decay rate at tree level.}\label{AllRatesTree}
\epsfig{file=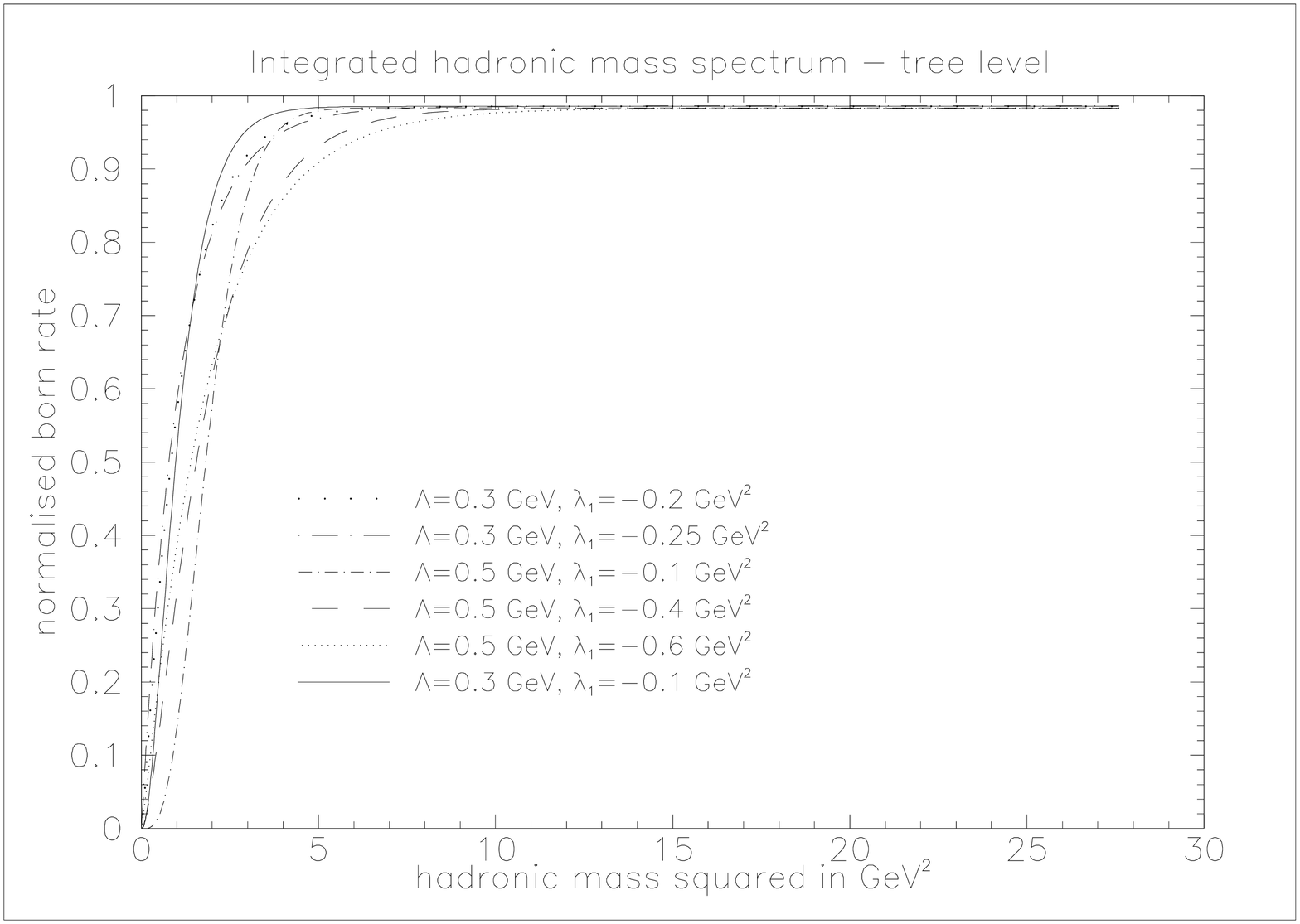,height=250pt,width=350pt}
\caption{$B\rightarrow X_u$ decay rate at tree level in the whole
range of hadronic mass.}\label{AllRatesTreeWide}
\end{figure}

\subsection{Results}\label{Extraction:Results}
\subsubsection{General result on $|V_{ub}|$ accuracy.}
 If we choose $M_X^2=4\GeV^2$ as the
hadronic mass squared cut, then the uncertainty of the extracted value 
of $|V_{ub}|$ will be about $10\%$. The source of this uncertainty
lies with the parameters that go into it.
The quantity $r$ which we have chosen for use in the determination of
the ratio of $|V_{ub}/V_{cb}|$ depends on the strong coupling
constant, the quark masses and the parameters $\LBar$ and $\Ljeden$
entering the shape function. This makes for quite an involved
structure of the various contributions. Before we specify and discuss
these, let us briefly recall the range of the survey that has been performed. 
 In order to make a prediction, we have examined the
various sets of parameters as defined by
Eq.(\ref{Set1}-\ref{Set6}). The six sets of 
nonperturbative parameters span a conservative range of the estimates
for the meson-quark mass difference ($\LBar$) and the kinetic energy
of the $b$ quark Fermi motion ($\Ljeden$). In 
addition, the strong coupling constant was taken to be $0.2$ or $0.3$ 
for each set, while the electron energy cut was 
kept constant at all times and set to $0.5 \GeV$.
This particular cut
is arguably of minor importance to the analysis and has been made to
provide for 
the identification of the semileptonic mode. More vital is the
remaining upper cut on the hadronic invariant mass. We look closer at
the range of $2\GeV^2$ to $4\GeV^2$ of this cut.
\par
 By examining the variation of the observable $r$ 
 due to the changes of the parameters we can obtain an estimate of the 
accuracy that can be achieved. The form of the function $r(M_X^2)$
obtained for all the studied sets has been presented in
Fig. \ref{All}. The
value of $r(M_{cut}^2=4\GeV)$ can 
be approximately written as 
\begin{equation}
r=1.74+0.3\frac{\Delta\alpha_s}{\tilde{\alpha_s}}+0.2\frac{\Delta{\Lbar}}
{\tilde{\Lbar}} +0.08\frac{\Delta\lam1}{\tilde{\lam1}}.
\end{equation}
These numbers give a flavour of the dependence on the parameters. Let
us now turn to the identification of the effects brought about by the
nonperturbative and perturbative corrections. Since it is the hadronic 
invariant mass cut that is variable in our analysis, we will look at
the value of the examined ratio $r$ as a function of this cut,
$r(M_X^2)$. We will look at the partonic result at the Born level and
at the corrections it receives from the perturbation theory as well as 
HQET leading twist approximation.
\par
%
\subsubsection{Partonic results.}
\begin{figure}[!]
\epsfig{file=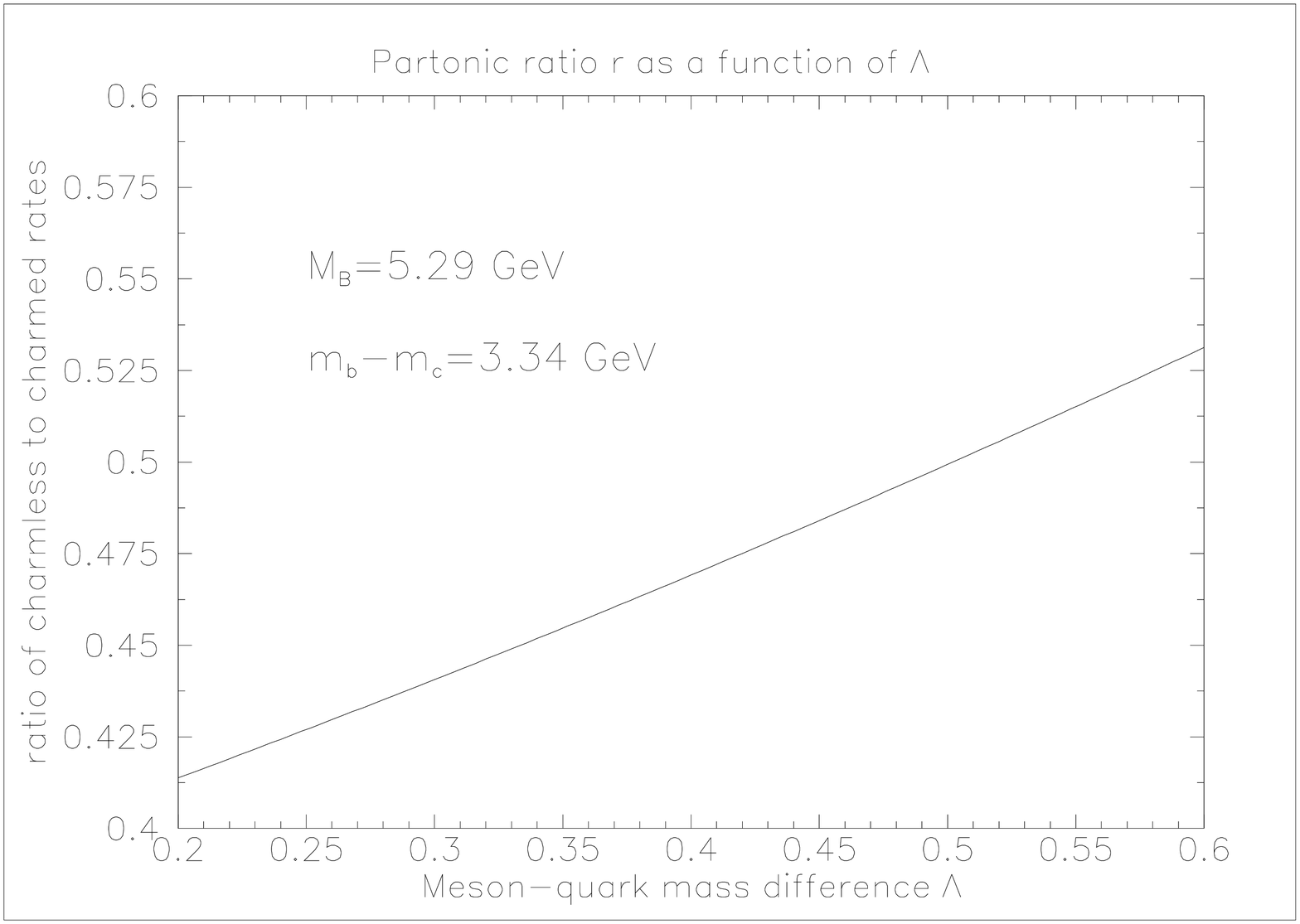,height=250pt,width=350pt}
\caption{The partonic value of the ratio {\em r} of charmless and
charmed decays as dependent on the $\LBar$ parameter.}\label{rpart}
\end{figure}

Although it might appear that partonic results, understood to treat
the decay as that of a free quark into a free quark and a pair of free 
leptons, have little in common with nonperturbative data, the latter
intrude via the quark mass ratio. This is a simple consequence of the
difficulty that arises as soon as one tries to treat quarks as free,
ignoring confinement. Thus, while the $b\rightarrow u$
decay rate is truly devoid of any reference to nonperturbative
quantities in this zeroth approximation, the decay to charmed states
will bear a mark of the $\LBar$ parameter, which sets the $b$ quark
mass. The values of the scaled decay rates have been quoted in
Eqs.(\ref{bc1},\ref{bc2}). With the  mass of the $b$ quark fixed, the ratio can also be
thought of as fixed  
due to the fairly well known mass difference between $b$ and $c$
quarks of $3.4 \GeV$. Curiously enough, we must conclude that if we
only considered the partonic result, the fact that we are calculating
ratio rather that the rate itself would be counterproductive to
precision. But of course such reasoning would be unrealistic as the
corrections are essential to accuracy and must be included. We do 
not present any plots for the partonic quantity of $r(M_X^2)$. Since the invariant
mass of the hadrons is always zero in this approach, such plots would have
to be series of horizontal lines. However, beneath we give the
parton model values of $r$ obtained assuming $\LBar=0.3\GeV$ and
$\LBar=0.5\GeV$. They are clearly the reciprocals of the scaled rates
stated in Eqs.(\ref{bc1},\ref{bc2}). 
\begin{eqnarray}
r&=&2.2694 \mbox{ for\ } \LBar = 0.3 \GeV, \\
r&=&2.0025 \mbox{ for\ } \LBar = 0.5 \GeV.
\end{eqnarray}
The dependence of the partonic value of $r$ on the parameter $\LBar$
is nearly linear in any reasonable interval, as seen in
Fig. \ref{rpart}.
\begin{figure}
\epsfig{file=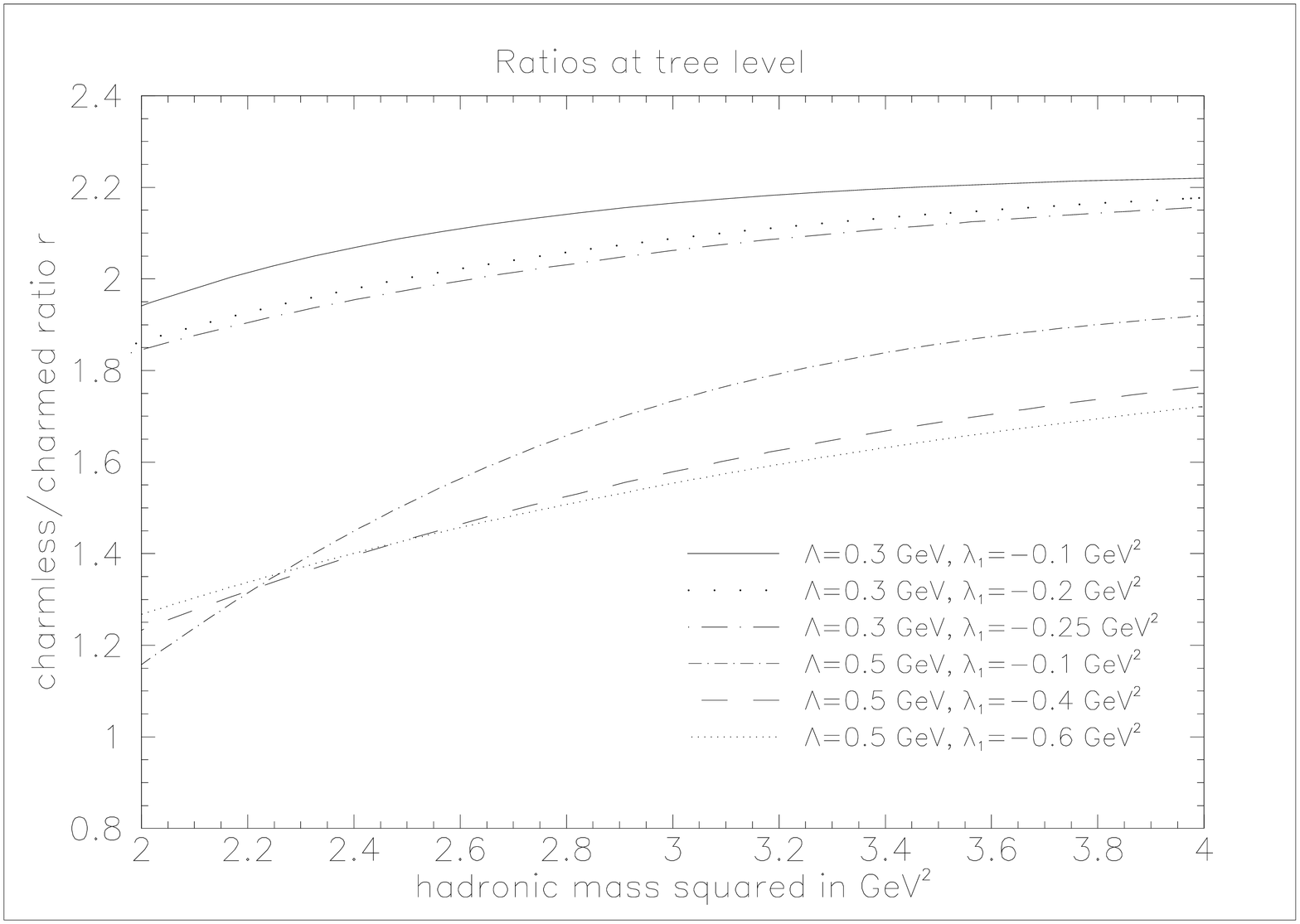,height=250pt,width=350pt}
\caption{Ratios charmless/charmed at tree level.}\label{AllRatiosTree}
\end{figure}
\subsubsection{Convoluted partonic results.} 
\par
The rather dull hadronic mass spectrum delivered by the partonic model 
gains colour when the Fermi motion is accounted for. The distribution
gets indented at the lower end of the spectrum because now the
hadronic mass is taken to incorporate also the 'brown muck' or the light
degrees of freedom, or, couched in the language of phenomenological
models, the spectator. This kinematical argument is relevant  even
though we are dealing with a formally developped approximation to QCD, 
which renounces model dependence. This is because the kinematics of
this approximation is clear. The $b$ quark is static with its mass, as 
it were, bulging or shrinking. The conservation of four-momentum then
leads to a 
unique expression for the hadronic invariant mass,
\begin{equation}
M_X^2=m^{*2}z-x_p{\LBar}^* m_b^* - \LBar^{*2}.
\end{equation}
This formula implies that the typical scale on which the effects of
including the Fermi motion are expected to be noticeable is set by
$\LBar m_b$, which is around $2\GeV$ and whose variations are discernible.
The extent to which the light constituents of the $B$ meson exert
influence over the hadronic mass distribution is gauged by the
parameters of the shape function. In particular, the departure of the
$b$ quark mass from the mass of the meson, as indicated by $\LBar$,
shows positive 
correlation with the shift of the distribution, see
Fig. \ref{AllRatesTree} showing the rates and Fig. \ref{AllRatiosTree} 
for the ratio $r(M_X^2)$. Note that the probabilistic nature of the
convolution implies that barring the effects of the imposed cuts one
should recover the partonic results upon phase space integration. That 
this does happen is visible from Fig.\ref{AllRatesTreeWide} where the
different sets tend to a common value as the hadronic mass is cut
sufficiently high. 
\subsubsection{Convoluted one-loop corrections.}
\par
\begin{figure}
\epsfig{file=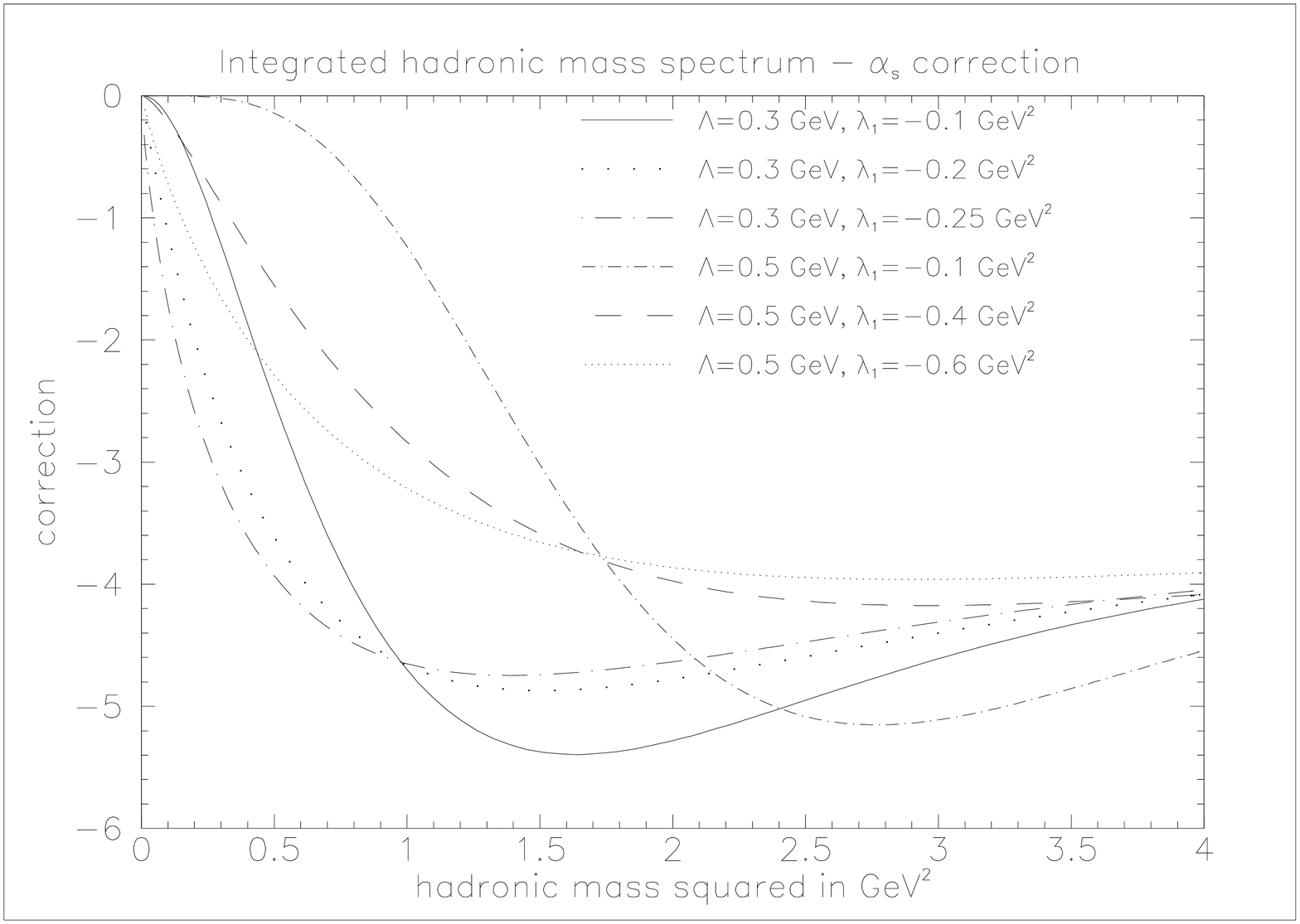,height=250pt,width=350pt}
\caption{Perturbative correction to integrated hadronic mass spectrum.}\label{CorrRates}
\epsfig{file=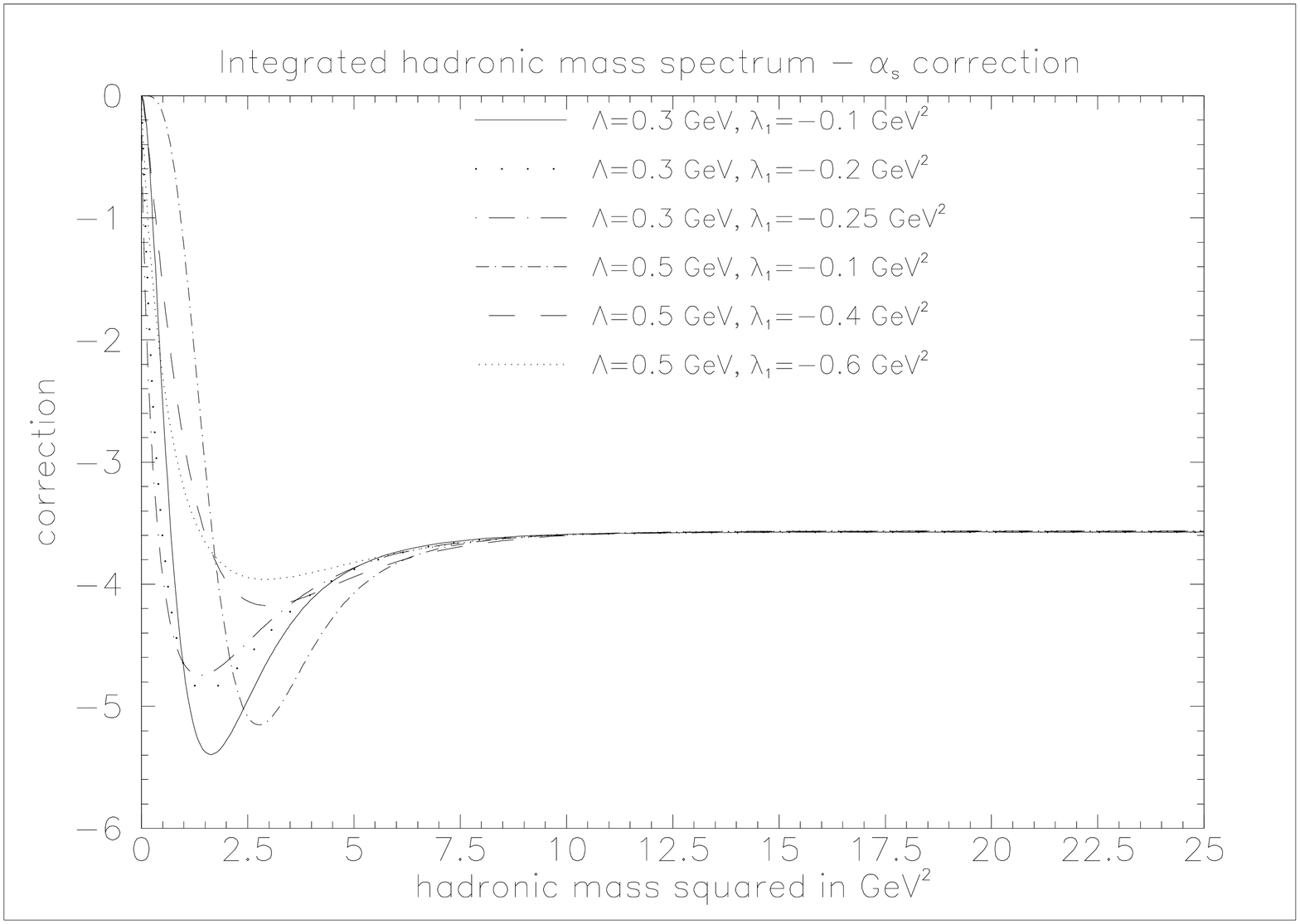,height=250pt,width=350pt}
\caption{Perturbative correction to integrated hadronic mass spectrum.}\label{CorrRatesWide}
\end{figure}

As explained in Sec. \ref{Combine}, the perturbative corrections are
combined with the nonperturbative ones in essentially the same way as
the tree level partonic term. It is well known that one-loop
corrections themselves yield a singular hadronic mass distribution,
characterised by a Dirac delta peak at zero mass and a continuous but
divergent part. The Kinoshita-Lee-Nauenberg theorem
\cite{Kinoshita:1962ur,Lee:1964is}, ensures that the
divergences cancel once they are properly regularised. In a strict
perturbation theory, final results are only obtained upon integration
over some final range of the invariant mass, enclosing an interval
around zero. The convolution prescribed in the leading twist
approximation performs this integration so that no infinity enters the 
one-loop corrected and convoluted terms, as seen in
Fig. \ref{CorrRates}.
However, the large negative correction remains in the
convoluted spectra as a dip close to the lower end. As with the tree
level result, the corrections are affected by the Fermi motion in a
degree that depends on the size of the applied hadronic mass
cut. Clearly, the low end of the spectrum is very sensitive to the
details of the shape function, while this influence is washed out
above the rough scale of $\LBar m_b$. Again, we recover the partonic
result after inclusion of all phase space, see Fig. \ref{CorrRatesWide}.
\par
\subsubsection{Perturbative and nonperturbative corrections.}
Summarizing, it is seen that although low cuts on the hadronic mass
are favoured by the requirement of cutting off the charmed admixture,
they leave the resulting distribution uncomfortably dependent on the
Fermi motion. This motion is 
known via the moments of the shape function and since this knowledge
is not very accurate we cannot rely on the details of it. Therefore we 
suggest that to be conclusive one must take the cut at least at the
level of $M_X^2=2\GeV$. On the other hand, to prevent the charmed
states from excessive proliferation and swamping the charmless events, 
we do not consider values higher than $M_X^2=4\GeV$.
\par
\begin{figure}
\epsfig{file=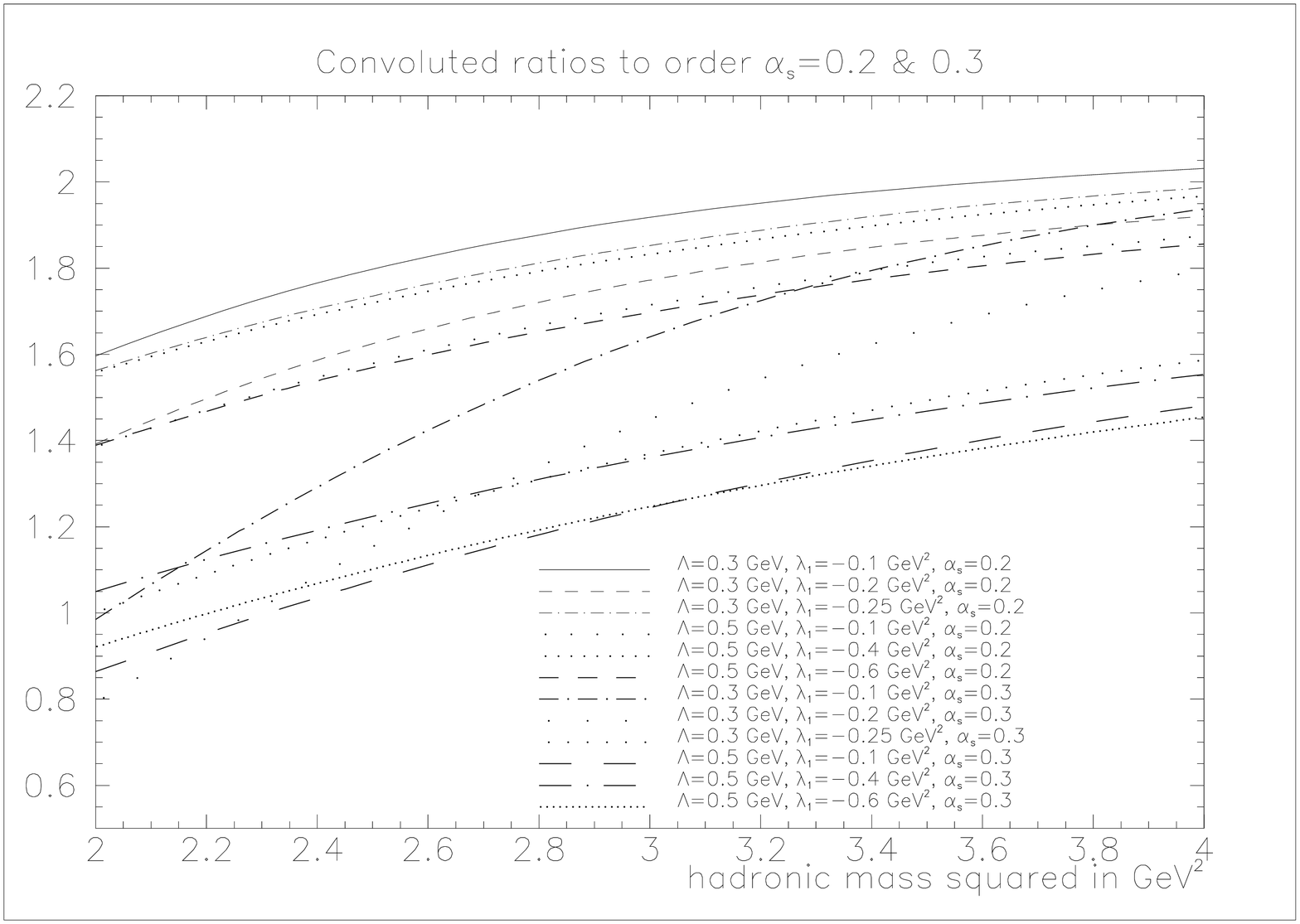,height=250pt,width=350pt}
\caption{Ratios $r$ for all the examined sets of parameters.}\label{All}
\end{figure}

 Deciding upon these ranges, we can come to conclusions about the
attainable accuracy of $|V_{ub}|$ as measured in the way we
propose. To this end, we have collected the calculated values of the
ratio $r(M_X^2)$ in Fig. \ref{All}. The spread of the ratio for a
given cut reflects the uncertainty ascribed to the measurement of the
square of the matrix element. Thus, if one sets the cut to $4\GeV$,
one can hope of extracting $|V_{ub}|$ with an uncertainty of around
$10\%$.
\subsubsection{Discussion of uncertainty.}
The above conclusion regarding the precision we assign to our method
aims to be conservative and it is based on a scan of a wide range of
parameters entering the distributions. However, it is possible to
improve on the accuracy of these parameters. It might be useful to
exploit the analogy to the $b\rightarrow s\gamma$ decays and extract
the nonperturbative information from them, as suggested in
\cite{Mannel:1999gs}. On the other hand, the question remains whether
one can use other observables of the semileptonic decays to help
establish parameters. This latter method has the advantage of
eliminating possible systematic errors due to different kinds of
considered processes.
\par
\begin{figure}[!]
\epsfig{file=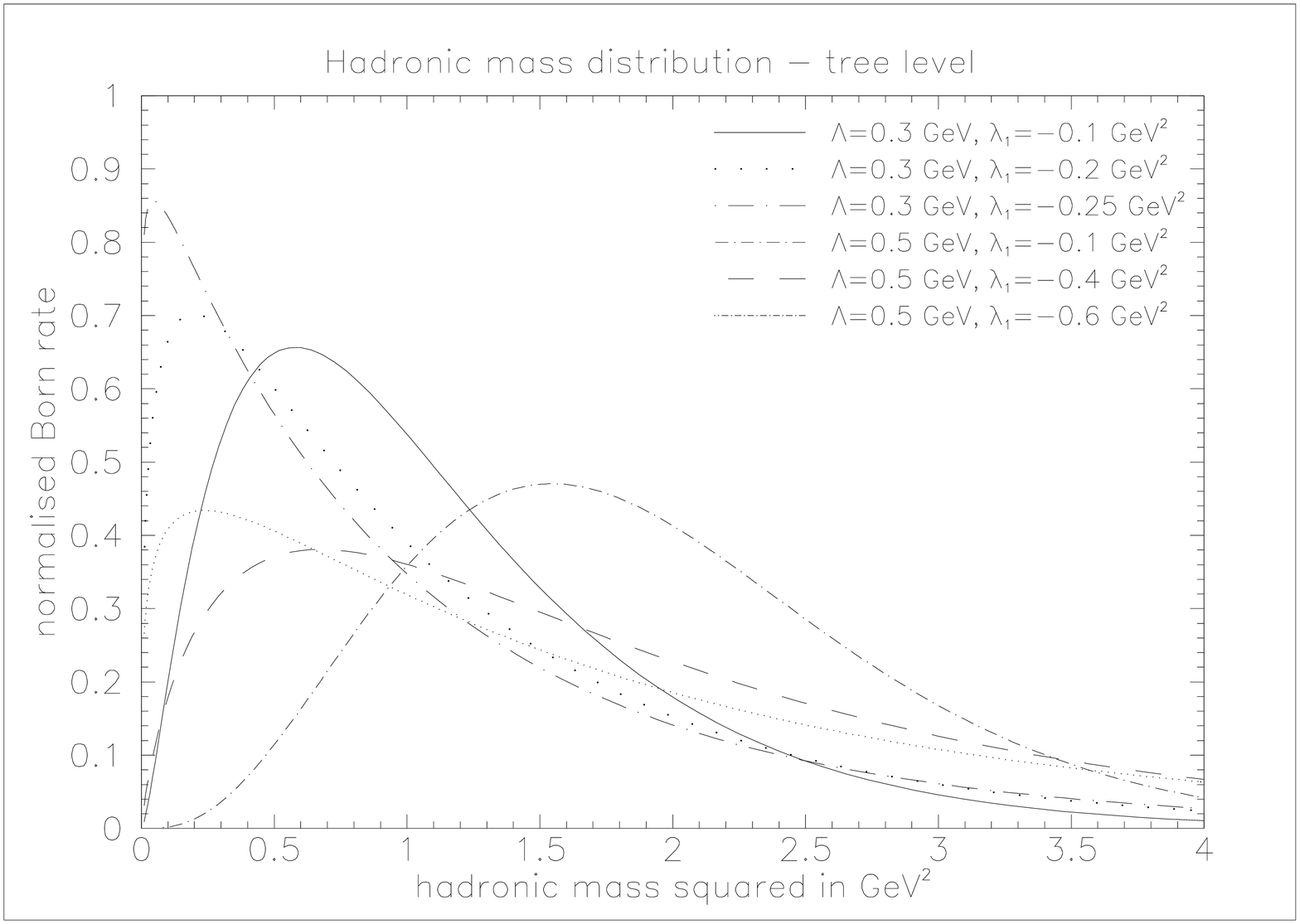,height=250pt,width=350pt}
\caption{Hadronic mass distributions at tree level.}\label{AllTreeDGDM2}
\epsfig{file=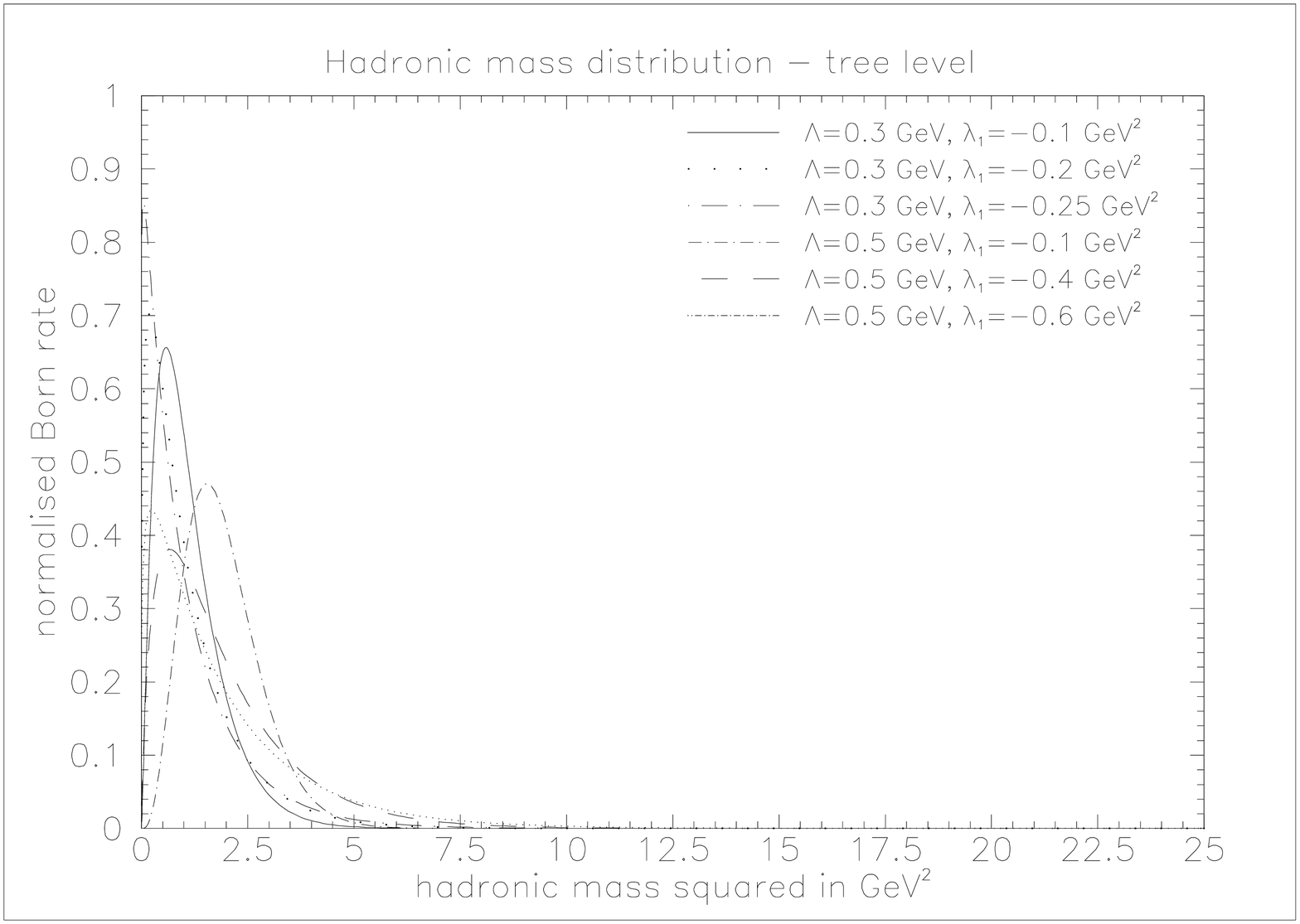,height=250pt,width=350pt}
\caption{Hadronic mass distributions at tree level.}\label{AllTreeDGDM2Wide}
\end{figure}
\begin{figure}[!]
\epsfig{file=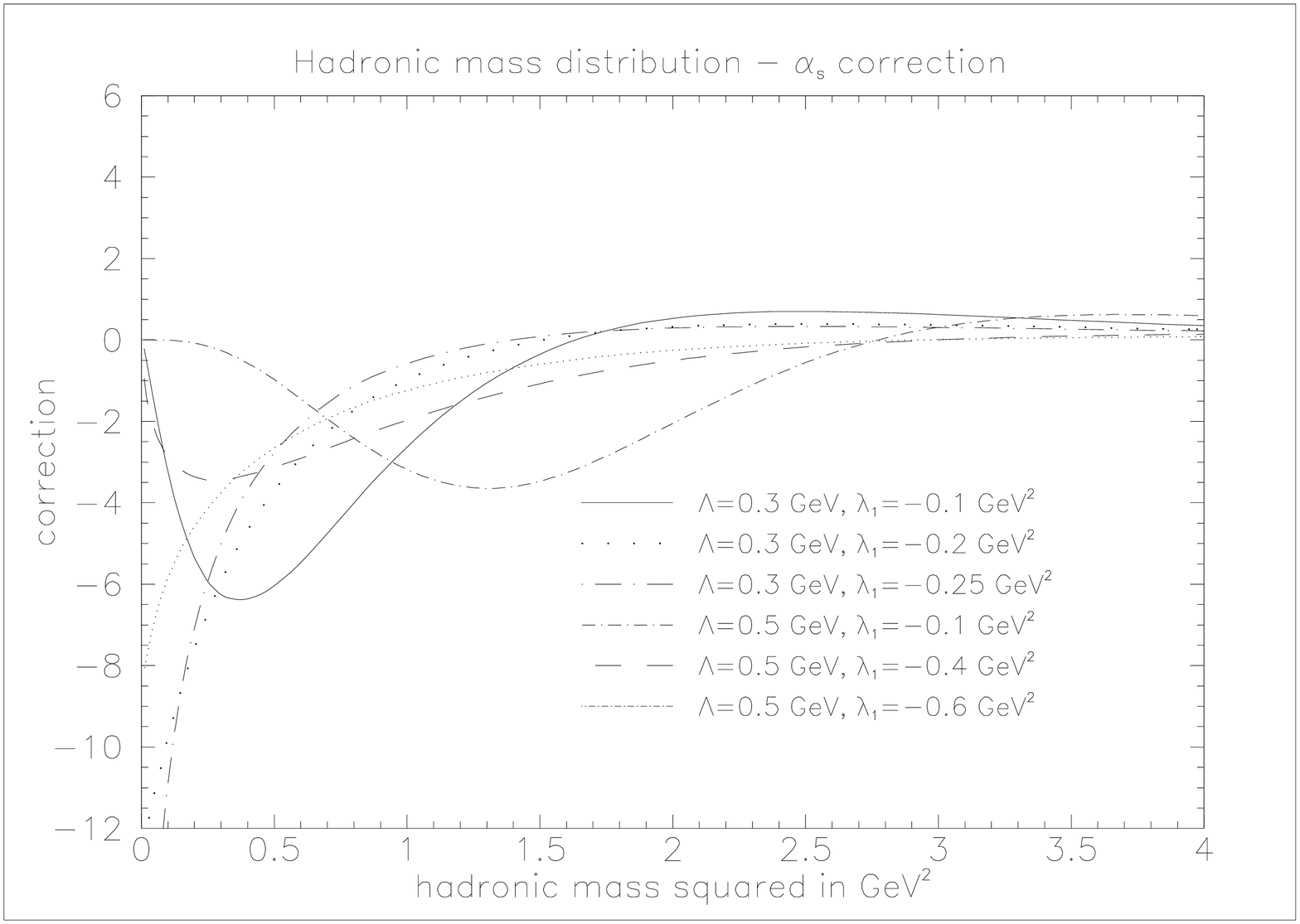,height=250pt,width=350pt}
\caption{${\cal O}(\alpha_s)$ correction to hadronic mass distribution.}\label{AllCorrDGDM2}
\epsfig{file=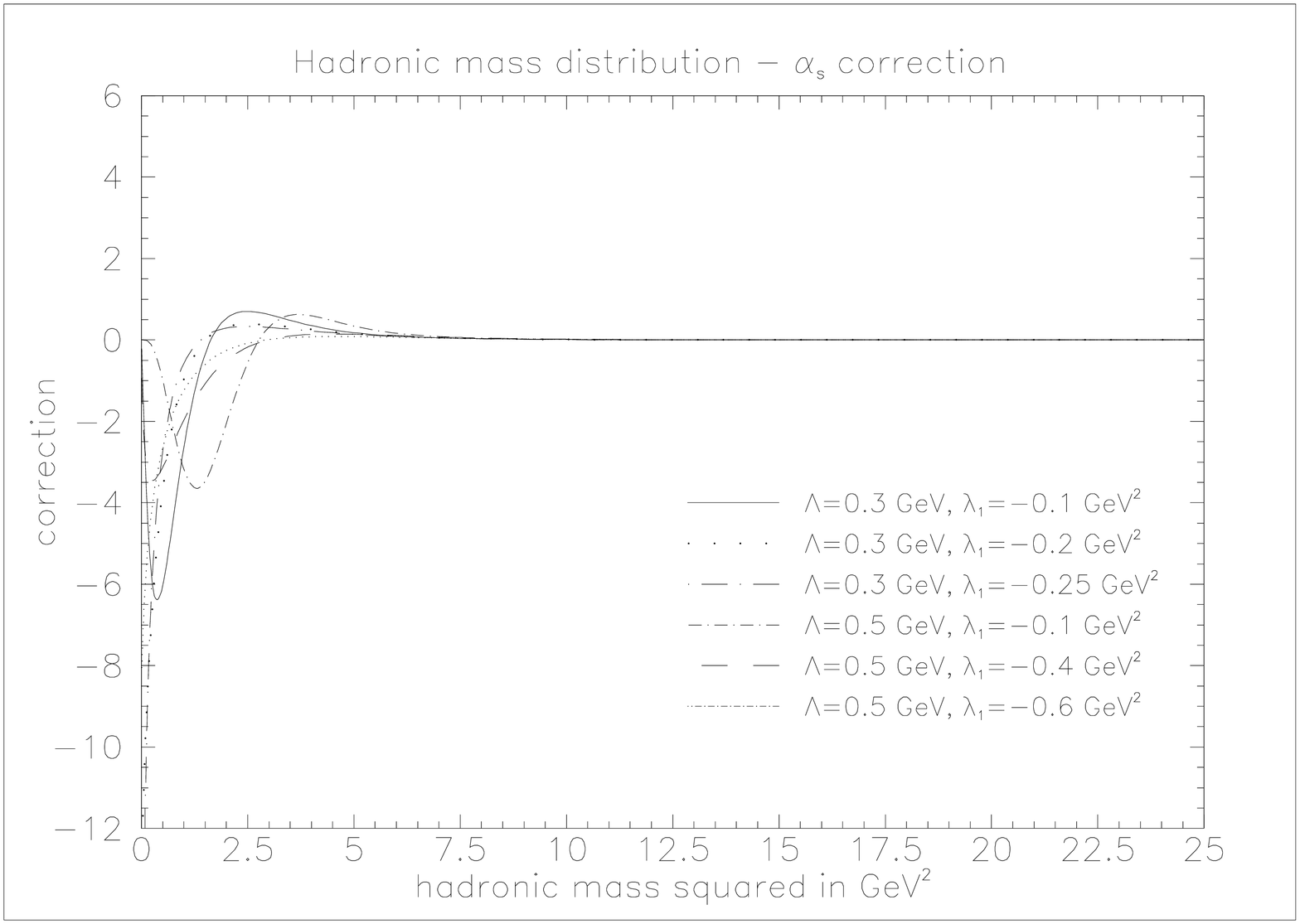,height=250pt,width=350pt}
\caption{${\cal O}(\alpha_s)$ correction to hadronic mass distribution.}\label{AllCorrDGDM2Wide}
\end{figure}
\begin{figure}
\epsfig{file=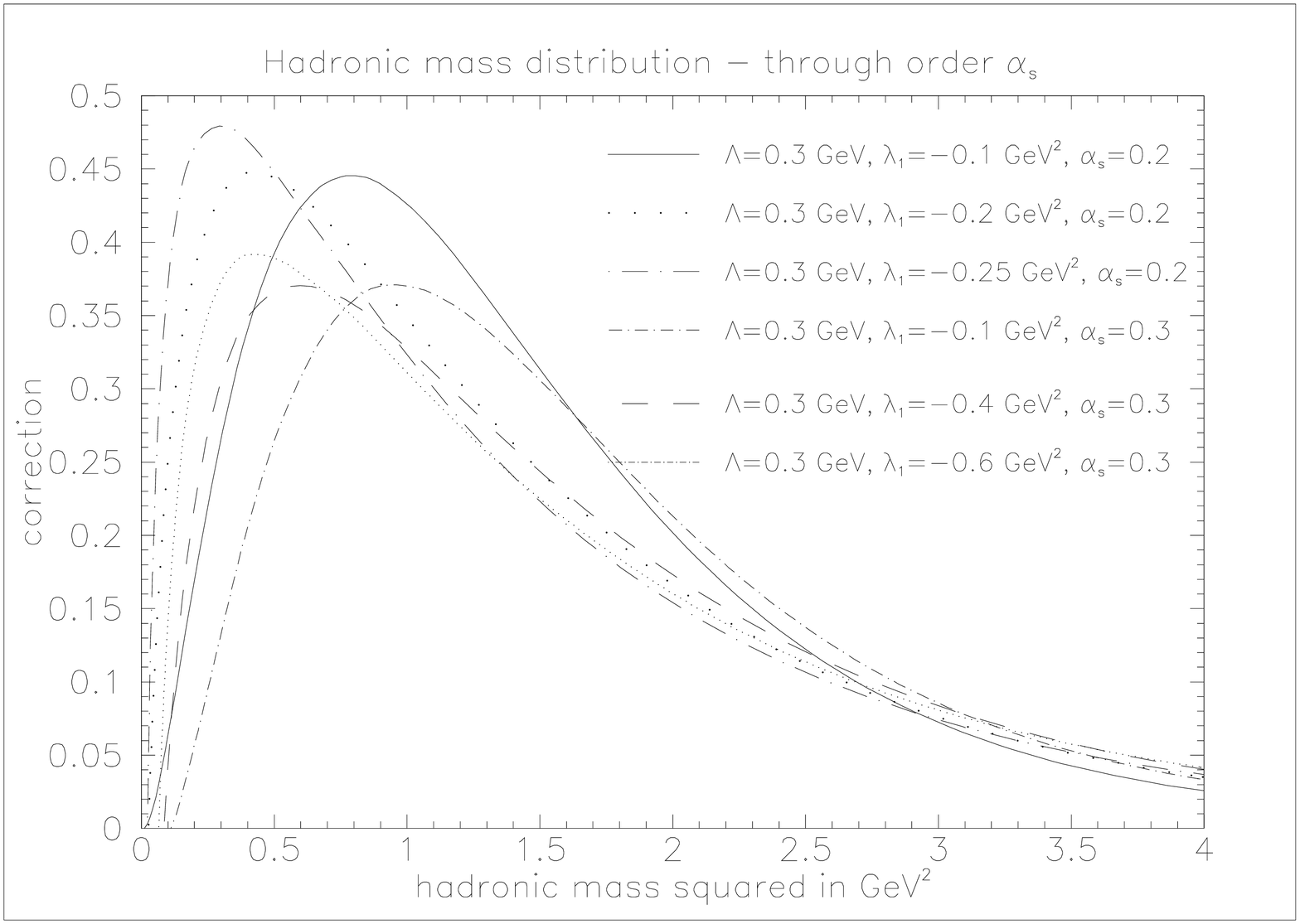,height=250pt,width=350pt}
\caption{Hadronic mass distributions to order $\alpha_s$ for $\Lbar=0.3$ GeV.}\label{DGDM2CompleteL03}
\epsfig{file=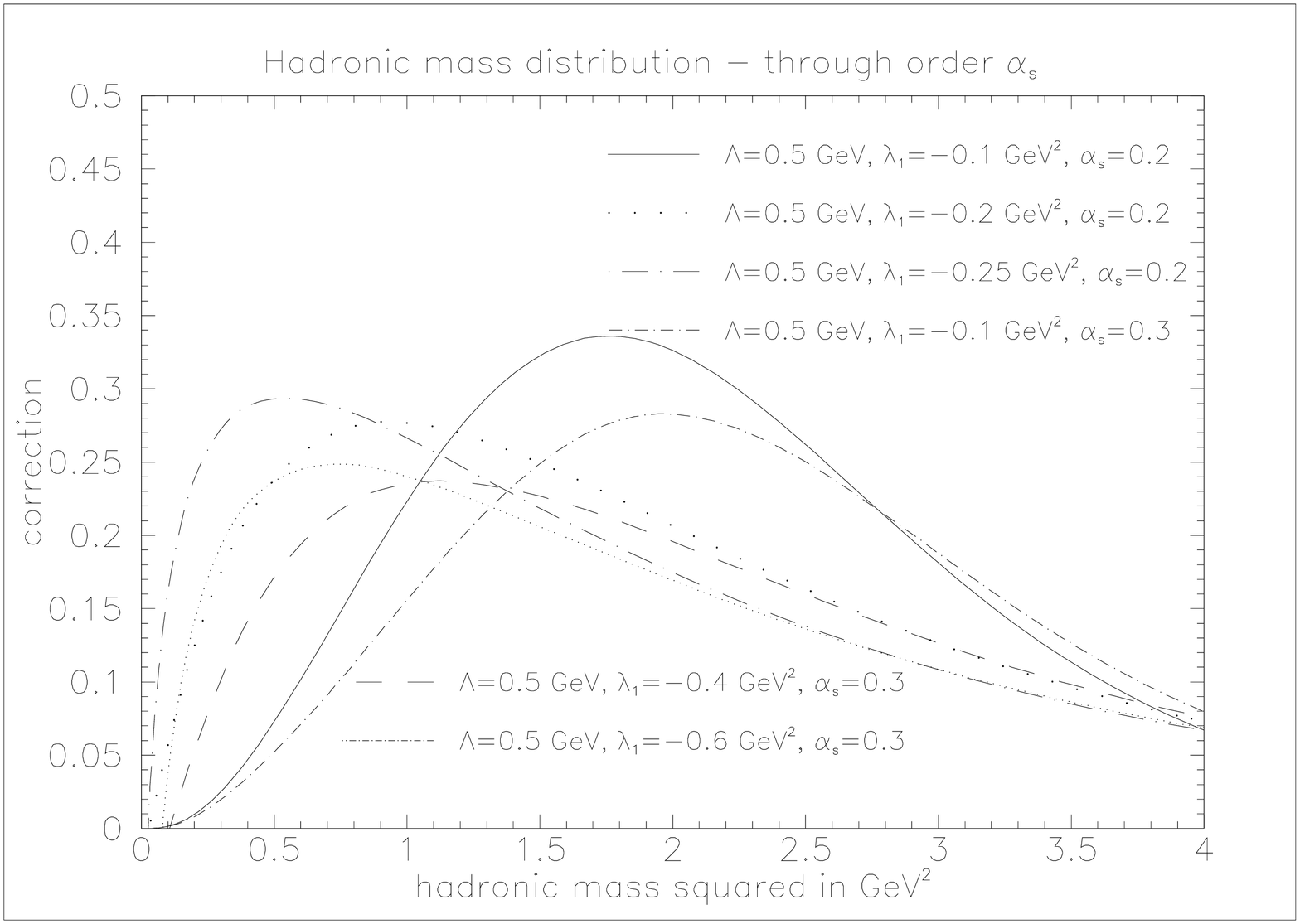,height=250pt,width=350pt}
\caption{Hadronic mass distributions to order $\alpha_s$ for $\Lbar=0.5$ GeV.}\label{DGDM2CompleteL05}
\end{figure}

We have taken a look at the differential distribution of the hadronic
mass. This is to be distinguished from the integrated distribution
discussed so far. As already seen, this quantity is sensitive to the
details of the shape function. One can see this for oneself in Fig.
\ref{DGDM2CompleteL03} for $\Lbar=0.3$ 
GeV and Fig. \ref{DGDM2CompleteL05} for $\Lbar=0.5$ GeV. The
contribution from the tree level term is presented in
Figs. \ref{AllTreeDGDM2} and \ref{AllTreeDGDM2Wide} for a broader
range of the hadronic mass. Similarly, the one-loop corrections
themselves have been shown in Figs. \ref{AllCorrDGDM2},
\ref{AllCorrDGDM2Wide}.

%
%
\subsection{Summary}
We calculate corrections of perturbative and nonperturbative nature to 
the hadronic mass spectrum in the semileptonic $B$ decays. One-loop
results are convoluted with the shape functions to give predictions of 
the $|V_{ub}|$ matrix element with an uncertainty of $10\%$. This
error estimate relies on a the current knowledge of  parameters
characterising the $B$ mesons. We state the functional dependence of
the errors in terms of the involved parameters so that an estimate can 
be made in case the precision of the latter is improved on. We also
suggest an observable to be employed in determinig these parameters
from the semileptonic decays and present the prediction for it through 
one-loop level supplemented by leading twist corrections.
\pagebreak
\section{Longitudinal polarisation of $\tau$ lepton}\label{Longitudinal}
\subsection{Introduction}
The longitudinal polarisation of the $\tau$ lepton is a non-trivial quantity
due to the large mass of the tauon. This is contrary to the electron or
even muon, whose masses are so small that the $V-A$ structure of the 
interaction causes them to be practically completely left-handed upon 
production in $B$ decays. The polarisation itself is an interesting quantity
as it can provide means of measuring the quark masses, especially the 
difference between $b$ and $c$ quark masses. Unlike the rate itself, the 
polarisation is regular througout the range of the charged lepton energy
even after inclusion of one-loop corrections. Furthermore, the dependence on
the poorly known $CKM$ matrix elements is eliminated. The need for calculating
the correction to this quantity is apparent given that the decay rate receives
as much as $20\%$ contribution from the one-loop diagrams. It is therefore
all the more remarkable that the situation is in the end different for
the polarisation. It turns out, as the calculation below shows, that the
polarisation receives only a tiny correction. In view of the poor knowledge
of the strong coupling constant, this kind of cancellation is also interesting
when one considers extracting quark masses from this quantity. 
The first indication of the smallness of the radiative correction to
the $\tau$ longitudinal polarisation was found in
\cite{Czarnecki:1995bn} where this quantity was calculated in the rest 
frame of the pair of leptons. 
\par
In order to find the polarisation, one may proceed as follows. First we get
the result for the unpolarized distribution, as found in \cite{\JM}. Then we
are left with the task of finding the decay rate into either state of 
polarisation of the charged lepton. It seems natural, taking into account the
structure of the weak interactions, that we calculate the rate of decay
into a negatively polarized electron. Thus we present formulae for the 
double distribution in terms of the lepton energy and the invariant mass
of the intermediate $W$ boson. Combining these with the formulae for the
unpolarized rate we readily find the polarisation. We also integrate the
rate over the invariant mass of $W$ to obtain the charged lepton energy 
spectrum. This allows us to evaluate the polarisation in terms of the
tauon energy, which is shown in a plot.
\par
When giving the results for the polarisation we also use the
nonperturbative corrections of order $1/m_b^2$, found in
\cite{Falk:1994gw} for the polarised lepton. That paper also gives the 
unpolarised distributions in the mass expansion, which were first
calculated in \cite{Balk:1994sz} and \cite{Koyrakh:1994pq}. The HQET
corrections appear in the form of extra terms proportional to the
phenomenological constants $\lambda_1$ and $\lambda_2$. While the
second of them can be related to the mass spliting between vector and
pseudoscalar $B$ mesons due to chromomagnetic interaction, and is
therefore fairly well established, the parameter $\lambda_1$ refers to 
the kinetic energy of the Fermi motion of the $b$ quark within the
meson and only rough estimates can be made of it. It is thus
convenient to express any corrections in terms of factors multiplying
these parameters, which we do.
This part of the paper  starts with a description of the evaluation of the
polarisation in Sec. \ref{Long:Evaluation}, which is followed by the
results given in an analytical form in Sec. \ref{Long:Results}. The
method of incorporating the QCD corrections has been discussed in
Sec. \ref{PertCorr}. Later on the moments of lepton energy distribution
are studied, Sec.\ref{Long:Moments}, and we summarize the results.
\subsection{Evaluation}\label{Long:Evaluation}
Let us start with the tree level matrix element for the unpolarized decay rate:
\begin{equation}\label{tauliniowy}
{\cal M}_{0,3}^{un}(\tau)= q\cdot\tau Q\cdot\nu. 
\end{equation}
With this kind of linear dependence on the four-momentum $\tau$, it is worth noting
that the matrix element with the $\tau$ polarisation taken into account is,
\begin{equation}\label{Kliniowy}
{\cal M}_{0,3}^{pol}={1\over 2}{\cal M}_{0,3}^{un}(K=\tau -ms)={1\over 
2}(q\cdot K) (Q\cdot\nu), \end{equation}
where $m$ stands for the lepton's mass and we have introduced the four-vector $K$
\begin{equation}
K=\tau -ms.
\end{equation}
This relation is rendered useful by expressing the polarisation
four-vector $s$ in the following way:
\begin{equation}\label{stolQ}
s={\cal A}\tau +{\cal B}Q
\end{equation}
This is correct due to the fact that now only the temporal component of $Q$ does not
vanish, whereas the spatial parts of $s$ and $\tau$ are parallel. The coefficients
${\cal A},{\cal B}$ appearing in the formula above can be evaluated using the 
conditions defining the polarisation four-vector $s$:
\begin{eqnarray}
s^2&=&-1\\
s\cdot\tau &=&0.
\end{eqnarray}
Upon this one arrives at the following expressions:
\begin{eqnarray}
{\cal A}^{\pm}&=&\pm {1\over\sqrt{\eta}}{x\over{\tau_+ -\tau_-}},\\
{\cal B}^{\pm}&=&\mp {{2\sqrt{\eta}}\over{\tau_+ -\tau_-}}.
\end{eqnarray}
where the superscripts at ${\cal {A,B}}$ denote the polarisation of the lepton.

Applying now the representation (\ref{stolQ}) of the polarisation $s$ we readily
obtain the following useful formula for the matrix element with the lepton polarized:
\begin{equation}\label{abexp}
{\cal M}_{0,3}^{\pm}=\mp{{\tau_{\mp}} \over {\tau_+ -\tau_-}}{\cal 
M}_{0,3}^{un}(\tau)\pm{\eta \over{\tau_+ -\tau_-}}{\cal M}_{0,3}^{un}(Q).
\end{equation}
The first term on the right hand side of (\ref{abexp}) can be calculated immediately
once we know the result for the unpolarized case. Thus the problem reduces to
performing this calculation again with the only difference amounting to replacing
the four-momentum $\tau$ of the lepton with that of the decaying quark, $Q$.
The Born-approximated $\tau$ energy distribution can be written explicitly as
\begin{equation}
{{d\Gamma ^{\pm}}\over{dx}}=12{\Gamma}_0 f_0^{\pm}(x),
\end{equation}
 
 where
\begin{equation}\label{G0}
{\Gamma}_0={{G_F^2m_b^5}\over{192{\pi}^3}}|V_{CKM}|^2.
\end{equation}
The Born level function $f_0(x)$ reads, for the unpolarized case,
\begin{equation}
f_0(x)={1\over 6}\zeta ^2 \tau _3 \left\{ \zeta [x^2-3x(1+\eta)+8\eta]
+(3x-6\eta)(2-x)\right\},
\end{equation}
while the polarized cases are obtained using the function $\Delta f_0$:
\begin{equation}
\Delta f_0(x) = {1\over{12}}\tau _3^2\zeta ^2\{\zeta (3-x-\eta)+3(x-2)\},
\end{equation}
in the following way:
\begin{equation}
f_0^{\pm}(x)={1\over 2}f_0(x)\pm \Delta f_0(x).
\end{equation}
In the formulae above,
\begin{equation}
\tau _3=\sqrt{x^2-4\eta},\qquad \zeta =1-{\rho \over {1-x+\eta}}.
\end{equation}
At the tree level, one can express the polarisation integrated over the 
energy of the charged lepton as well:

  \begin{eqnarray}
  P &=& 1 - 1/18\left\{-24\eta^2+(x_m^3-8\eta^{3/2})(3 + \eta - 3\rho) +\right. 
  \nonumber\\&&     12\eta(x_m^2) - 3x_m^4/2+ 3(1 - \eta)^3(\rho-\rho^3/s^4) + 
  \nonumber\\&&      3(-1 + \eta)\rho(1-\rho/s^2)[3(1-\eta)^2 + \rho(3 + 5\eta)]
                     +3T_3S/2+
  \nonumber\\&&      3(x_m-2\sqrt{\eta})(-12\eta - 4\eta^2 + 12\eta\rho - 3\rho^2 + 
                      3\eta\rho^2 + \rho^3) -
  \nonumber\\&&      12\eta\rho^3\ln(s^2/\rho)- 18(2\eta^2 - \rho^2 - \eta^2\rho^2)
                       \ln\left[2\sqrt{\eta}/(x_m+T_3)\right] + 
  \nonumber\\&&     18(1 - \eta^2)\rho^2                    
\left.\ln\left(2s^2\sqrt{\eta}/[(1-\eta)(1-\eta-T_3)- \rho - 
\eta\rho]\right)\right\} 
  \nonumber\\&&/ \left\{T_3S/12+(2\eta^2 - \rho^2 - 
\eta^2\rho^2)\ln[(x_m+T_3 )/(2\sqrt{\eta})] \right.
  \nonumber\\&&   \left. - (1 - 
\eta^2)\rho^2\ln\left\{2\sqrt{\eta}\rho/\left[(1-\eta)(1-\eta+T_3)- 
\rho - \eta\rho\right]\right\} \right\} ,
 \end{eqnarray}
where
\begin{equation}
s=1-\sqrt{\eta},\qquad T_3=\sqrt{x_m^2-4\eta},
\end{equation}
\begin{equation}
S=1-7[(1+\eta)(\eta+\rho ^2)+\rho(1+\eta ^2)]+\eta ^3 +\rho ^3 
+12\eta\rho .
\end{equation}
Noting that the decomposition of the polarisation four-vector affects
the leptonic tensor only, one sees immediately that it can also be
applied to the QCD corrections.
\subsection{Analytical results}\label{Long:Results}
The following formula gives the differential rate of the decay  
$b\rightarrow \tau{\bar 
\nu}X$, $X$ standing for a $c$ quark or a pair of $c$ and a gluon, once the 
lepton is taken to be negatively polarized:  
\begin{equation}\label{formula12}
 {{d{\Gamma}^-}\over{dx\,dt}}=\left\{ 
 \begin{array}{ll}
 12{\Gamma}_0\left[ F^-_0(x,t)-{{2{\alpha}_s}\over{3\pi}}F_{1,A}^-(x,t)\right] & {\rm for\: }(x,t)\:{\rm in\: A}\\
 12{\Gamma}_0{{2{\alpha}_s}\over{3{\pi}}}F_{1,B}^-(x,t) & {\rm for\: }(x,t){\rm \: in\: B}
 \end{array}
 \right.
\end{equation}
$F_{1}^-$ differs according to which region (A or B) it belongs to. 
Region A is available for the 3- and 4-body decay, while Region B with a gluon only.
The following formulae are given for the negative polarisation of the lepton,
that is, we take 
\begin{eqnarray}
{\cal A}^{-} &=&- {1\over\sqrt{\eta}}{x\over{\tau_+ -\tau_-}},\\
{\cal B}^{-}&=& {{2\sqrt{\eta}}\over{\tau_+ -\tau_-}}.
\end{eqnarray}
The factor $\Gamma _0$ is defined in Eq.(\ref{G0}), while

\begin{equation}
F_0^-(x,t)=
{{(1-\rho-x+t)}\over {\tau _+-\tau _-}}\left[\tau _+(x-t-\eta)-\eta 
(1+\rho-t)\right] 
\end{equation}
and
\begin{eqnarray}\label{F1main}
 F_{1,A}^-(x,t)= F_0^-{\Phi}_0+{\sum_{n=1}^5 D_n^A{\Phi}_n+D_6^A},\\
 F_{1,B}^-(x,t)= F_0^-{\Psi}_0+{\sum_{n=1}^5 D_n^B{\Psi}_n+D_6^B}.
\end{eqnarray}
The factor 12 in the formula (\ref{formula12}) is introduced to meet the widely used 
\cite{Czarnecki:1995bn,Falk:1994gw,Voloshin:1995cy} convention for
$F_0(x)$ and ${\Gamma}_0$. The symbols present in (\ref{F1main}) are defined as follows:
\begin{eqnarray}\label{Phis}
{\Phi}_0={{2p_0}\over{p_3}}[Li_2(1-{{1-{\tau}^+}\over{p_+}})
+Li_2(1-{{1-t/{\tau}^+}\over{p_+}})\nonumber\\
-Li_2(1-{{1-{\tau}^+}\over{p_-}})-Li_2(1-{{1-t/{\tau}^+}\over{p_-}})\nonumber\\
+Li_2(w_-)-Li_2(w_+)+4Y_p\ln\sqrt{\rho}]\nonumber\\
+4(1-{{p_0}\over{p_3}}Y_p)\ln(z_{max}-\rho)-4\ln z_{max},\\
{\Phi}_1=Li_2(w_-)+Li_2(w_+)-Li_2({\tau}_+)-Li_2(t/{\tau}_+)\\
{\Phi}_2={{Y_p}\over{p_3}},\\
{\Phi}_3={1\over 2}\ln\sqrt{\rho},\\
{\Phi}_4={1\over 2}\ln(1-{\tau}_+),\\
{\Phi}_5={1\over 2}\ln(1-t/{\tau}_+),\\
{\Psi}_0=4({{p_0}\over{p_3}}Y_p-1)\ln({{z_{max}-\rho}\over{z_{min}-\rho}})+4\ln({{z_{max}}\over{z_{min}}})\nonumber\\
+{{2p_0}\over{p_3}}[Li_2(1-{{1-{\tau}_+}\over{p_-}})+Li_2(1-{{1-t/{\tau}_+}\over{p_-}})\nonumber\\
-Li_2(1-{{1-{\tau}_+}\over{p_+}})-Li_2(1-{{1-t/{\tau}_+}\over{p_+}})\nonumber\\
+Li_2(1-{{1-{\tau}_-}\over{p_+}})+Li_2(1-{{1-t/{\tau}_-}\over{p_+}})\nonumber\\
-Li_2(1-{{1-{\tau}_-}\over{p_-}})-Li_2(1-{{1-t/{\tau}_-}\over{p_-}})],\\
{\Psi}_1=Li_2({\tau}_+)+Li_2(t/{\tau}_+)-Li_2({\tau}_-)-Li_2(t/{\tau}_-),\\
{\Psi}_2={1\over 2}\ln(1-{\tau}_-),\\
{\Psi}_3={1\over 2}\ln(1-t/{\tau}_-),\\
{\Psi}_4={1\over 2}\ln(1-{\tau}_+),\\
{\Psi}_5={1\over 2}\ln(1-t/{\tau}_+).
\end{eqnarray}
We introduce $C_1...C_5$ to simplify the formulae for $D_n^A$ and $D_n^B$:
\begin{eqnarray}
C_1 &=& {1\over\sqrt{x^2-4\eta}} \left[\rho (-\eta x+4 \eta t-\eta {\tau}_+-4 \eta+x {\tau}_+-t {\tau}_+)-2 \rho^2 \eta\right.
     +\eta (x t+x   \nonumber\\
&&     +4 t-2 t^2+6+2 \eta)
     \left.    +{\tau}_+ (-\eta x+\eta t+5 \eta-2 x t+x+t+t^2)\right] ;
\end{eqnarray}
\begin{eqnarray}
C_2 &=& \rho (\eta-t-2 {\tau}_+)+ (x t^2-2 t^2 {\tau}_+)/(2\eta)+ \eta (-x-2 t+2 {\tau}_++30)/2  \nonumber\\
&&+t (1+t-x)+{\tau}_+ (-2 x+2 t+6);\\
C_3 &=& {1\over\sqrt{x^2-4\eta}} \left[\rho (-10 \eta x+18 \eta t-7 \eta {\tau}_++x t+7 x {\tau}_+-11 t {\tau}_++4 {\tau}_+)\right.\nonumber\\
&&+\rho^2 (-9 \eta+{\tau}_+)
       +(2 x t^2 {\tau}_+-x^2 t^2/2-t^2 {\tau}_+^2)/\eta+\eta^2 (-{\tau}_++10)
\nonumber\\
&&       +x t (x-t-1)
                \eta (4 x t-x {\tau}_+-2 x-x^2/2+t {\tau}_++4 t\nonumber\\
&&                -4 t^2+11 {\tau}_+-{\tau}_+^2-12)+
      \left.    +{\tau}_+ (-6 x t-x+2 x^2-3 t+5 t^2-5)\right];\\
C_4 &=& {1\over\sqrt{x^2-4\eta}} \left[\rho (2 \eta x 
{\tau}_+/t-8 \eta x+8 \eta {\tau}_+/t+18 \eta t-11 \eta {\tau}_+
-4 \eta^2 {\tau}_+/t^2\right.\nonumber\\
&&     -x t+5 {\tau}_+ x-7 t {\tau}_+)+\rho^2 (2 \eta {\tau}_+/t-9 \eta-\eta^2 {\tau}_+/t^2)+x t (1+t-x)\nonumber\\
&&                +(-2 x t^2 {\tau}_++x^2 t^2/2+t^2 {\tau}_+^2)/\eta+\eta^2 (5 {\tau}_+/t^2-6 {\tau}_+/t-16/t+6)\nonumber\\
&&                +\eta (2 x {\tau}_+/t+2 x t-3 x {\tau}_+-4 x+x^2/2-10 {\tau}_+/t+3 t {\tau}_++12 t\nonumber\\
&&  \left.               -4 t^2+3 {\tau}_++{\tau}_+^2+4)+{\tau}_+ (-4 x t-3 x+2 x^2+11 t+2 t^2) \right];\\
C_5&=& [\rho /(x-t-\eta/t)] (-\rho \eta^2 {\tau}_+/t^2+\eta {\tau}_+/t-\eta-\eta^2 {\tau}_+/t^2+\eta^2/t+\rho \eta {\tau}_+/t\nonumber\\
&&-\rho \eta
   +\rho \eta^2/t)/2+ [\rho /(1-x+\eta)] (-\eta t-\eta {\tau}_++\eta^2+t {\tau}_+)/2\nonumber\\
&&      +\rho (3 \eta {\tau}_+/t+9 \eta-9 t-3 {\tau}_+)/2+(x t^2/4-t^2 {\tau}_+)/\eta+\eta (3 x-10 {\tau}_+/t\nonumber\\
&&-12 t-2 {\tau}_+-24+2 \eta)/4+(3 t+5) {\tau}_+/2-x t+6 t+5 t^2/2;     \\
D^A_1&=&C_1;\\
D^A_2&=& {1\over{4 \sqrt{x^2-4\eta}}}\left\{\rho \eta (34-6 x {\tau}_+/t^2-4 x/t^2+3 x/t+5 x t-2 x {\tau}_++2 x
 \right.\nonumber\\
&& +3 x^2/t^2
 +x^2+8 {\tau}_+/t^2+18 {\tau}_+/t-2/t+2 t {\tau}_++38 t-14 t^2+20 {\tau}_+)\nonumber\\
&& +\rho \eta^2 (16-2 x/t^2-5 {\tau}_+/t^2-2 {\tau}_+/t+16/t-{\tau}_+)+\rho {\tau}_+ (-4 x t-4 x+6 t\nonumber\\
&&+7 t^2+11)
   +\rho^2 \left[\eta (-10+6 x {\tau}_+/t^2+6 x/t^2+4 x {\tau}_+/t-3 x-3 x^2/t^2\right.\nonumber\\
&&   -2 x^2/t-12 {\tau}_+/t^2
  -18 {\tau}_+/t+6/t+18 t-6 {\tau}_+)+\eta^2 (x/t^2+{\tau}_+/t^2\nonumber\\
&&-{\tau}_+/t-2/t)
  +{\tau}_+ (2 x-5 t-7)+
          +\rho \eta (-10-2 x {\tau}_+/t^2-4 x/t^2-x/t\nonumber\\
&&+x^2/t^2+8 {\tau}_+/t^2
          +6 {\tau}_+/t-6/t)+\rho \eta^2 {\tau}_+/t^2 
 +\rho {\tau}_++\rho^2 \eta (x/t^2-2 {\tau}_+/t^2\nonumber\\
&&\left.+2/t)\right]+{\tau}_+ (-4 x t+2 x t^2+2 x+7 t+t^2-3 t^3-5)
   +\eta (-14+2 x {\tau}_+/t^2\nonumber\\
&& +x/t^2-4 x {\tau}_+/t-2 x/t+4 x t-2 x t^2+2 x {\tau}_+-x-x^2/t^2
        +2 x^2/t-x^2\nonumber\\
&&-2 {\tau}_+/t^2
-6 {\tau}_+/t-10 t {\tau}_++32 t-22 t^2+4 t^3+18 {\tau}_+)
    +\eta^2 (28+x/t^2\nonumber\\
&&-2 x/t
 \left.+x+3 {\tau}_+/t^2-5 {\tau}_+/t-14/t+t {\tau}_+-14 t+{\tau}_+)\right\};\\
D^A_3 &=&{1\over\sqrt{x^2-4\eta}}\left\{\rho \left[\eta (-6-4 x {\tau}_+/t^2-3 x/t^2-2 x {\tau}_+/t+x/t-13 x+2 x^2/t^2\right.\right.\nonumber\\
&&  +x^2/t+6 {\tau}_+/t^2+12 {\tau}_+/t-2/t+26 t-10 {\tau}_+)+\eta^2 
(-x/t^2-2 {\tau}_+/t^2\nonumber
\\&&+2/t)  \left.+{\tau}_+ (6 x+6-12 t)\right]+\rho^3 \eta (-x/t^2+2 {\tau}_+/t^2-2/t)\nonumber\\
&&  +\rho
^2 \left[\eta (-10+2 x {\tau}_+/t^2+3 x/t^2-x^2/t^2-6 
{\tau}_+/t^2-4 {\tau}_+/t+4/t)\right.\nonumber\\
&&  \left.-\eta^2 {\tau}_+/t^2-{\tau}_+\right]+\eta^2 (14+x/t^2-x/t+3 {\tau}_+/t^2-2 {\tau}_+/t-14/t-{\tau}_+)\nonumber\\
&&  +\eta (-14+2 x {\tau}_+/t^2+x/t^2-2 x {\tau}_+/t-x/t+2 x t-2 x-x^2/t^2+x^2/t\nonumber\\
&&  \left.-2 {\tau}_+/t^2-8 {\tau}_+/t+18 t-4 t^2+10 {\tau}_+)+{\tau}_+ (-2 x t+2 x+2 t+3 t^2-5)\right\};  \nonumber\\
\\
D^A_4&=&-C_3-C_2;\\
D^A_5&=&-C_4+C_2;\\
D^A_6&=&-{1\over 2}C_5+{1\over{4 \sqrt{x^2-4\eta}}} \left\{[\rho /(1-x+\eta)]  \left[\eta (t {\tau}_++t+{\tau}_+)-\eta^2 (1+t+{\tau}_+)\right.\right.\nonumber\\
&&\left.+\eta^3-t {\tau}_+)\right]
 +[\rho /(x-t-\eta/t)] (1+\rho) \left[\eta (t-{\tau}_+)+\eta^2 (-1+{\tau}_+/t^2+\right.\nonumber\\
&&  {\tau}_+/t -1/t)
 \left.+\eta^3 (1/t^2-{\tau}_+/t^3)\right]+\rho \eta (43-x {\tau}_+/t-4 x/t-5 x+2 x^2/t\nonumber\\
&& +9 {\tau}_+/t  +13 t+5 {\tau}_+)
 +\rho \eta^2 (-5-{\tau}_+/t^2+2 {\tau}_+/t+1/t)+{\tau}_+ (9 x t+5 x \nonumber\\
&& +4 t-6 t^2) +12 x t+5 x t^2-2 x^2 t
 +\rho (-9 x t-3 x {\tau}_++5 t {\tau}_+) +\nonumber\\
&&  \rho^2 \eta (-10+2 x/t   -3 {\tau}_+/t)
 +\rho^2 \eta^2 (-{\tau}_+/t^2+1/t)+ (-6 x t^2 {\tau}_++x^2 t^2 \nonumber\\
&& +2 t^2 {\tau}_+^2)/(2\eta)
+\eta (-35-x {\tau}_+/t+2 x/t-2 x t+4 x {\tau}_++26 x \nonumber\\
&&   -2 x^2/t-3 x^2/2-4 {\tau}_+/t
-2 t {\tau}_+-34 t-t^2-22 {\tau}_+-3 {\tau}_+^2)  \nonumber\\
&& \left. +\eta^2 (-2+2 x/t-x+6 {\tau}_+/t+2 t)\right\}; 
\end{eqnarray}

\begin{eqnarray}
 D_1^B &=&  C_1\\
 D_2^B &=&  C_2 -  C_3\\
 D_3^B &=& -C_2 -  C_4\\
 D_4^B &=&  C_2 +  C_3\\
 D_5^B &=& -C_2 +  C_4\\ 
 D_6^B &=&  C_5\\
\end{eqnarray}
 One can perform the limit $\rho\rightarrow 0$, which corresponds to the decay
 of the bottom quark to an up quark and leptons. The formulae, which are much 
 simpler in this case, are presented in the same manner as the full results:
\begin{equation}
{{ d\widetilde{\Gamma}^-}\over{dx\,dt}}=\left\{
 \begin{array}{ll}
 12{\Gamma}_0\left[ 
{\widetilde{F}}_0^-(x,t)-{{2{\alpha}_s}\over{3\pi}}{\widetilde{F}}_{1,A}^-(x,t)\right] & \rm for (x,t) in A\\
 12{\Gamma}_0{{2{\alpha}_s}\over{3{\pi}}}{\widetilde{F}}_{1,B}^-(x,t) & \rm for (x,t) in B
 \end{array}
 \right.
\end{equation}
with 
\begin{equation}
\widetilde{F}_0^-(x,t)=
{{(1-x-t)}\over {\tau _+-\tau _-}}\left[\tau _+(x-t-\eta)-\eta (1-t)\right]
 \end{equation}
and
\begin{eqnarray}
\widetilde{F}_{1,A}^-(x,t)=\widetilde{F}_0{\widetilde{\Phi}}_0+{\sum_{n=1}^5 {\cal D}_n^A{\widetilde{\Phi}}_n+{\cal D}_6^A},\\
\widetilde{F}_{1,B}^-(x,t)=\widetilde{F}_0{\widetilde{\Psi}}_0+{\sum_{n=1}^5 {\cal D}_n^B{\widetilde{\Psi}}_n+{\cal D}_6^B}.
\end{eqnarray}
where
\begin{eqnarray}
{\widetilde{\Phi}}_0=2\left[Li_2({{{\tau}_+-t}\over{1-t}})+Li_2({{1/{{\tau}_+}-1}\over{1/t-1}})+Li_2(t)\right]+{1\over 2}{\pi}^2\nonumber\\
\nonumber +\ln ^2(1-{\tau}_+)_2\ln ^2(1-t)+\ln ^2(1-t/{{\tau}_+})\\
 -2\ln (1-t)\ln z_{max},\\ \nonumber
{\widetilde {\Phi}}_1={{\pi ^2}\over 12}+Li_2(t)-Li_2(\tau _+)-Li_2(t/{\tau _+}),\\
{\widetilde {\Phi}}_{2\Diamond 3}={{2\ln (1-t)}\over {1-t}},\\ 
{\widetilde {\Phi}}_4={\Phi}_4,\\ 
{\widetilde {\Phi}}_5={\Phi}_5,
\end{eqnarray}
and
\begin{eqnarray}
{\widetilde{\Psi}}_0=2\left[ Li_2({ {{\tau_+}-t} \over{1-t} })+Li_2({ {{1/{\tau_-}}-1} \over{1/t-1}})-Li_2({ {{\tau_+}-t}\over {1-t} })\right. \nonumber\\
\left.-Li_2({{1/{{\tau}_+}-1}\over {1/t-1}})\right]+\ln (1-t)\ln ({{z_{max}}\over{z_{min}}})\nonumber\\ \nonumber
-\ln ({{1-{{\tau}_+}}\over{1-{{\tau}_-}}})\ln \left[ (1-{{\tau}_+})(1-{{\tau}_-})\right] \\ 
-\ln ({{1-t/{{\tau}_+}}\over{1-t/{{\tau}_-}}})\ln \left[ (1-t/{{\tau}_+})(1-t/{{\tau}_-})\right],\\ 
{\widetilde{\Psi}}_n={\Psi}_n, \qquad  (n=1...5).
\end{eqnarray}
\begin{eqnarray}
{\cal C}^A_1&=&{1\over\sqrt{x^2-4\eta}} \left[\eta (6+x t-x {\tau}_++x+t {\tau}_++4 t-2 t^2+5 {\tau}_+)+2 \eta^2+\right.
\nonumber\\&&              \left.    {\tau}_+(x-2 x t +t +t^2 )\right];
\end{eqnarray}
\begin{eqnarray}
{\cal C}^A_2&=& (x t^2-2 t^2 {\tau}_+)/(2\eta)+ \eta (30-x-2 t+2 {\tau}_+)/2-x t-2 x {\tau}_++2 t {\tau}_+
\nonumber\\&&   +t+t^2+6 {\tau}_+;   \\
{\cal C}^A_3&=&{1\over\sqrt{x^2-4\eta}} \left[(4 x t^2 {\tau}_+-x^2 t^2-2 t^2 {\tau}_+^2)/(2\eta)+
+\eta^2 (10-{\tau}_+)    \right. \nonumber   \\
&&              \eta (-24+8 x t-2 x {\tau}_+-4 x-x^2+2 t {\tau}_++8 t-8 t^2+22 {\tau}_+-2 {\tau}_+^2)/2 \nonumber    \\
&&     \left.   -6 x t {\tau}_+-x t-x t^2-x {\tau}_++x^2 t+2 x^2 {\tau}_+
-3 t {\tau}_++5 t^2 {\tau}_+-5 {\tau}_+\right];     \\
{\cal C}^A_4&=&{1\over\sqrt{x^2-4\eta}} \left[  (-4 x t^2 {\tau}_+
+x^2 t^2+2 t^2 {\tau}_+^2)/(2\eta)+\eta (4+2 x {\tau}_+/t   +2 x t       \right.  \nonumber  \\
&&-3 x {\tau}_+-4 x+x^2/2    -10 {\tau}_+/t+3 t {\tau}_++12 t-4 t^2+3 {\tau}_++{\tau}_+^2) \nonumber \\
&&  +\eta^2 (6+5 {\tau}_+/t^2-6 {\tau}_+/t
-16/t)  -4 x t {\tau}_++x t+x t^2-3 x {\tau}_+  \nonumber \\
&& \left.  -x^2 t+2 x^2 {\tau}_++11 t {\tau}_++2 t^2 {\tau}_+\right];    \\
{\cal C}^A_5&=& (x t^2-4 t^2 {\tau}_+)/(4\eta)+ \eta (-24+3 x-10 {\tau}_+/t-12 t-2 {\tau}_+)/4 \nonumber
\\&&+\eta^2/2-x t+3 t {\tau}_+/2+6 t+5 t^2/2+5 {\tau}_+/2;    \\
{\cal D}^A_1&=&{\cal C}_1    \\
{\cal D}^A_{2\Diamond 3}&=&{1\over{4\sqrt{x^2-4\eta}}} 
\left[ \eta (-14+2 x {\tau}_+/t^2+x/t^2-4 x {\tau}_+/t-2 x/t+4 x t-2 x t^2\right.  \nonumber
\\&&    +2 x {\tau}_+-x-x^2/t^2+2 x^2/t-x^2-2 {\tau}_+/t^2-6 {\tau}_+/t-10 t {\tau}_++32 t  \nonumber
\\&&-22 t^2+4 t^3
      +18 {\tau}_+)+ \eta^2 (28+x/t^2-2 x/t+x+3 {\tau}_+/t^2-5 {\tau}_+/t  \nonumber
\\&&-14/t
      +t {\tau}_+
   \left.-14 t+{\tau}_+)+ (-4 x t {\tau}_++2 x t^2 {\tau}_++2 x 
{\tau}_++7 t {\tau}_++t^2 {\tau}_+\right.  \nonumber
\\&&-3 t^3 {\tau}_+
 \left.   -5 {\tau}_+)\right];    \\
{\cal D}^A_4&=&-{\cal C}_2-{\cal C}_3,   \\
{\cal D}^A_5&=&{\cal C}_2-{\cal C}_4,    \\
{\cal D}^A_6&=&-{1\over 2}{\cal C}_5+{1\over{4\sqrt{x^2-4\eta}}} 
\left[ (-6 x t^2 {\tau}_++x^2 t^2+2 t^2 {\tau}_+^2)/(2\eta)\right.  \nonumber\\
&&     + \eta (-35-x {\tau}_+/t+2 x/t-2 x t+4 x {\tau}_++26 x-2 x^2/t-3 x^2/2-4 {\tau}_+/t  \nonumber\\
&&     -2 t {\tau}_+-34 t-t^2-22 {\tau}_+-3 {\tau}_+^2)  
      + \eta^2 (-2+2 x/t-x+6 {\tau}_+/t+2 t) \nonumber\\
&&     \left. + (9 x t {\tau}_++12 x t+5 x t^2+5 x {\tau}_+-2 x^2 t+4 t {\tau}_+-6 t^2 {\tau}_+)\right];
\end{eqnarray}
\begin{figure}[htb]
\begin{center}
\leavevmode
\epsfxsize = 270pt
\epsfysize = 270pt
\epsfbox[1 140 578 702]{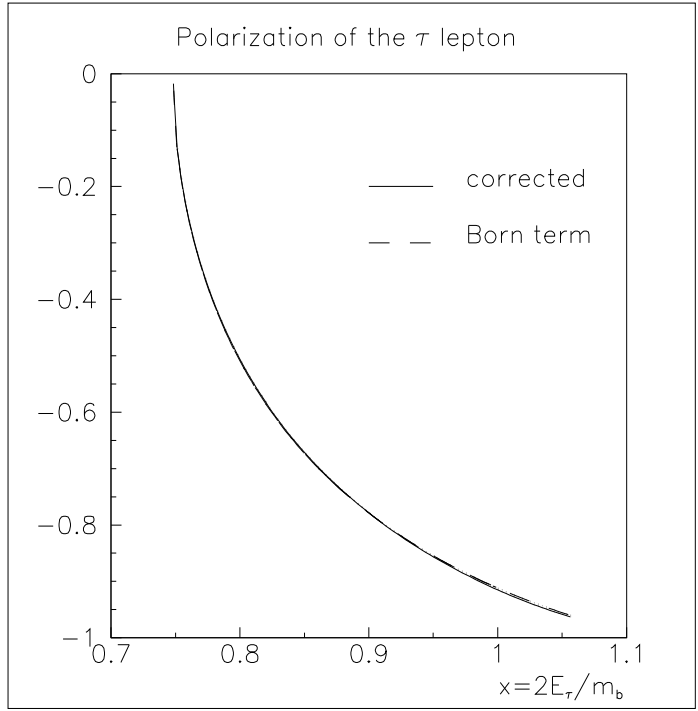}
\end{center}
\caption{Polarisation of $\tau$ lepton in the Born approximation (dashed line)
 and including the first order QCD correction (solid line) 
 as functions of the scaled $\tau$ energy x. The mass of the $b$ quark
 taken at $4.75$ GeV, $c$ quark $1.35$ GeV and the coupling constant $\alpha _s=0.2$}
\label{FIG1}
\end{figure}
 On integration over $t$, one obtains tau energy distributions according 
to the formula
\begin{equation}\label{DEFF1}
{1 \over {12{\Gamma}_0}}{{d\Gamma ^-}\over{dx}}=f_0^-(x)-{{2{\alpha}_s} \over {3\pi}}f_1^-(x).
\end{equation}
 where if we drop the superscript '$-$' we obtain the corresponding formula
for the unpolarized case.
\par
\begin{figure}[htb]
\begin{center}
\leavevmode
\epsfxsize = 250pt
\epsfysize = 250pt
\epsfbox[1 140 578 702]{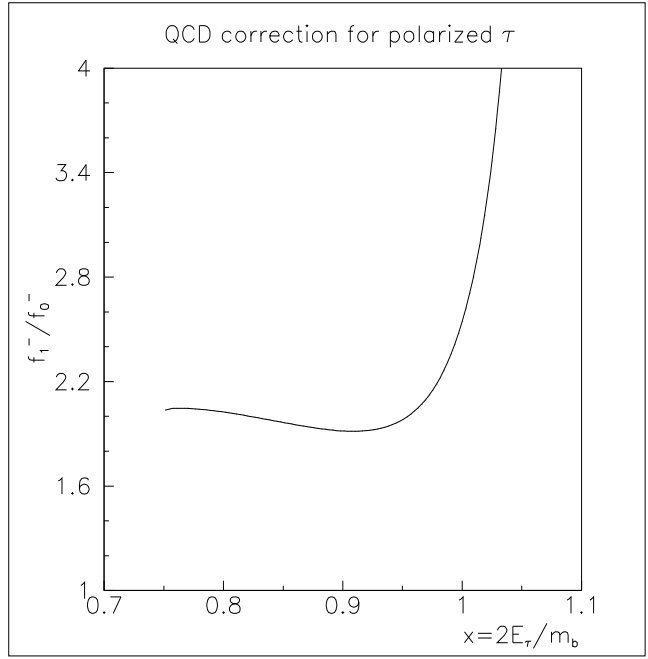}
\end{center}
\caption{ The ratio $f_1^-(x)/f_0^-(x)$ representing the radiative
correction for the negatively polarized state as dependent on the scaled
$\tau$ lepton energy x.}
\label{FIG2}
\end{figure}
\begin{figure}[htb]
\begin{center}
\leavevmode
\epsfxsize = 250pt
\epsfysize = 250pt
\epsfbox[1 140 578 702]{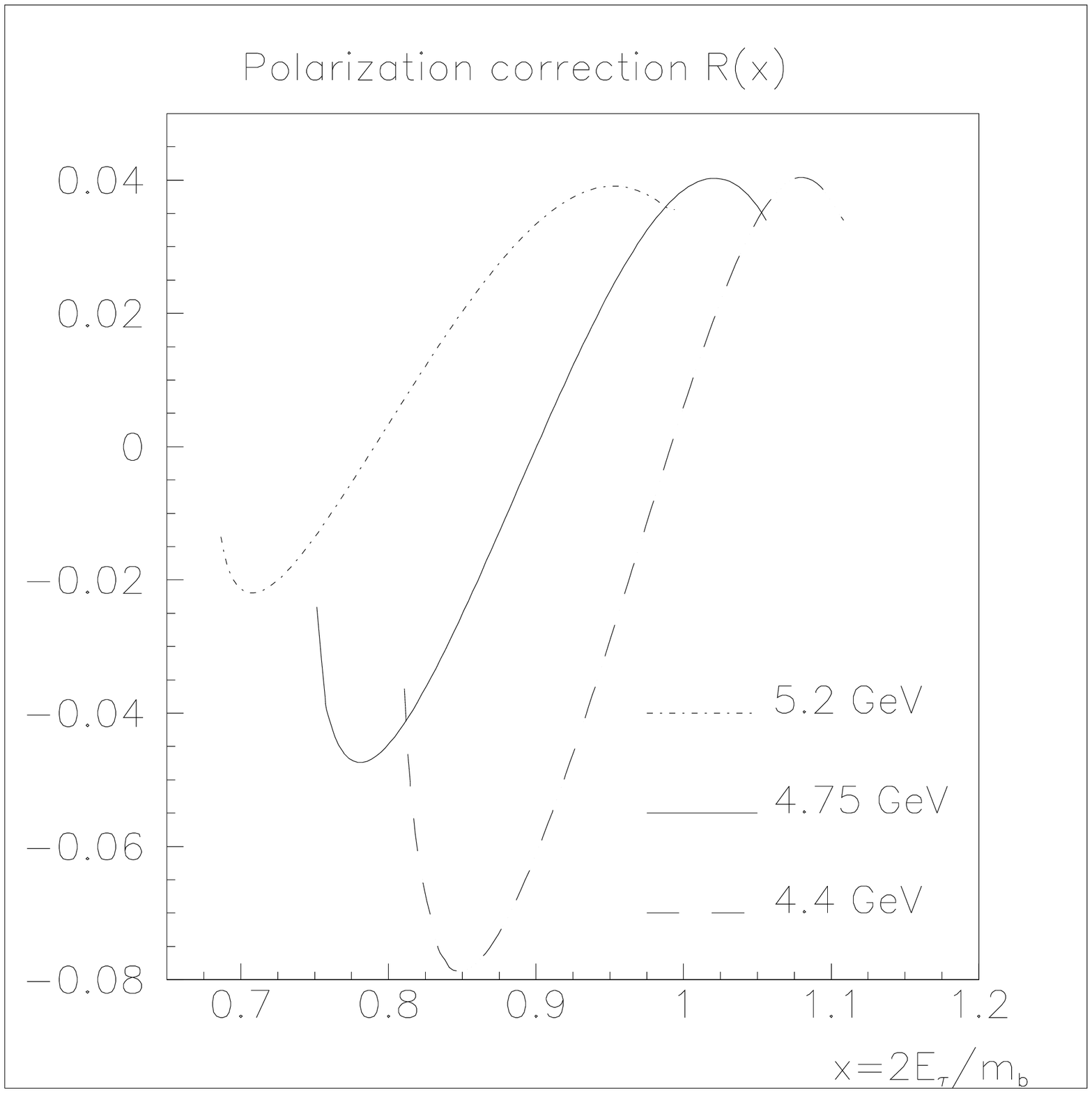} \vspace{1em} 
\end{center}
\caption{ The QCD-correction 
function R(x)  for the pole mass values of 
the $b$ quark taken to be $4.4$ GeV (dashed), $4.75$ GeV (solid) and 
$5.2$ GeV (dash-dotted) as dependent on the scaled $\tau$ lepton energy x.}
\label{FIG3}
\end{figure}

 The results are presented in Fig. \ref{FIG1}, where the polarisation
is plotted versus the $\tau$ lepton energy. Both the Born and first order 
approximation are showed. 
In order to make the correction more explicit, we
also present another diagram, where the function $R(x)$ is drawn, defined
as following:
\begin{equation}\label{PviaR}
1-P(x)=[1-P_0(x)][1+{{2\alpha _s}\over {3\pi}}R(x)]
\end{equation}
which gives the expression for $R(x)$:
\begin{equation}
R(x)={{f_1(x)}\over{f_0(x)}}-{{f_1^-(x)}\over{f_0^-(x)}},
\end{equation}
thus its meaning is how the radiative corrections differ with respect to
the state of polarisation they act on. In (\ref{PviaR}) $P_0$ denotes the
zeroth order approximation to the polarisation. The function $f_1^-(x)/f_0^-(x)$ has 
been presented in Fig. \ref{FIG2}, while Fig. \ref{FIG3} shows the
function $R(x)$, so that
one can immediately see how small the correction is in comparison to the
correction to the energy distribution.

\par
 Integrating over the charged lepton energy one can obtain its total 
polarisation as well as the corrections to which it is subject. If we 
take $m_b=4.75$ GeV as the central value for the decaying quark mass and 
$m_b=4.4$ and $m_b=5.2$ for the limits, we arrive at the following (the 
mass difference $m_b - m_c=3.4$ GeV everywhere): 
\begin{equation}
1-P=(1-P_0)\left( 1+{{2\alpha _s}\over{3\pi}}R_s +{{\lambda _1^2}\over 
{m_b^2}}R_{np}^1 + {{\lambda _2^2}\over{m_b^2}}R_{np}^2\right)
\end{equation}
with
\begin{eqnarray*}
P_0&=&-0.7388^{-0.0105}_{+0.0109}\\
R_s&=&-0.016^{-0.023}_{+0.017}\\
R_{np}^1&=&\ 0.421^{+0.027}_{-0.025}\\
R_{np}^2&=&-2.28^{-0.22}_{+0.16}\\
\end{eqnarray*}

\subsection{Moments of $\tau$ energy distribution}\label{Long:Moments}
 The moments of $\tau$ energy distribution, which are useful sources of information
on the physical parameters regarding the discussed decay  can be evaluated according
to the formula:
\begin{eqnarray}
M_n^{\pm}=\int_{E_{min}}^{E_{max}}E^n_{\tau}{{d\Gamma 
^{\pm}}\over{d{E_{\tau}}}}dE_{\tau},\\
r_n^{\pm}={{M_n^{\pm}}\over{M_0^{\pm}}},
\end{eqnarray}
where $E_{min}$ and $E_{max}$ are the lower and upper limits for $\tau$ energy and
$M_n$ include both perturbative and nonperturbative QCD corrections to $\tau$ energy
spectrum. The superscripts denote the polarisation states. Since one obviously 
has 
\begin{equation}
M_n=M_n^++M_n^-
\end{equation}
where $M_n$ stands for the unpolarized momenta, we only give the values of
the momenta for the negative polarisation case. The unpolarized distributions
were given in \cite{\JM}. The nonperturbative corrections to the charged lepton spectrum from
semileptonic $B$ decay have been derived in the HQET framework  
up to order of $1/m_b^2$ \cite{Balk:1994sz,Koyrakh:1994pq,Koyrakh:1996fr,Falk:1994gw}. The corrected heavy lepton
energy spectrum can be written in the following way:
\begin{equation}\label{HQET1}
{1 \over {12{\Gamma}_0}}{{d\Gamma}\over{dx}}=f_0(x)-{{2{\alpha}_s} \over {3\pi}}f_1(x)+{{{\lambda}_1}\over{m_b^2}}f_{np}^{(1)}(x)+{{{\lambda}_2}\over{m_b^2}}f^{(2)}_{np}(x),
\end{equation}
where ${\lambda}_1$ and ${\lambda}_2$ are the HQET parameters corresponding to the $b$
quark kinetic energy and the energy of interaction of the $b$ quark magnetic moment
with the chromomagnetic field produced by the light quark in the meson $B$. The functions
$f_{np}^{(1,2)}$ can be easily extracted from the formula (2.11) in \cite{\FLNN}.The 
formula (\ref{HQET1}) looks identically if one considers definite polarisation
state of the final $\tau$ lepton. The appropriate calculation within the
HQET scheme has also been performed \cite{\FLNN}, see formula (2.12) therein.
\par
  Following ref.\cite{Voloshin:1995cy} we expand the ratios $r_n$:
\begin{equation}
r_n^-=r_n^{(0)-}\left(1-{{2{\alpha}_s}\over {3\pi}}{\delta}_n^{(p)-}+{{\lambda _1}\over{m_b^2}}{\delta}_n^{(1)-}+{{\lambda _2}\over {m_b^2}}{\delta}_n^{(2)-}\right),
\end{equation}
where $r_n^{(0)}$ is the lowest approximation of $r_n$,
\begin{equation}
r_n^{(0)-}=({{m_b}\over 2})^n{{\int_{2\sqrt{\eta}}^{1+\eta 
-\rho}f_0^-(x)x^ndx}\over{\int_{2\sqrt{\eta}}^{1+\eta -\rho}f_0^-(x)dx}}.
\end{equation}
Each of the ${\delta}_n^{(i)-}$ is expressed by integrals of the corresponding correction
function $f^{(i)}(x)$ and the tree level term $f_0(x)$
\begin{equation}
{\delta}_n^{(i)-}={{\int_{2\sqrt{\eta}}^{1+\eta 
-\rho}f^{(i)-}(x)x^ndx}\over{\int_{2\sqrt{\eta}}^{1+\eta 
-\rho}f_0^-(x)x^ndx}}-{{\int_{2\sqrt{\eta}}^{1+\eta 
-\rho}f^{(i)-}(x)dx}\over{\int_{2\sqrt{\eta}}^{1+\eta -\rho}f_0^-(x)dx}},
\end{equation}
where the index $i$ denotes any of the three kinds of corrections discussed
above. The coefficients $\delta _n^{(i)}$~ depend only on the two ratios
of the charged lepton and the $c$~ quark to the mass $m_b$. Following 
ref.\cite{\JM} we employ the functional dependence of the form
\begin{equation}
\delta^{(i)-}_n(m_b,m_c,m_{\tau})=\delta^{(i)-}_n\left({{m_b}\over{m_{\tau}}},
{{m_c}\over{m_b}}\right).
\end{equation}
The quark masses are not known precisely so we have calculated the coefficients
in a reasonable range of the parameters, that is, $4.4$~GeV$\leq m_b \leq 5.2$~
GeV and $0.25 \leq m_c/m_b \leq 0.35$ and then fitted to them functions of
the following form:
\begin{eqnarray}\label{deltapq}
\delta (p,q)&=&a+b(p-p_0)+c(q-q_0)+d(p-p_0)^2 \nonumber\\
            &=&+e(p-p_0)(q-q_0) + f(q-q_0)^2,
\end{eqnarray}
where $p=m_b/m_{\tau},p_0=4.75$~GeV$/1.777$~GeV$=2.6730,q=m_c/m_b,q_0=0.28$
and the polynomial coefficients can be fitted for each of the $\delta 
^{(i)}_n$ separately with a relative error of less than 2\%. 
Our choice of the central values reflects the realistic masses
of quarks: $m_b=4.75$~GeV and $m_c=1.35$~GeV,for which $\delta ^{(i)}_n=a_n^{(i)}$.
\par
To bring out the difference in the extent to which 
the corrections affect the two different
polarisation states it is useful to compare these coefficients with
the ones obtained with the polarisation summed over. This can be done
along the lines suggested by the treatment of the polarisation itself,
see Eq.(\ref{PviaR}). The corresponding expansion takes the form
\begin{equation}
{{r_n^-}\over{r_n}}={{r^{-(0)}_n}\over{r_n^{(0)}}}\left[1-{{2\alpha_s}\over{3\pi}}(\delta _n^{(p)-}-\delta _n^{(p)})
+{{\lambda _1}\over{m_b^2}}(\delta _n^{(1)-}-\delta _n^{(1)})
+{{\lambda _2}\over{m_b^2}}(\delta _n^{(2)-}-\delta _n^{(2)})
\right].
\end{equation}
With these, one can readily find the actual relative correction of each kind, assuming
reasonable values of $\alpha _s,\lambda _1$~and$\lambda _2$. Here we take
$\alpha _s=0.2$, $0.15$~GeV$^2 \leq -\lambda_1 \leq 0.60$~GeV$^2$,
$\lambda_2=0.12$~GeV$^2$,
keep the $b$~quark mass fixed at $4.75$~GeV and the mass of the $c$ quark
equal to $1.35$~GeV. The corrections then read for $n=1(n=5)$
\begin{center}
\begin{tabular}{|c|c|c|}                       \hline
        & \multicolumn{2}{c|} {Correction to} \\ \hline
Type           & $r_n^-/r_n^{(0)-} $                & $r_n/r_n^{(0)}$
\\ \hline
perturbative   & $-0.00084(-0.0048)$                & $-0.0009(-0.0052)$
\\ \hline
kinetic energy & $0.008\pm 0.005(0.06\pm 0.04)  $   & $0.008\pm
0.005(0.06\pm 0.04)$ \\ \hline
chromomagnetic & $-0.0097(-0.053) $                 & $-0.0092(-0.0511)$
\\ \hline
\end{tabular}
\end{center}
\subsection{Summary}
We have calculated the one-loop QCD corrections to the longitudinal
polarisation of the $\tau$ lepton in semileptonic $B$ decays. This
calculation has been published in \cite{Jezabek:1997rk}. As was
expected from earlier results \cite{Czarnecki:1995bn}, the effect of
the correction on the polarisation is negligible, which makes it a
promising quantity for the determination of the quark
masses. Discussing the moments of the lepton energy distribution, we
have also included the nonperturbative corrections.
\pagebreak
\section{Polarisation of $\tau$ lepton with respect to $W$}\label{WPOL}
\subsection{Introduction}
 The method used in the calculation of the longitudinal polarisation
can easily be modified to give other polarisations.
 This fact matters insomuch that 
experimentally it is the polarisation along the intermediating 
$W$ boson direction that is easier to measure \cite{Rozanska:1998me}, see also
\cite{Kiers:1997zt}. 
The reason is that the direction of $\tau$ lepton can be determined
at B factories with rather poor accuracy. On the other hand the direction
of $W$ is opposite to the direction of hadrons in semileptonic B decays.
The latter can be well measured at least for the exclusive 
$B\to D\tau\bar\nu_\tau$ and $B\to D^*\tau\bar\nu_\tau$ channels
which probably contribute the dominant contribution to the inclusive
decay rate.
\par
In the calculation below, we find the tree level term and the ${\cal
O}(1/m_b^2)$ HQET corrections to the doubly differential rate with the
final $\tau$ 
lepton polarized along the momentum of the intermediate $W$ boson. The one-loop
correction is presented in a plot as it has been evaluated numerically. Also 
we give the tree level charged lepton distribution in an analytic
form. 
\par
The HQET mass power corrections to the lepton spectra in semileptonic
$B$ decays have been applied to the calculation of the polarisation of 
the lepton. The mass of the lepton had of course to be taken into
account, which was done for unpolarised distributions in
\cite{Balk:1994sz,Koyrakh:1994pq}. The kinematical complications
arising due to the massive lepton are also discussed therein. The
longitudinal polarisation was calculated in \cite{Falk:1994gw}. Some
further problems emerged in the calculation of the corrections to the
transverse polarisations found in \cite{Gremm:1995rv}. The case we
consider leads to apparent divergences, but they cancel so that no
special regularisation has to be employed.
\par
After a brief note on the kinematics in Sec.\ref{W:Kinematics} we
discuss the way the polarisation is evaluated, Sec. \ref{W:Evaluation}.
An analytic formula for the double differential distribution is
presented including the mass power corrections in
Sec. \ref{W:Anal}. Due to its complicated form, we only show numerical 
form of the lepton energy distribution in Sec.\ref{W:Spectra}. This
section contains also the one-loop perturbative corrections. All of
these corrections are presented in a plot. Finally we summarize the results.
\subsection{Kinematics}\label{W:Kinematics}
Recall that the phase space is defined by the ranges of the kinematical variables:
\begin {eqnarray}
2\sqrt{\eta}&\le x \le & 1+\eta-\rho=x_m\quad ,\\
t_{min}={\tau}_- \left(1-{\rho \over {1-{\tau}_-}}\right) &\le t\le &
{\tau}_+\left( 1-{\rho \over 1-{\tau}_+}\right)=t_{max}\quad .
\end{eqnarray}
The limits above are obtained within the parton model approximation.
They change if we allow for Fermi motion, which we must in order to be
able to discuss the HQET corrections to the decay 
widths\cite{\BKPS,\FLNN,Koyrakh:1994pq,Koyrakh:1996fr,Gremm:1995rv,Manohar:1994qn}. 
Also, contrary to
the parton model case, the energy of neutrino can vary within limits 
which depend in a non-trivial manner on the values of the 
variables $x,t$. The details of the subject 
were discussed in \cite{\BKPS},
so we will only state here that the integrations involving delta functions 
and their derivatives have the effect of confining the range of the variables
$x,t$ to that of the parton model. This is easily seen from the form
of the corrections, see Eq.(\ref{d3a}).
\subsection{Polarisation evaluation}\label{W:Evaluation}
The polarisation is found by evaluating the unpolarized decay width and 
any of the two corresponding to a definite polarisation, 
according to the definition,
\begin{equation}
P={{\Gamma ^+-\Gamma ^-}\over{\Gamma ^++\Gamma^-}}
=1-2{{\Gamma ^-}\over{\Gamma}}\quad ,
\end{equation}
where $\Gamma=\Gamma ^++\Gamma ^-$. 
In this calculation we follow the method employed when dealing with the
longitudinal polarisation of the charged lepton.
Thus in the rest frame of the decaying quark, one can decompose:
\begin{equation}\label{stolQa}
s={\cal A}Q +{\cal B}W \quad .
\end{equation}
The coefficients ${\cal A},{\cal B}$ can be evaluated using the conditions 
defining the polarisation four-vector $s$, which reduce to the following 
when the parton model value of the neutrino energy is assumed:
\begin{eqnarray}
{\cal A}_0^{\pm}&=&\mp {{t+\eta}\over\sqrt{t(x-x_-)(x_+ - x)}}\quad ,\\
{\cal B}_0^{\pm}&=&\pm {x\over\sqrt{t(x-x_-)(x_+ - x)}}\quad , 
\end{eqnarray}
where the superscripts at ${\cal {A,B}}$ denote the polarisation of the 
lepton,while
\begin{equation}
x_{\pm}=(1+\eta /t)w_{\pm} \quad .
\end{equation}
 However, 
the coefficients in the decomposition (\ref{stolQ}) need to be re-evaluated, 
taking into account the Fermi motion and working in the rest frame of 
the decaying meson.
We will presently construct the four-vector $s$ representing a charged 
lepton polarized along the direction of the $W$ boson. 
The defining properties of $s$ are
\begin{equation}\label{s1}
s^2=-1
\end{equation}
and
\begin{equation}\label{s2}
s\cdot\tau=0\ ,
\end{equation}
complemented by the relation $\vec{s} \parallel \vec{W}$. 
Since we are working in the rest frame of the decaying 
meson, the four-vector $s$ can be decomposed as
\begin{equation}\label{sdecomp}
s={\cal A}v+{\cal B}W\ ,
\end{equation}
where $v$ and $W$ stand for the four-velocity of the $B$ 
meson and the four-momentum of the intermediate $W$ boson, 
respectively. While this form automatically satisfies 
$\vec{s}\parallel \vec{W}$, the other two relations 
(\ref{s1}) and (\ref{s2}) have to be imposed, hence yielding 
the expressions for the coefficients appearing in (\ref{sdecomp}). 
With $v=(1,0,0,0)$, one readily identifies:
\begin{equation}
v\cdot\tau=x/2\ ,\qquad v\cdot\nu = x_{\nu}/2\ ,\qquad v^2=-1\ .
\end{equation}
These combined with the other dot products lead to 
the following formulae:
\begin{eqnarray}
{\cal A}_{\pm}&=&{{\mp (t+\eta)}\over{\sqrt{x(t+\eta)(x+x_{\nu})
-x^2t-(t+\eta)^2}}}\quad ,\\  
{\cal B}_{\pm}&=&{{\pm x}\over{\sqrt{x(t+\eta)(x+x_{\nu})
-x^2t-(t+\eta)^2}}}\quad .
\end{eqnarray}
The evaluation of the HQET corrections involves differentiation 
over the neutrino energy, once or twice. The denominator in the 
above expressions is easily seen to vanish at the point where 
the $W$ boson is at rest. It is known that the kinematics of 
the process, together with the delta functions and their 
derivatives, finally reduces to integration over the partonic 
phase space. Then there is one such point where the denominator 
vanishes,
\begin{eqnarray}
x&=&1-\sqrt{\rho}+{\eta\over{1-\sqrt{\rho}}}\quad ,\\
t&=& (1 -\sqrt\rho)^2 \quad .
\end{eqnarray}
One might thus raise the question of analyticity of the expressions 
obtained in this way. However, the divergences cancel and moreover 
the resulting distribution is continuous if we ignore the endpoint 
behaviour. That this is so indeed, may be verified by changing the 
variables from $t$ to the square of the three-momentum of the $W$ 
boson. Then the singularity makes its presence only on integration 
over $w_3^2$ rather than affecting the analytical structure of the 
distributions. It turns out that using this variable one obtains an 
analytic expression. This is made clear once one notices that the 
only terms that occur in the course of the calculation are the dot 
products of the four-vector $s$ and the other four-vectors. Writing 
them out explicitly,
\begin{eqnarray}
s_+\cdot v={{\tau _3 \cos \theta}\over{\sqrt{x^2-\tau _3\cos ^2\theta}}}
\quad ,\\
s_+\cdot W={{(x+x_{\nu})\tau _3\cos\theta 
- 2xw_3}\over{2\sqrt{x^2-\tau _3\cos ^2\theta}}}\quad ,
\end{eqnarray}
with
\begin{equation}
\cos\theta={{w_3^2-\eta-(x_{\nu}-x)/4}\over{w_3\tau _3}}\quad ,
\end{equation}
where the subscript denotes the polarisation direction, 
we easily verify that the triple differential distribution 
is analytic in the neutrino energy. 
 Lastly, let us note that another change of variable can be 
useful for evaluating the distribution. Namely, using the 
cosine of the angle subtended by the tau lepton and the 
neutrino\cite{Gremm:1995rv}  eliminates the singular terms from the 
double differential distribution. We have checked numerically 
that the resulting distribution is the same.
Applying now the representation (\ref{stolQa}) of the polarisation four-vector
$s$ we obtain the following formula for 
the matrix element with the lepton polarized:
\begin{eqnarray}\label{abexpW}
{\cal M}^{\pm}&=&{1\over 2}{\cal M}^{un}(\tau)
\pm {{\sqrt{\eta}}\over\sqrt{x(t+\eta)(x_{\nu}+x)-x^2t-(t+\eta)^2}}
    \left[x{\cal M}^{un}(W)\right.\nonumber \\
&&\left. -(t+\eta){\cal M}^{un}(Q)\right]\ .
\end{eqnarray}
The above expression is valid for the HQET corrections, too.
The first term on the right hand side of (\ref{abexpW}) can be calculated 
immediately once we know the result for the unpolarized case. Then the 
other terms require the formal replacement of the four-momenta 
$\tau \rightarrow W$ and $\tau \rightarrow Q$ in the argument.
\subsection{Double differential distribution}\label{W:Anal}
The polarized distribution can be written in the form,
\begin{equation}\label{HQET2}
{1\over\Gamma_0}\;
{{d\Gamma ^{\pm}}\over{\,dx\,dt}}={1\over2}F^{un}\pm
\left(\widetilde{F}
+\widetilde{F}_+ - \widetilde{F}_-\right)\ ,
\end{equation}
  where
\begin{equation}\label{G0a}
{\Gamma}_0={{G_F^2m_b^5}\over{192{\pi}^3}}|V_{CKM}|^2 \ .
\end{equation}
The first term on the right hand side of Eq.(\ref{HQET2}) stands 
for the unpolarized
distribution, given e.g. in \cite{\BKPS}, Eq.$(30)$. 
Here we will only give  the new other term:\footnote{A 
FORTRAN 
code for this formula 
is available from piotr@charm.phys.us.edu.pl } 
\begin{eqnarray}\label{LW2}
\widetilde{F}&=& \sqrt{\eta}\;{\cal W}\,
\left[ 6 f_1 + {K_b}{\cal W} 
\left( f_2 + f_3 {\cal W}^2 +
{\textstyle{3\over 2}} f_4 {\cal W}^4 \right)
+ {G_b} \left( f_5 + f_6 {\cal W}^2 \right)  \right]\ ,
\end{eqnarray}
where
\begin{eqnarray}
f_1 &=& - xt (   1 + \rho - \eta ) + 2 xt^2 
- x (   1 + \rho\eta - 2\rho + \rho^2 + \eta ) - x^2t \nonumber \\
&&+ x^2  ( 1 - \rho )
 +\, t  ( 1 + 2\rho\eta - \rho^2 )
+ t^2  ( 2\rho - \eta ) - t^3 - \rho^2\eta + \eta \ ,
  \\
f_2 &=& - 8xt  (  1 - \rho + \eta ) -16xt^2 
+8 x  ( 1 + \rho\eta - 2\rho + \rho^2 - \eta ) + 6x^2t 
\nonumber  \\&&
-\, 6x^2( 1- \rho ) + 2t ( 1 - 6\rho\eta 
- 4\rho + 3\rho^2 + 12\eta )
+ 2t^2  ( 8 - 6\rho + 3\eta ) 
\nonumber  \\&&   
+\, 6t^3
- 8\rho\eta + 6\rho^2\eta + 2\eta + 8\eta^2 \ ,
\end{eqnarray}
\begin{eqnarray}
f_3 &=& xt  (  - 10\rho\eta + 6\rho\eta^2 
               + 14\rho^2\eta + 9\rho^2\eta^2 
               - 6\rho^3\eta + 2\eta 
                + 9\eta^2 - 4\eta^3 )
\nonumber  \\&&
      +\, xt^2  ( 1 - 9\rho\eta^2 - 5\rho 
                + 18\rho^2\eta + 7\rho^2 
                - 3\rho^3 + 6\eta - 13\eta^2 )
\nonumber  \\&&
       +\, xt^3  ( 1 - 18\rho\eta - 2\rho + 
                  9\rho^2 - 14\eta + 3\eta^2 )
       - xt^4  (  5 + 9\rho - 6\eta )
\nonumber  \\&&
    +\, 3xt^5    + x  (  - 5\rho\eta^2 + 4\rho\eta^3 
              + 7\rho^2\eta^2 - 3\rho^3\eta^2 
                  + \eta^2 + 4\eta^3 )
\nonumber  \\&&
       +\, x^2t  ( 1 + 14\rho\eta + 2\rho\eta^2 
                  - 3\rho - \rho^2\eta + 3\rho^2 
                  - \rho^3 - 13\eta + 8\eta^2 )
\nonumber  \\&&
       +\, x^2t^2  (  - 6 + 5\rho\eta 
                    + 6\rho + 15\eta - \eta^2 )
       + x^2t^3  ( 7 + 3\rho - 3\eta )
        - 2x^2t^4
\nonumber  \\&&
        +\, x^2 \eta  (  - 3\rho + 8\rho\eta
                + 3\rho^2 -  \rho^2\eta
                - \rho^3  - 7\eta + 1 )
        + x^3t  ( 1 - \rho )^2 
        - x^3t^2  \eta 
\nonumber   \\&&
        -\,  x^3t^3 
        + x^3 \eta  ( 1 - \rho )^2 \ ,
\\
f_4 &=& x^2t  (  - 6\rho\eta^2 - 2\rho^2\eta^3 
          + 6\rho^3\eta^2 + 4\rho^3\eta^3 
          - 3\rho^4\eta^2 + 3\eta^2 - 2\eta^3 )
\nonumber  \\&&
    +\, x^2t^2  (  - 6\rho\eta - 2\rho\eta^3 - 6\rho^2\eta^2 
             - 6\rho^2\eta^3 + 6\rho^3\eta + 12\rho^3\eta^2 
             - 3\rho^4\eta + 3\eta - 6\eta^2  ) 
\nonumber  \\&&
    +\, x^2t^3  ( 1 - 6\rho\eta^2 + 4\rho\eta^3 - 2\rho 
              - 6\rho^2\eta - 18\rho^2\eta^2 + 12\rho^3\eta 
              + 2\rho^3 - \rho^4 - 6\eta + 2\eta^3 )
\nonumber  \\&&
    +\, x^2t^4  (  - 2 - 6\rho\eta + 12\rho\eta^2 - 18\rho^2\eta 
                - 2\rho^2  + 4\rho^3 + 6\eta^2 - \eta^3 )
\nonumber  \\&&
    +\, x^2t^5  ( 12\rho\eta - 2\rho - 6\rho^2 + 6\eta - 3\eta^2 )
    + x^2t^6  ( 2 + 4\rho - 3\eta )
    - x^2t^7 
\nonumber  \\&&
    +\, x^2  (  - 2\rho\eta^3 + 2\rho^3\eta^3 - \rho^4\eta^3 + \eta^3 )
    + x^3t  ( 8\rho\eta + \rho\eta^2 + 2\rho\eta^3 - 12\rho^2\eta 
              + \rho^2\eta^2
\nonumber  \\&&
          \qquad     + 3\rho^2\eta^3 + 8\rho^3\eta 
              - \rho^3\eta^2 
              - 2\rho^4\eta - 2\eta - \eta^2 + 3\eta^3 )
\nonumber  \\&&
    +\, x^3t^2  (  - 1 - \rho\eta + 6\rho\eta^2 - 3\rho\eta^3 
              + 4\rho - \rho^2\eta + 9\rho^2\eta^2 - 6\rho^2 
              + \rho^3\eta  + 4\rho^3
\nonumber  \\&&
              \qquad  - \rho^4 + \eta + 9\eta^2 - 3\eta^3 )
\nonumber  \\&&
    +\, x^3t^3  ( 1 + 6\rho\eta - 11\rho\eta^2 - \rho + 9\rho^2\eta 
              - \rho^2 + \rho^3 + 9\eta - 11\eta^2 + \eta^3 )
\nonumber  \\&&
    +\, x^3t^4  ( 3 - 13\rho\eta + 2\rho + 3\rho^2 - 13\eta + 4\eta^2 )
    + x^3t^5  (  - 5 - 5\rho + 5\eta )
    + 2x^3t^6 
\nonumber  \\&&
    +\, x^3  ( 4\rho\eta^2 + \rho\eta^3 - 6\rho^2\eta^2 + \rho^2\eta^3 
          + 4\rho^3\eta^2 - \rho^3\eta^3 - \rho^4\eta^2 - \eta^2 
          - \eta^3 )
\nonumber  \\&&
    +\, x^4t  (  - 6\rho\eta + 2\rho\eta^2 + 6\rho^2\eta + 
             \rho^2\eta^2 - 2\rho^3\eta + 2\eta - 3\eta^2 )
\nonumber  \\&&
    +\, x^4t^2  ( 1 + 4\rho\eta + \rho\eta^2 - 3\rho 
               + 2\rho^2\eta + 3\rho^2  - \rho^3 - 6\eta + 3\eta^2 )
\nonumber  \\&&
    +\, x^4t^3  (  - 3 + 2\rho\eta + 2\rho + \rho^2 + 6\eta - \eta^2 )
    + x^4t^4  ( 3 + \rho - 2\eta )
\nonumber  \\&&
    -\, x^4t^5 
    + x^4  (  - 3\rho\eta^2 + 3\rho^2\eta^2 - \rho^3\eta^2 + \eta^2 ) \ ,
  \\
f_5 &=& 4xt  ( 3 + 5\rho - 5\eta )
       - 40xt^2 
       + 4x  ( 1 + 5\rho\eta - 6\rho + 5\rho^2 + \eta ) + 10x^2t 
\nonumber  \\&&
       -\, 2x^2  (  1  - 5\rho )
       - 6t  (  1 + 10\rho\eta - 5\rho^2 )
       - 2t^2  (  4 + 30\rho - 15\eta ) + 30t^3 
\nonumber  \\&&
       +\, 30\rho^2\eta - 6\eta + 8\eta^2   \  ,
  \\
f_6 &=& 
         xt  (  - 18\rho\eta - 2\rho\eta^2 
             + 6\rho^2\eta - 15\rho^2\eta^2
           + 10\rho^3\eta  + 2\eta + 17\eta^2 - 4\eta^3 )
\nonumber  \\&&
   +\, xt^2  ( 1 - 16\rho\eta + 15\rho\eta^2 - 9\rho - 30\rho^2\eta 
             + 3\rho^2 + 5\rho^3 + 10\eta - \eta^2 )
\nonumber  \\&&
   +\, xt^3  ( 1 + 30\rho\eta - 10\rho - 15\rho^2 + 10\eta - 5\eta^2 )
   + xt^4  ( 7 + 15\rho - 10\eta )
\nonumber  \\&&
   -\, 5xt^5 
   + x  (  - 9\rho\eta^2 + 4\rho\eta^3 + 3\rho^2\eta^2 
          + 5\rho^3\eta^2 + \eta^2 + 8\eta^3 )
\nonumber  \\&&
   +\, x^2t  (  - 1 + 30\rho\eta - 10\rho\eta^2 + 7\rho + 5\rho^2\eta 
           - 11\rho^2 + 5\rho^3 - 11\eta - 4\eta^2 )
\nonumber  \\&&
   +\, x^2t^2  (  - 2 - 25\rho\eta + 18\rho - 19\eta + 5\eta^2 )
   - 15x^2t^3  (  1 + \rho - \eta )
\nonumber  \\&&
   +\, 10x^2t^4 
   + x^2  ( 7\rho\eta + 12\rho\eta^2 - 11\rho^2\eta + 5\rho^2\eta^2 
             + 5\rho^3\eta - \eta - 9\eta^2 )
\nonumber  \\&&
   +\, x^3t  ( 1 - 6\rho + 5\rho^2 + 8\eta )
   + x^3t^2  ( 8 - 5\eta )
   - 5x^3t^3 
\nonumber  \\&&
    +\, x^3  (  - 6\rho\eta + 5\rho^2\eta + \eta )  \ ,
\end{eqnarray}
\begin{eqnarray}
\widetilde{F}_{\pm}&=& \sqrt{\eta}\; {\cal W}_\pm\, 
\left\{ \left[
{K_b} \left( h_{1,\pm} + h_{2,\pm} {{\cal W}_\pm}^2 \right) 
+  {G_b} h_{3,\pm} \right] 
{\delta (z_{\pm})} \right.\nonumber\\
&&\left.+ {K_b} h_{4,\pm} {\delta^\prime(z_{\pm})} \right\}\ ,
\end{eqnarray}
where
\begin{eqnarray}
h_{1,\pm} &=& 
        -8xt\eta + 8x\eta^2 + 4x^2t + 12x^2\eta + 2x^3t 
              - 2x^3\eta - 2x^4 -16t\eta - 16\eta^2
\nonumber\\&&
       -\, 4\sigma_{\pm}\;  ( 6xt - 2x\eta + 3x^2t + x^2\eta 
              - x^3 - 8t\eta - 4t^2 - 4\eta^2 )
\nonumber\\&&
       -\, 8\sigma_{\pm}^2\;  ( 3xt - x\eta - 5x^2 + 4t + 4\eta )
       \; +\; 16\sigma_{\pm}^3  ( 3x + t + \eta )  \ ,
\\
h_{2,\pm} &=& 
        2\sigma_{\pm}\; x\, ( 8t\eta^2 + 16t^2\eta + 8t^3 - 4xt\eta^2
             + 4xt^3 - 8x^2t\eta - 6x^2t^2 - 2x^2\eta^2 
\nonumber\\&& 
            -\, x^3t^2 + x^3\eta^2 + x^4t + x^4\eta )
         \; +\; 4\sigma_{\pm}^2\; x\, (  - 4t\eta^2 - 8t^2\eta - 4t^3
\nonumber\\&& 
            - 12xt\eta - 8xt^2 - 4x\eta^2 + 2x^2t\eta 
            - x^2t^2 
           +\, 3x^2\eta^2 + 3x^3t + 3x^3\eta )
\nonumber\\&&
       +\, 8\sigma_{\pm}^3\; x\, (  - 4t\eta - 2t^2 - 2\eta^2 
            + 4xt\eta + xt^2 + 3x\eta^2 + 3x^2t + 3x^2\eta )
\nonumber\\&&
       +\, 16\sigma_{\pm}^4\; x\, ( 2t\eta + t^2 + \eta^2 
             + xt + x\eta )  \  ,
\\
h_{3,\pm} &=& 
       - 16xt\eta - 8xt + 8xt^2 + 8x\eta + 8x\eta^2 - 8x^2t 
              + 8x^2\eta + 24t^2 - 24\eta^2
\nonumber\\&&
       +\, 4\sigma_{\pm}\; (  - 10xt - 14x\eta - 5x^2t + 5x^2\eta 
             - 4x^2 + 5x^3 + 12t - 4t^2 + 12\eta
\nonumber\\&&
         +\, 4\eta^2 )
       \; +\; 16\sigma_{\pm}^2\; ( 5x\eta - 2x + 5x^2 - 9t - 9\eta )
\nonumber\\&&       \;+\; 80\sigma_{\pm}^3\; ( x + t + \eta ) \  ,
   \\
h_{4,\pm} &=& 
        4\sigma_{\pm}\; ( 4xt\eta - 4xt^2 + 6x^2t + 2x^2\eta 
             + x^3t - x^3\eta - x^4 - 8t\eta - 8t^2 )
\nonumber\\&&
       +\, 8\sigma_{\pm}^2\; ( 8xt + 4x\eta + x^2t - 3x^2\eta 
             - 3x^3  + 4t\eta + 4t^2 )
\nonumber\\&&
       -\, 16\sigma_{\pm}^3\; ( xt + 3x\eta + 3x^2 - 2t - 2\eta )
       \;-\;  32\sigma_{\pm}^4\;  ( x + t + \eta ) \  ,
\end{eqnarray}
and
\begin{equation}
{\cal W}_{\pm}={1\over\sqrt{x(t+\eta)(2\sigma_{\pm}+x)-x^2t-(t+\eta)^2}}\ ,
\end{equation}
\begin{equation}
{\cal W}={1\over\sqrt{x(t+\eta)(\xnuz+x)-x^2t-(t+\eta)^2}} \ ,
\end{equation}
\begin{equation}
\sigma _{\pm}=(t-\eta)/(2\tau _{\pm})\ ,
\qquad z_{\pm}=1+t-\rho-x-2\sigma _{\pm}\ .
\end{equation}
The parameters $K_b,G_b$, representing the kinetic energy 
and the chromomagnetic
energy, are defined according to \cite{Manohar:1994qn}
\subsection{Lepton energy distribution}\label{W:Spectra}
As regards the HQET correction terms, we only give the energy distribution
in the form of a diagram evaluated numerically. Beneath we also give the
Born level approximation analytically. The analytic formulae for the polarized 
distribution can be simplified if we split the kinematical range of $y$ into
two parts, separated by the value of the charged lepton energy where the 
virtual $W$ boson can stay at rest. This value is
\begin{equation}
x_W = 1 -\sqrt{\rho}+{\eta\over{1-\sqrt{\rho}}}\quad .
\end{equation}
In the formulae below, the superscripts $A,B$ refer to the appropriate
regimes:
\begin{eqnarray}
x < x_W && \qquad {\rm region}\ A \ ,\\
x > x_W && \qquad {\rm region}\ B \ .
\end{eqnarray}
The energy distribution of polarized $\tau$ lepton reads,
\begin{equation}
{{d\Gamma ^{\pm}}\over{dx}}=12{\Gamma}_0 
\left[{1\over 2}f(y)\pm \Delta f(x)\right]\ .
\end{equation}
The function $f(x)$ represents the unpolarized case,
\begin{equation}
f(x)={1\over 6}\zeta ^2 \sqrt{x^2-4\eta}\, \left\{ \zeta 
\left[x^2-3y(1+\eta)+8\eta\right]
+(3x-6\eta)(2-x)\right\}\quad ,
\end{equation}
with
\begin{equation}
\zeta=1-{\rho\over{1+\eta -x}}\quad .
\end{equation}
The function $\Delta f(x)$ reads,
\begin{equation}
\Delta f(x)= {3\over8}\sqrt{\eta|x-1|}\;\phi _1\Psi
\;+\; {1\over 4}\eta\;\phi_2\ ,
\label{eq:deltaf}
\end{equation}
with  
\begin{eqnarray}
\phi _1&=& -5{\lambda}^3/(x-1)^4
  +3{\lambda}(4{\eta}-{\lambda}-{\lambda}^2)/(x-1)^3 +\, 
(4{\eta}{\lambda}-4{\eta}+{\lambda} \nonumber \\
&& +7{\lambda}^2+{\lambda}^3)/(x-1)^2
  +(-1+4{\eta}{\lambda}-28{\eta}+15{\lambda}-{\lambda}^2
-{\lambda}^3)/(x-1)\nonumber \\
&&  -1+12x{\eta}-11x{\lambda}+7x-x^2+12{\eta}{\lambda}
-24{\eta}+14{\lambda}-11{\lambda}^2 \  ,
\end{eqnarray}
\begin{eqnarray}
\phi _2^A&=& \sqrt{x^2-4\eta}\, \left[\, 15{\lambda}^2{\xi}/(x-1)^3 
 +(10{\eta}{\lambda}{\xi}^2-16{\eta}{\xi}+24{\lambda}{\xi}
-10{\lambda}{\xi}^2 \right. \nonumber \\
&&   -20{\lambda}-6{\lambda}^2{\xi})/(x-1)^2
 +(-4-14{\eta}{\lambda}{\xi}^2-48{\eta}{\xi}+66{\eta}{\xi}^2
-24{\eta}{\xi}^3 \nonumber \\
&&  +8{\eta}^2{\xi}^3-76{\lambda}{\xi}+14{\lambda}{\xi}^2+48{\lambda}
  +3{\lambda}^2{\xi}+25{\xi}-26{\xi}^2+8{\xi}^3)/(x-1) \nonumber \\
&& \left.
  +3-3x+57{\eta}{\xi}-22{\eta}{\xi}^2-12{\lambda}{\xi}-21{\lambda}
  +34{\xi}-18{\xi}^2+8{\xi}^3\, \right] \ ,
\end{eqnarray}
\begin{eqnarray}
\phi _2^B&=&15{\zeta}{\lambda}^2/(x-1)^3
 +  ( 60{\eta}{\zeta}{\lambda} - 16
    {\eta}{\zeta} - 30{\eta}{\zeta}^2{\lambda} 
- 16{\zeta}{\lambda} - 21{\zeta}{\lambda}^2\nonumber\\
&&          + 10{\zeta}^2{\lambda} )/(x-1)^2
 +    (  - 104{\eta}{\zeta}{\lambda}
  - 84{\eta}{\zeta} + 52{\eta}{\zeta}^2{\lambda} 
+ 122{\eta}{\zeta}^2 - 40{\eta}{\zeta}^3\nonumber\\
&& + 160{\eta}^2{\zeta} - 160{\eta}^2{\zeta}^2 + 40
         {\eta}^2{\zeta}^3 - 24{\zeta}{\lambda} 
+ 9{\zeta}{\lambda}^2 + 17{\zeta} - 4
         {\zeta}^2{\lambda}- 22{\zeta}^2  \nonumber\\
&& + 8{\zeta}^3 )/(x-1) +  18 - 29x{\eta} 
+ 27x{\lambda} - 21x + 3x^2 + 46{\eta}{\zeta}{\lambda} 
   - 59{\eta}{\zeta} \nonumber\\
&& - 14{\eta}{\zeta}^2{\lambda} + 78{\eta}{\zeta}^2 
- 16{\eta}{\zeta}^3 - 71{\eta}{\lambda} + 59{\eta}
          - 43{\eta}^2{\zeta} + 26{\eta}^2{\zeta}^2 
- 8{\eta}^2{\zeta}^3\nonumber\\
&& + 46{\eta}^2 - 6{\zeta}{\lambda} - 3{\zeta}{\lambda}^2 
+ 69{\zeta} - 6{\zeta}^2{\lambda} 
 - 52{\zeta}^2 + 16{\zeta}^3 - 42{\lambda} + 24{\lambda}^2 \  ,
\end{eqnarray}
where
\begin{equation}
 \xi = 2 - \zeta\ ,\qquad 
\lambda = \rho + \eta \quad .
\end{equation}
The function $\Psi$ can be written in the form,
\begin{equation}
\Psi =\left\{
\begin{array}{ll}
\arccos \omega _{min} \, - \, \arccos \omega _{max}\,  ,\quad & x<1 \\
 \mbox{arcosh}\, \omega _{max} -\mbox{arcosh}\, \omega _{min} \, ,\quad  & x>1
\end{array}\right.   
\end{equation}
with
\begin{equation}
\omega _{min,max}={{2(x-1)t_{min,max}+x(x_m -x)-2\eta}
\over{x\sqrt{(x_m -x)^2+4\eta\rho}}} \quad .
\end{equation}
Due to terms containing inverse powers of $(x-1)$ the expression 
(\ref{eq:deltaf}) for $\Delta f(x)$ is apparently divergent
at $x=1$. However, expanding $\Delta f(x)$ in powers of $(x-1)$
for $x<1$ and $x>1$ one can check that this function is regular
at $x=1$.
\begin{figure}
\begin{center}
\epsfig{file=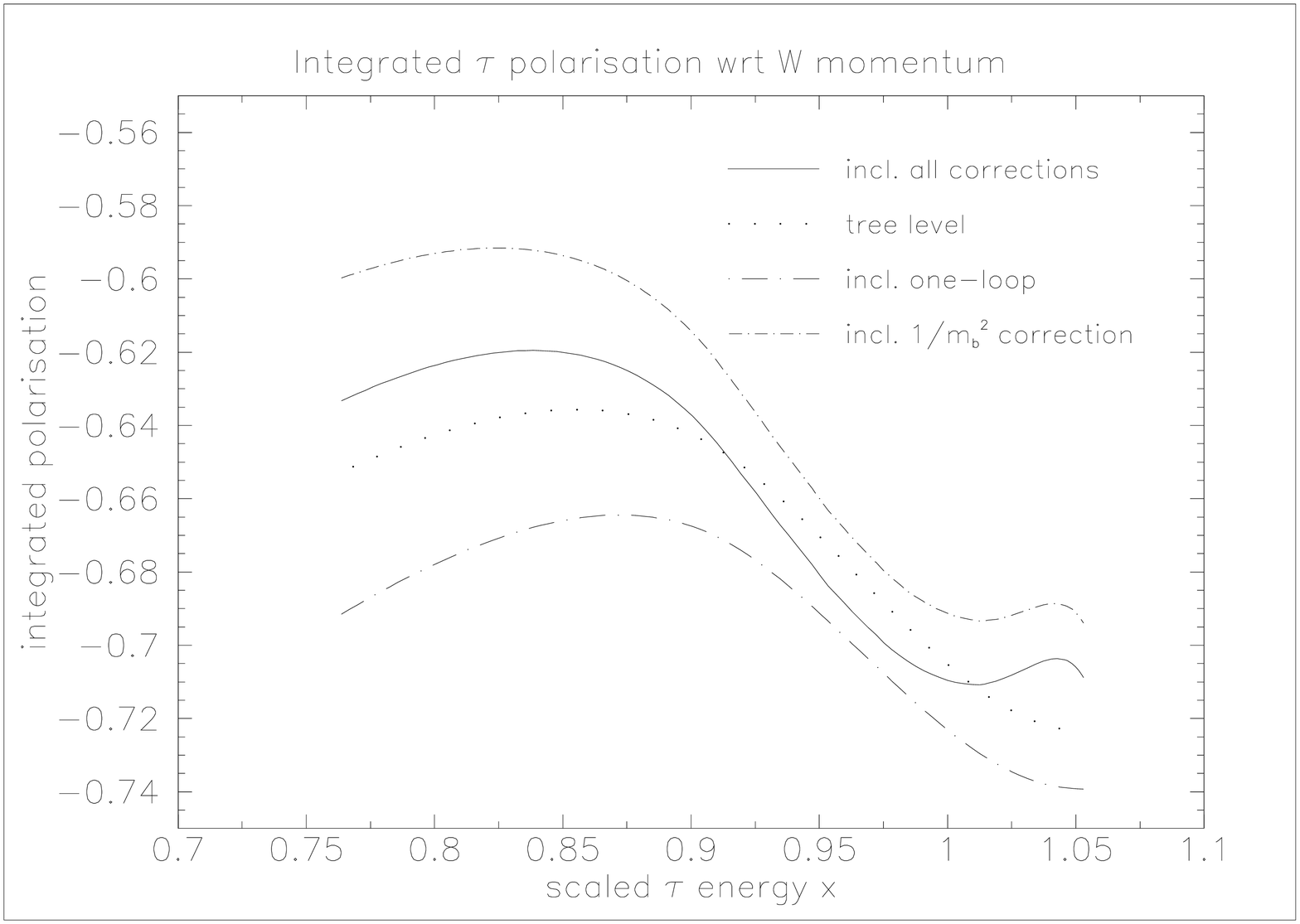,height=200pt,width=250pt}
\end{center}
\caption{Integrated polarisation of $\tau$ lepton along the direction 
of virtual W 
in the Born approximation (dot-dashed), including HQET corrections 
(long-dashed), perturbative corrections (short-dashed) and with both 
corrections included (solid)
 as functions of the scaled $\tau$ energy $x$. The mass of the $b$ quark
 taken to be $4.75$ GeV, $c$ quark $1.35$ GeV. The strong coupling
constant $\alpha_s=0.2$.}
\label{cwall}  
\end{figure}
\begin{figure}
\begin{center}
\leavevmode
\epsfig{file=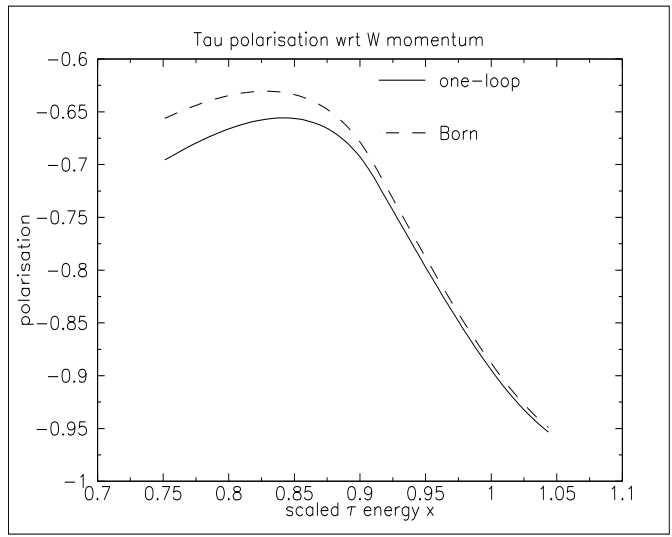,width=250pt,height=200pt}
\end{center}
\caption{Polarisation of $\tau$ lepton along the direction 
of virtual W 
in the Born approximation (dashed) and including one loop corrections (solid)
 as functions of the scaled $\tau$ energy $x$. The mass of the $b$ quark
 taken to be $4.75$ GeV, $c$ quark $1.35$ GeV. $\alpha_s=0.2$.}
\label{FIG1a}  
\end{figure}
\par
The HQET contribution to the decay distributions is known to render them 
unreliable near the endpoint values of the tauon energy. This ambiguity 
reveals itself in the polarisation as well. Similar problems appear 
also in calculations of perturbative corrections\cite{Jezabek:1989ja,Czarnecki:1994pu}.
All these problems are cured if instead of distributions their moments
are considered\cite{Czarnecki:1994gt,Voloshin:1995cy,Czarnecki:1995bn}. In the case of $\tau$ polarisation 
a better defined quantity is 
the integrated polarisation
\begin{equation}
P_{int}(\bar x)=
\int _{x_{min}}^{\bar x}\, {dx}\, 
\left(\; {d\Gamma^+\over dx}\; - \; {d\Gamma^-\over dx} \;\right)\; 
\left/ \;
\int _{x_{min}}^{\bar x}\, {dx}\,
\left(\; {d\Gamma^+\over dx}\; + \; {d\Gamma^-\over dx} \;\right) \right.
\label{eq:intpol}
\end{equation}
where both the lowest-order perturbative and the HQET terms are included.
In Fig.\ref{cwall} the integrated polarisation 
is shown as a function of the scaled energy $x$ of the $\tau$ lepton.
The lowest order prediction corresponds to the dot-dashed line and the
solid line is obtained including HQET and one-loop perturbative corrections.
The latter have been evaluated numerically. The diagram also shows the effect of the perturbative corrections only. The perturbative correction to the 
polarisation is regular so there is no need to integrate over the scaled energy
$x$. Fig. \ref{FIG1a} shows the differential polarisation at tree level and including the one-loop correction. It is remarkable that this correction, though
smaller in weight than the correction to the decay rate, is not suppressed 
to the extent that it is for the longitudinal polarisation. On the other hand,
we see the two polarisations coincide at the endpoint where the energy of 
the tauon is large so that its direction is almost equal to that of
the pair of leptons. This is of course expected. 
\par
On integration over the whole range of the charged lepton energy one 
arrives at 
the total polarisation at the tree level corrected for the $O(1/m_b^2)$ 
effects as predicted by HQET. 
For $ m_b = 4.75$ GeV and $ m_c = 1.35$ GeV, we obtain
\begin{equation}
P=-0.7235+4.21{{K_b}\over{m_b^2}}+1.48{{G_b}\over{m_b^2}}\quad .
\end{equation}
Taking $K_b=0.15$ GeV$^2$, $G_b=-0.18$ GeV$^2$ and $\alpha_s=0.2$ we obtain $P=-0.709$.
\par
Although we are mostly concerned with the tau lepton polarisation here,
the formulae derived in the present work
may well be used in evaluating the polarisation of the light
leptons. Interestingly, in the limit of a vanishing mass of the charged 
lepton the polarisation falls to zero
apart from the endpoints, c.f. (\ref{eq:deltaf})  and (\ref{LW2}). 
It is due to the chiral $V-A$ structure of the weak charged current that, 
according to Eq.(\ref{Kliniowy}), the decay widths with a definite 
polarisation differ by a term proportional to $m_{\tau}s^\mu$. The 
polarisation four-vector of the charged lepton can be decomposed as follows: 
\begin{equation}
s^{\mu} = \left(\, s^0,\vec{s}\, \right) =
\left(\,{p\over m}\sqrt{1-(\vec{s}_{\perp})^2},
\vec{s}_{\perp},{E\over m}\sqrt{1-(\vec{s}_{\perp})^2}\,\right)\quad ,
\label{eq:smu}
\end{equation} 
where $\vec{s}_{\perp}$ is understood to mean the part 
of the three-vector $\vec{s}$ perpendicular to the direction
of the charged lepton. The quantities 
$E$ and $p$ denote, respectively, the energy and the three-momentum 
value of the charged lepton. 
This form can easily be seen to meet the definition of 
the polarisation four-vector.
As the lepton mass approaches zero Eq.(\ref{eq:smu})  gives 
\begin{equation}
ms^{\mu}\approx \sqrt{1-(\vec{s}_{\perp})^2}\; {\tau}^{\mu}
\;+\; m\, \left(0,\vec{s}_{\perp},0\right)\quad .
\label{eq:msmu}
\end{equation}
However, if we want to keep the angle subtended by the 
polarisation vector and the lepton momentum constant 
the parallel part of the polarisation should be proportional
to the perpendicular one, thereby forcing the factor of 
$\sqrt{1-(\vec{s}_{\perp})^2}$ to be of order of $m/E$.
Then the r.h.s of
Eq.(\ref{eq:msmu}) tends to zero for $m\to 0$. 
For the vanishing charged lepton mass the polarisation can be 
non-zero only where the virtual $W$ boson is collinear with the charged 
lepton. In general the contribution to polarisation 
is appreciable only for $W$ direction within the cone defined by
the condition
\begin{equation}
|\vec{s}_{\perp}|/|\vec s| = {\cal O}(m/E) \quad .
\label{eq:cond}
\end{equation}
In particular this happens if $p$ is much larger than the energy 
of the neutrino. For semitauonic B decays the condition (\ref{eq:cond}) 
is satisfied in the whole phase space and the resulting polarisation is
fairly large. 
\subsection{Summary}
We have calculated the Born term and HQET first power correction to
the polarisation of the $\tau$ meson with respect to the momentum of
the intermediate $W$ boson. They are presented in an analytical
form. Apart from that, the one-loop perturbative corrections have been 
evaluated numerically. In contrast to the longitudinal one, this
polarisation receives noticeable corrections from perturbation
expansion. All of the considered contributions have been presented in
a diagram. 
\pagebreak
\section{Summary}
We have calculated a few quantities characterising the semileptonic decay
of the $B$ mesons. Perturbative and nonperturbative corrections have been
found to the double distribution in terms of hadronic mass and electron energy
for the decay $B\rightarrow X_u$ and to the polarisation of the $\tau$ lepton 
in the decay $B\rightarrow X_c$. 
\par
The longitudinal polarisation of the $\tau$ lepton receives only a tiny 
correction at one-loop level and it is nearly equal to the Born result
to an excellent approximation. As such, this quantity may be of use in 
determining the quark mass differences.
\par
The polarisation of $\tau$ is best measured with respect to the 
momentum of the intermediate $W$ boson, so we have studied the appropriate
distributions. We have found the leading HQET corrections of order $1/m_b^2$ 
analytically as well as computed numerically the one-loop perturbative 
corrections. The latter have turned out not to be negligible, contrary 
to the situation with the longitudinal polarisation.
\par
The problem of measuring the $V_{ub}$ matrix element has been addressed.
We have proposed a quantity to be compared with experimental data and 
found the theoretical prediction for it including one-loop perturbative
and leading twist nonperturbative corrections. The estimated error on the
extracted value is $10\%$.
\par
The results presented in this paper, save for the HQET corrections to the
longitudinal polarisation, have been found for the first time in the
papers \cite{Jezabek:1997rk}-\cite{Jezabek:1999mn}.
\section{Acknowledgements}
I would like to thank my advisor, Professor Marek Je\.zabek, for the guidance 
and many discussions. Thanks go also to Professor Maria R\'o\.za\'nska for
pointing out the need for calculating the $\tau$ polarisation, and Professor 
Thomas Mannel for elucidating aspects of HQET.

\pagebreak
\par

\bibliography{spires}
\bibliographystyle{aip}
\end{document}